\documentclass[a4paper]{aa} 
\usepackage{graphics} 
\usepackage{graphicx} 
\usepackage[]{pict2e}
\usepackage{color, colortbl}
\usepackage{tikz}
\usepackage{pgfplots}
\usepackage{mathtools,amssymb}

\pgfplotsset{compat=1.16}

\usepackage{txfonts}
\def\ammo{\rm NH_3}
\def\dammo{\rm NH_2D}

\def\diaz{\rm N_2H^+}
\def\ddiaz{\rm N_2D^+}
\def\htwo{\rm H_2}
\def\htwodplus{\rm H_2D^+}
\def\dtwohplus{\rm D_2H^+}
\def\dthreeplus{\rm D_3^+}
\def\hthreeplus{\rm H_3^+}
\def\toththree{\widetilde{\rm H_3^+}}
\def\dcoplus{\rm DCO^+}
\def\hcoplus{\rm HCO^+}
\def\meth{\rm CH_3OH}

\def\percc{\rm cm^{-3}}
\def\htwodplusline{\rm H_2D^+(1_{10}-1_{11})}
\def\kms{\rm km\,s^{-1}}
\def\ms{\rm m\,s^{-1}}
\def\Kkms{\rm K\,km\,s^{-1}}
\def\persqcm{\rm cm^{-2}}
\def\pers{\rm s^{-1}}

\thicklines

\begin{document}

\title{A low cosmic-ray ionisation rate in the prestellar core Ophiuchus/H-MM1.}
\subtitle{Mapping of the molecular ions ortho-$\htwodplus$, $\diaz$,
  and $\dcoplus$
\thanks{This publication is based on data acquired with the Atacama
    Pathfinder EXperiment (APEX). APEX is operated by the European
    Southern Observatory (ESO) on behalf of the Max-Planck-Institut
    für Radioastronomie (MPIfR).}}

   \author{J. Harju
          \inst{1,2}
          \and
          C. Vastel \inst{3}
          \and
          O. Sipil{\"a} \inst{1}
                \and 
          E. Redaelli \inst{1}
          \and
          P. Caselli \inst{1}
          \and
          J. E. Pineda \inst{1}
                    \and
          A. Belloche \inst{4}
          \and
          F. Wyrowski \inst{4}
          }

 \institute{Max-Planck-Institut f{\"u}r extraterrestrische Physik, Gie{\ss}enbachstra{\ss}e 1, D-85748 Garching, Germany 
   \and
   Department of Physics, P.O. Box 64, FI-00014, University of Helsinki, Finland
   \and
   IRAP, Universit{\'e} de Toulouse, CNRS, CNES, UPS, Toulouse, France
   \and
    Max-Planck-Institut f{\"u}r Radioastronomie,
    Auf dem H{\"u}gel 69, D-53121 Bonn, Germany
    }

   \date{Received 2 April 2024; accepted 29 May 2024}

 
  \abstract
   {}
   {We aim to test the use of three common molecular ions, ortho-$\htwodplus$ (o$\htwodplus$), $\diaz$, and $\dcoplus$ as probes of the internal structure and kinematics of a dense, starless molecular cloud core.}
   {The prestellar core H-MM1 in Ophiuchus was mapped in the o$\htwodplusline$, $\diaz(4-3)$, and $\dcoplus(5-4)$ lines with the Large APEX sub-Millimeter Array (LAsMA) multibeam receiver of the Atacama Pathfinder EXperiment (APEX) telescope. We also ran a series of chemistry models to predict the abundance distributions of the observed molecules, and to estimate the effect of the cosmic-ray ionisation rate on their abundances.}
   {The three line maps show different distributions. The o$\htwodplus$ map is extended and outlines the general structure of the core, while $\diaz$ mainly shows the density maxima, and the $\dcoplus$ emission peaks are shifted towards one edge of the core where a region of enhanced desorption has been found previously. According to the chemical simulation, the fractional o$\htwodplus$ abundance remains relatively high in the centre of the core, and its column density correlates strongly with the cosmic-ray ionisation rate, $\zeta_{\htwo}$. Simulated line maps constrain the cosmic-ray ionisation rate to be low, between $5\times10^{-18}\,\pers$ and $1\times10^{-17}\,\pers$ in the H-MM1 core. { This estimate agrees with the gas temperature measured in the core.}}
   {The present observations show that very dense, cold gas in molecular clouds can be traced by mapping in the ground-state line of o$\htwodplus$, and high-$J$ transitions of $\dcoplus$ and $\diaz$, despite the severe depletion of the latter two molecules. Modelling line emission of o$\htwodplus$ provides a straightforward method of determining the cosmic-ray ionisation rate in dense clouds, where the primary ion, $\hthreeplus$, is not observable.}

   \keywords{Astrochemistry -- ISM: clouds, molecules -- ISM: abundances, cosmic rays  -- ISM: individual objects: H-MM1}

   \maketitle
%

\section{Introduction}
\label{introduction}

Simple molecular ions, such as $\hcoplus$ and $\diaz$, found in dense interstellar clouds are manifestations of gas-phase chemistry driven by cosmic-ray ionisation. The ionisation of $\htwo$, the principal constituent of these clouds, leads immediately to the formation of the trihydrogen cation, $\hthreeplus$. This triangular cation is observable by its rovibronic spectrum in diffuse clouds in absorption against strong infrared background sources, but is difficult to detect in highly obscured, dense regions. On the other hand, $\hthreeplus$ is easily destroyed in exothermic proton transfer reactions with the common neutral species CO, N$_2$, and HD, to give $\hcoplus$, $\diaz$, and $\htwodplus$, which are observable in cold, dark clouds owing to their pure rotational spectra. Reactions of CO and N$_2$ with $\htwodplus$, or the doubly and triply deuterated forms $\dtwohplus$ and $\dthreeplus$, give then also $\dcoplus$ and $\ddiaz$. Because of the relatively simple chemical network connecting these molecules to the primary ions,  $\hcoplus$ and $\diaz$, and more recently also $\htwodplus$, have been used to estimate the electron abundance, $X({\rm e^-})$, and the cosmic-ray ionisation rate of $\htwo$, $\zeta_{\htwo}$, in dense interstellar clouds. The methods generally involve estimates of the abundances of CO, N$_2$, and other major destroyers of $\hthreeplus$, and often combine these with the fractionation ratios $\dcoplus/\hcoplus$ or $\ddiaz/\diaz$ to derive the electron fraction and the $\hthreeplus$ abundance in the cloud (e.g., \citealt{1977ApJ...217L.165G}; \citealt{1979ApJ...234..876W}; \citealt{1998ApJ...499..234C}; \citealt{2000A&A...358L..79V}; \citealt{2002ApJ...565..344C}; \citealt{2008A&A...492..703C}; \citealt{2020MNRAS.495L...7B}; \citealt{2021A&A...656A.109R}; \citealt{2023ApJ...947L..18S}). The usefulness of analytic formulas to derive $\zeta_{\htwo}$ from the observed fractionation ratios is discussed in \cite{2024A&A...685A..67R}. Ionising collisions with energetic particles have a fundamental influence on the physics and chemistry of interstellar molecular clouds (\citealt{2020SSRv..216...29P}). The ionisation of $\htwo$ is dominated by low-energy ($<1$\,GeV/nucleon) particles, mainly protons and their secondary electrons. Cosmic ray propagation models predict that the total ionisation rate of $\htwo$ decreases with an increasing $\htwo$ column density of the medium, from values estimated in diffuse clouds, $\sim 10^{-16}-10^{-15}\,\pers$ (\citealt{2012ApJ...745...91I}; \citealt{2017ApJ...845..163N}), down to $\sim10^{-17}\,\pers$ at column densities $N(\htwo)\sim10^{23}\,\persqcm$ characteristic of dense cores (\citealt{2009A&A...501..619P}; \citealt{2015ApJ...812..135I}; \citealt{2018A&A...614A.111P}; \citealt{2021ApJ...909..107I}; \citealt{2024A&A...682A.131P}). Besides its central role in astrochemistry, cosmic-ray ionisation is also the dominant direct heating mechanism of gas in quiescent regions shielded from UV radiation. The gas temperature in these regions is determined by equilibrium between the { cosmic-ray} ionisation rate, the gas cooling rate by line emission, and the energy transfer rate by collisions between gas particles and dust grains (\citealt{1983ApJ...265..223B}; \citealt{2001ApJ...557..736G}; \citealt{2002A&A...394..275G}). Conversely, gas temperature measurements can also be used to constrain the cosmic-ray ionisation rate (\citealt{2015arXiv150203380G}; \citealt{2019ApJ...884..176I}). 

In addition to their potential use for estimating the cosmic-ray ionisation rate, the deuterated isotopologues of $\hthreeplus$ are predicted to be useful probes of the interiors of molecular cloud cores (\citealt{2003A&A...403L..37C}; \citealt{2003ApJ...591L..41R}; \citealt{2004A&A...418.1035W}; \citealt{2006RSPTA.364.3081V}). The detection of $\htwodplus$ and $\dtwohplus$ in centrally condensed starless cores (\citealt{2003A&A...403L..37C}; \citealt{2004ApJ...606L.127V};  \citealt{2006ApJ...645.1198V}; \citealt{2011A&A...526A..31P}) show that these light deuterated ions with permanent electric dipole moments can retain high abundances in very dense, cold gas, where many C-, N-, and O-bearing molecules are frozen onto dust grains. 

Here we present maps of the prestellar core Ophiuchus/H-MM1 in the $\lambda=0.8$\,mm lines of ortho-$\htwodplus$ (hereafter o$\htwodplus$), $\diaz$, and $\dcoplus$, obtained using the Large APEX sub-Millimeter Array (LAsMA) of the Atacama Pathfinder EXperiment (APEX) telescope. The ground-state lines of o$\htwodplus$ and para-$\dtwohplus$ (hereafter p$\dtwohplus$) have been previously detected towards this core (\citealt{2011A&A...526A..31P}), and a VLA mapping shows severe depletion of ammonia in its centre (\citealt{2022AJ....163..294P}). The goal of the present observations is to derive the distributions of the o$\htwodplus$, $\diaz$, and $\dcoplus$ in the core, test the predictions of chemistry models concerning their abundances, and put constraints on the cosmic-ray ionisation rate in the region.

\section{Observations}
\label{observations}

The H-MM1 core was mapped with the LAsMA array on APEX using a frequency set-up that covered the rotational lines o$\htwodplus(1_{10}-1_{11})$, $\diaz(4-3)$, and $\dcoplus(5-4)$ at $\lambda=0.8$\,mm. This was achieved by tuning the upper sideband (USB) of the dual-sideband receiver to 372.6\,GHz. The frequencies, upper-state energies, and the critical densities of the observed transitions are listed in Table~\ref{transitions}. The backend was composed of Fast Fourier Transform Spectrometers (FFTS4G) that covered two 4\,GHz intermediate frequency (IF) bands resolved into 65 536 spectral channels of 61.03\,kHz in width, corresponding to $\sim 49\,\ms$ at 372.4\,GHz. The observations were done between June 3 and July 2, 2023, under the project number M-0111.F-9508A-2023.  The APEX telescope is located at an altitude of 5100\,m at Llano de Chajnantor, in Chile.

\begin{table}
\caption[]{Observed transitions}
\begin{tabular}{llrl} \hline
Transition & Frequency$^\dagger$ & $E_{\rm u}$ & $n_{\rm crit}(\htwo)^*$ \\
            & (MHz) & (K)  & ($\percc$) \\ \hline
\noalign{\smallskip}
o$\htwodplus(1_{10}-1_{11})$ & 372421.3558$^a$ & 104.3  & $1.3\times10^{5}$ \\
$\diaz(4-3)$ & 372672.4518$^b$ & 44.7 & $3.2\times10^{6}$ \\
$\dcoplus(5-4)$  & 360169.7783$^c$ & 51.9 & $3.3\times10^{6}$ \\ \hline
\noalign{\smallskip}
\end{tabular}

$^\dagger$ The frequencies are adopted from $^a$ \cite{2017JMoSp.332...33J}; $^b$ \cite{2009A&A...494..719P}; $^c$ \cite{2007ApJ...662..771L} 

$^*$at 10\,K according to Eq.~(4) of \cite{2015PASP..127..299S}, using the collision rate coefficients computed by \cite{2009JChPh.130p4302H}, \cite{2005A&A...432..369S}, and \cite{2020MNRAS.497.4276D}.

\label{transitions}
\end{table}

The LAsMA array has 7 pixels separated by approximately $40\arcsec$. The mapping was performed in the on-the-fly (OTF) mode. The scanning direction was alternated between equatorial east-west and north-south. The separation between rows and columns was $5\farcs6$, equalling to one third of the beam width at 372.4\,GHz. With a scanning speed of $5\farcs6\,\pers$ and a dump time of 1\,s, the dump step was equal to the separations between rows and columns. The coordinates of the map centre are R.A. $16^{\rm h}27^{\rm m}59\farcs6$, Dec. $-24\degr33\arcmin40\arcsec$ (J2000.0). This position lies $\sim 15\arcsec$ east of the density peak of the core, R.A. $16^{\rm h}27^{\rm m}58\farcs65$, Dec. $-24\degr33\arcmin41.2\arcsec$, which is used as (0,0) of the maps presented in this paper.

\begin{figure*}
\unitlength=1mm
\begin{picture}(160,70)(0,0)
\put(85,0){
\begin{picture}(0,0) 
\includegraphics[width=9.5cm,angle=0]{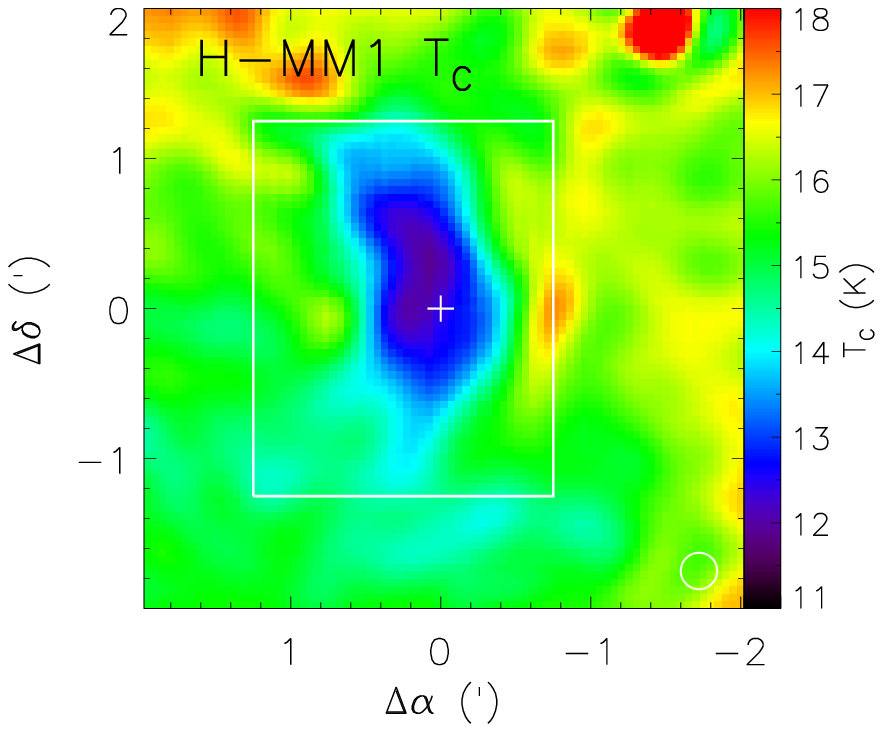}
\end{picture}}
\put(-5,0){
\begin{picture}(0,0) 
\includegraphics[width=9.5cm,angle=0]{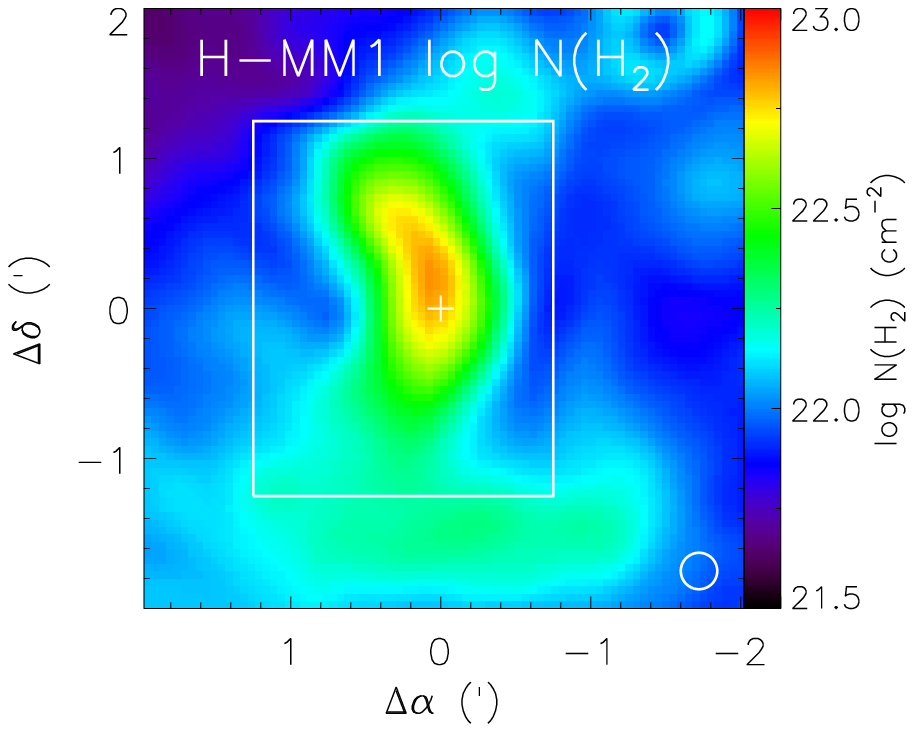}
\end{picture}}
\end{picture}  
\caption{$\htwo$ column density and dust colour temperature ($T_{\rm C}$) maps of H-MM1 derived from {\sl Herschel}/SPIRE HiRes data.  { The rectangle shows the area covered by  the grid used for the line maps derived from the OTF observations. The circle shows the effective beam size (FWHM) of the SPIRE HiRes maps at 500\,$\mu$m.  The plus sign shows the location of the $\htwo$ column density peak of the core, used as (0,0) in the molecular line maps.}}
   \label{spire_maps}
    \end{figure*}

The distribution of the sampled positions is pseudo-random, and to form image cubes, the observations were interpolated and averaged to a regular grid with a cell size of $5\farcs6$. Following the recommendation of \cite{2007A&A...474..679M}, we used as the gridding function a Bessel function of the first kind, $J_1$, tapered with a Gaussian. In practice, the observation representing a certain grid position was obtained by accumulating all observations with a radius of three cell widths (one FWHM of the telescope beam) from this position, weighted by the function $J_1(\pi r/a)/(\pi r/a) \times \exp(-(r/b)^2)$, where $r$ is the distance from the grid position in units of the cell size, and the constants are $a=1.55$, $b=2.52$. Even though the operation corresponds to a convolution with this function, it causes a minimal broadening of the effective beam (when the grid spacing is sufficiently small), and thus preserves the observed intensity (see Fig.~6 of \citealt{2008PASJ...60..445S}). In addition to the gridding function, the accumulated spectra were weighted by the inverse square of the system noise temperature, $T_{\rm SYS}$ (the integration time was 1\,s for all spectra). The size of the gridded area is approximately $120\arcsec\times150\arcsec$. This region is shown in Fig.~\ref{spire_maps} { as a rectangle} superposed on the $\htwo$ column density and dust colour temperature maps derived using {\sl Herschel}/SPIRE HiRes images.

\section{Spectra and maps}
\label{spectra_maps}

\begin{figure*}
\unitlength=1mm
\begin{picture}(160,45)(0,0)
\put(120,0){
\begin{picture}(0,0) 
\includegraphics[width=6cm,angle=0]{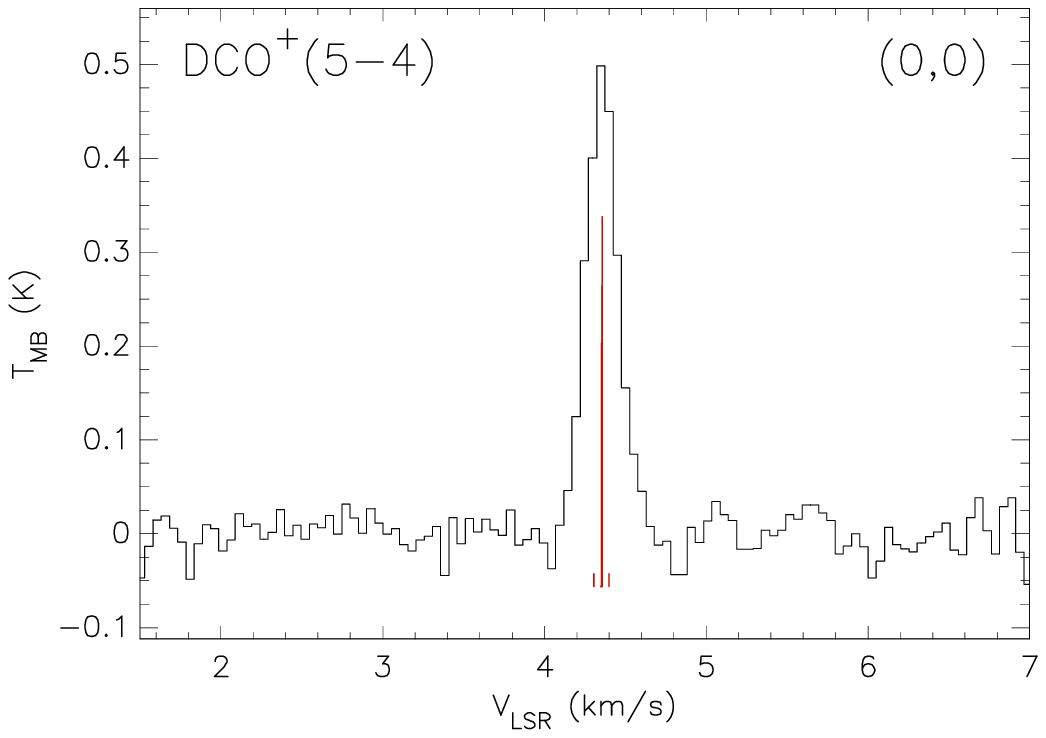}
\end{picture}}
\put(60,0){
\begin{picture}(0,0) 
\includegraphics[width=6cm,angle=0]{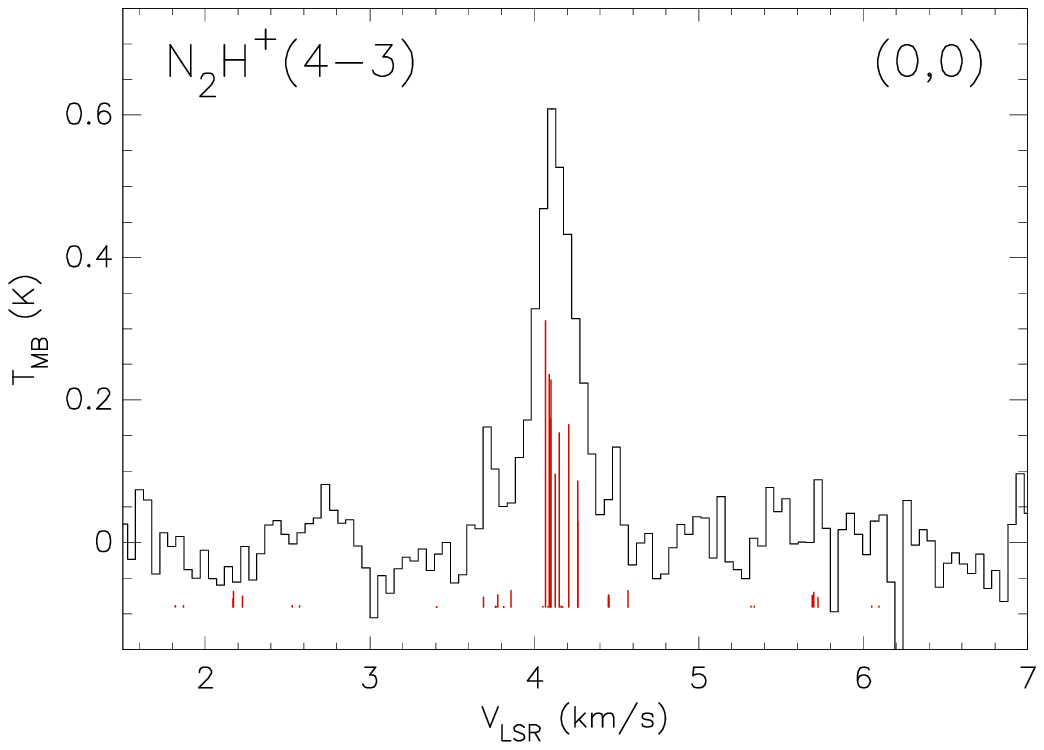}
\end{picture}}
\put(0,0){
\begin{picture}(0,0) 
\includegraphics[width=6cm,angle=0]{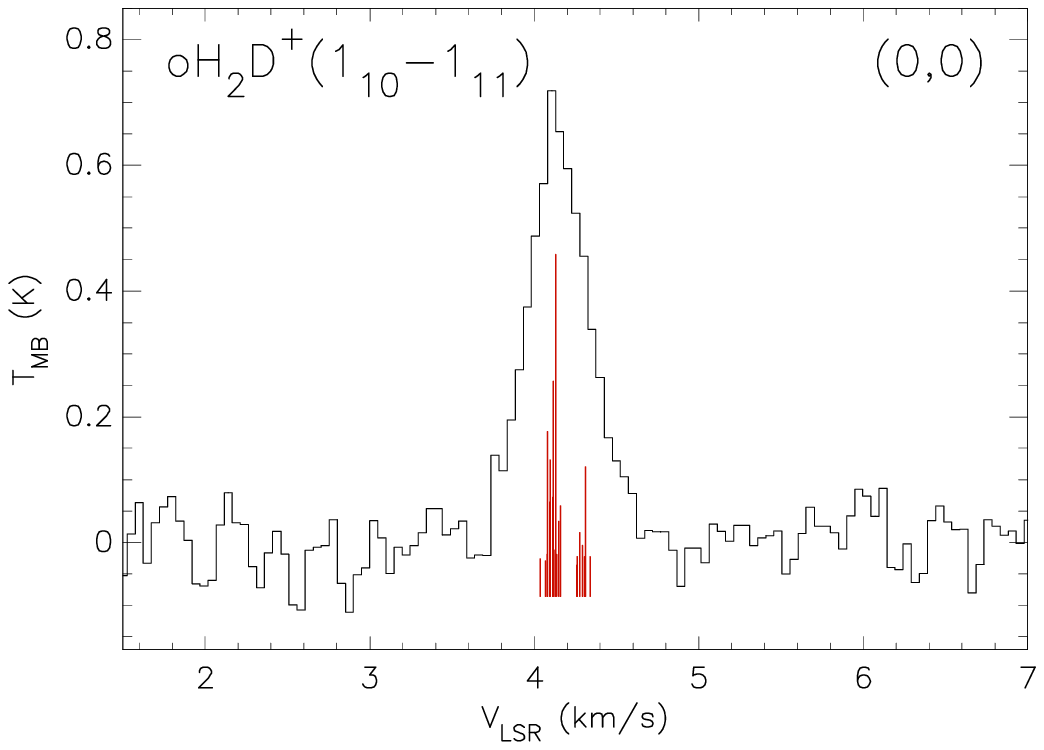}
\end{picture}}
\end{picture}
\caption{ o$\htwodplusline$, $\diaz(4-3)$, and $\dcoplus(5-4)$ spectra towards the density peak of the core. The red bars indicate the hyperfine structure components of the line.}
\label{select_spectra}  
\end{figure*}  

The o$\htwodplusline$, $\diaz(4-3)$, and $\dcoplus(5-4)$ spectra towards the core centre are shown in Fig.~\ref{select_spectra}. Each spectrum represents one cell of the image cube constructed from the OTF observations as described in Sect.~\ref{observations}, and corresponds to an integration towards the grid positions with a $\sim 18\arcsec$ beam. The spectra are presented on the $T_{\rm MB}$ scale. { Based on  measurements performed in 2023, available on the APEX website\footnote{https://www.apex-telescope.org/telescope/efficiency/}, we assume that the main-beam efficiency of the telescope, $\eta_{\rm MB}$, is 0.74 at the frequencies of the o$\htwodplus$ and $\diaz$ lines, and 0.75 at the frequency of the $\dcoplus$ line.} The hyperfine structures of the transitions are indicated with red bars. The frequencies and the relative intensities of the hyperfine components are adopted from \cite{2017JMoSp.332...33J} and \cite{1997MolPh..91..319J} for $\htwodplus$, and from \cite{2009A&A...494..719P} for $\diaz$. For $\dcoplus$, these are taken from the Jet Propulsion Laboratory (JPL) of California Institute of Technology service for molecular spectroscopy.\footnote{https://spec.jpl.nasa.gov} { The spectroscopic parameters this molecule are derived by \cite{2005A&A...433.1145C} and \cite{2007ApJ...662..771L}.} 

\begin{figure*}
\unitlength=1mm
\begin{picture}(160,65)(0,0)
\put(120,65){
\begin{picture}(0,0) 
\includegraphics[width=6.5cm,angle=270]{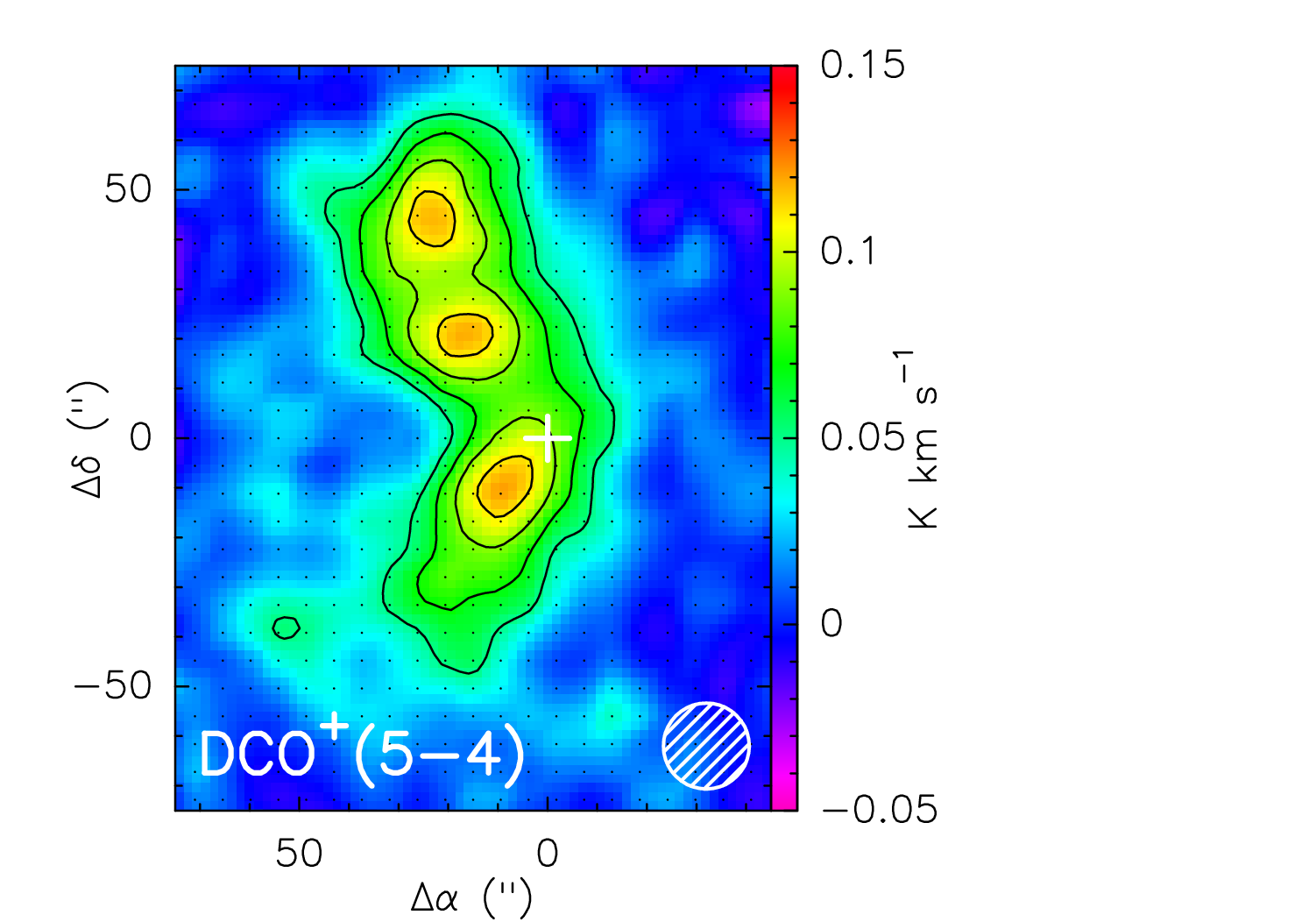}
\end{picture}}
\put(57.3,65){
\begin{picture}(0,0) 
\includegraphics[width=6.5cm,angle=270]{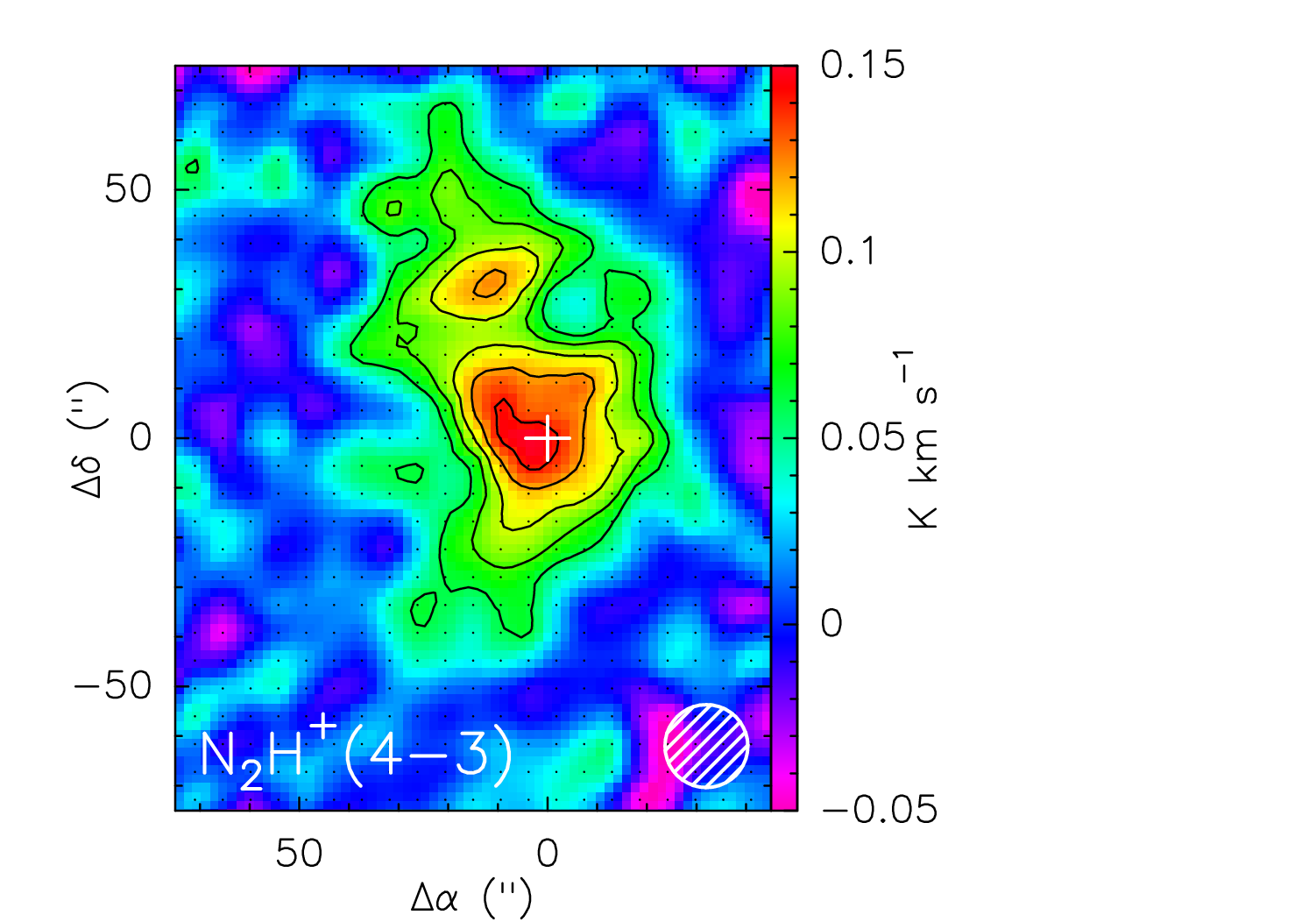}
\end{picture}}
\put(-5,65){
\begin{picture}(0,0) 
\includegraphics[width=6.5cm,angle=270]{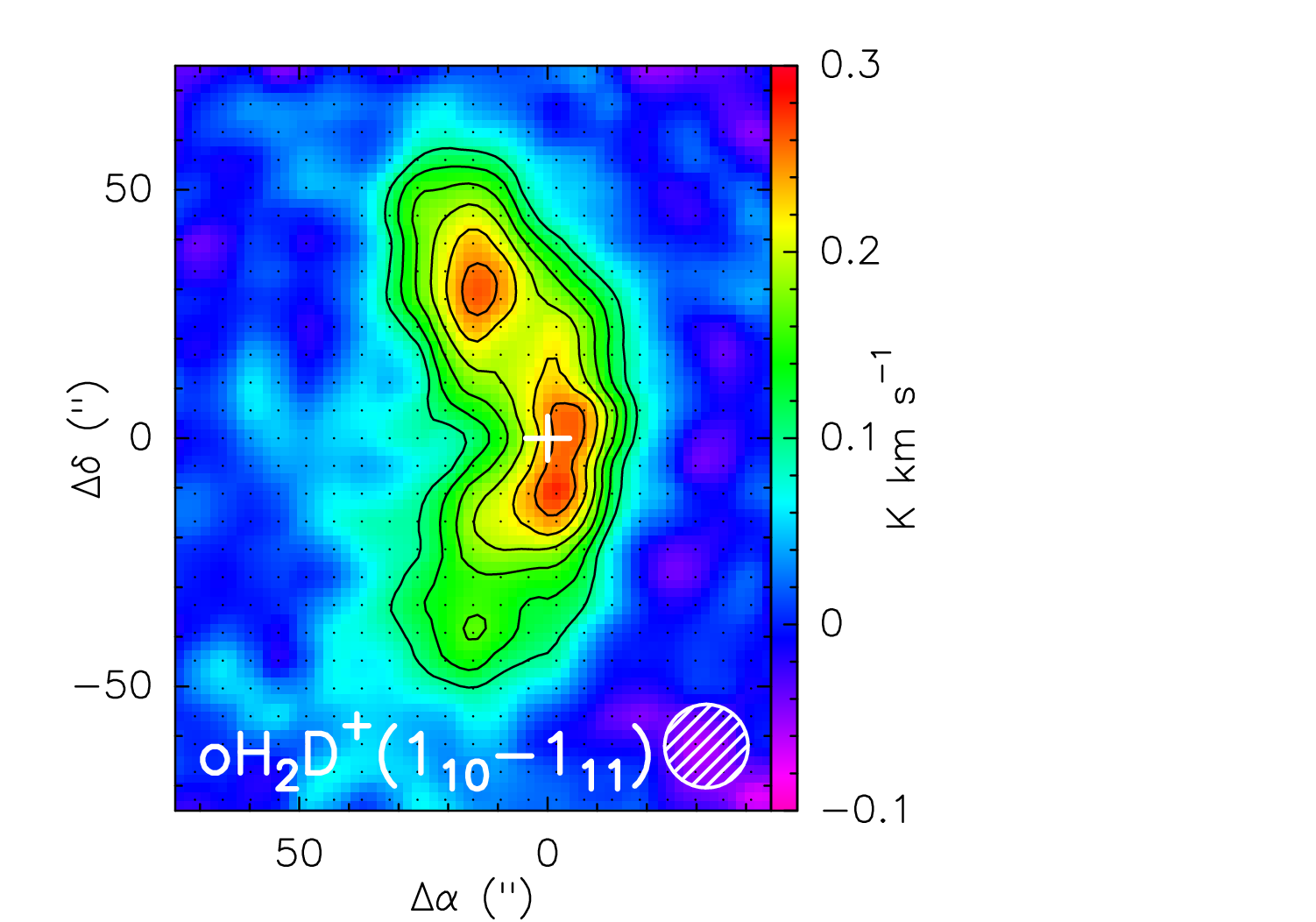}
\end{picture}}

\end{picture}  
\caption{Integrated o$\htwodplusline$, $\diaz(4-3)$, and   $\dcoplus(5-4)$ line intensity maps of H-MM1. The contours are: 0.10  to 0.25 by 0.03 $\Kkms$ (o$\htwodplus$); 0.06 to 0.14 by 0.02 $\Kkms$ ($\diaz$); and 0.05 to 0.11 by 0.02 $\Kkms$ ($\dcoplus$). The intensity scale is $T_{\rm A}^*$.}
   \label{line_maps}
\end{figure*}

\begin{figure}
\unitlength=1mm
\begin{picture}(80,70)(0,0)
\put(5,70){
\begin{picture}(0,0) 
\includegraphics[width=7cm,angle=270]{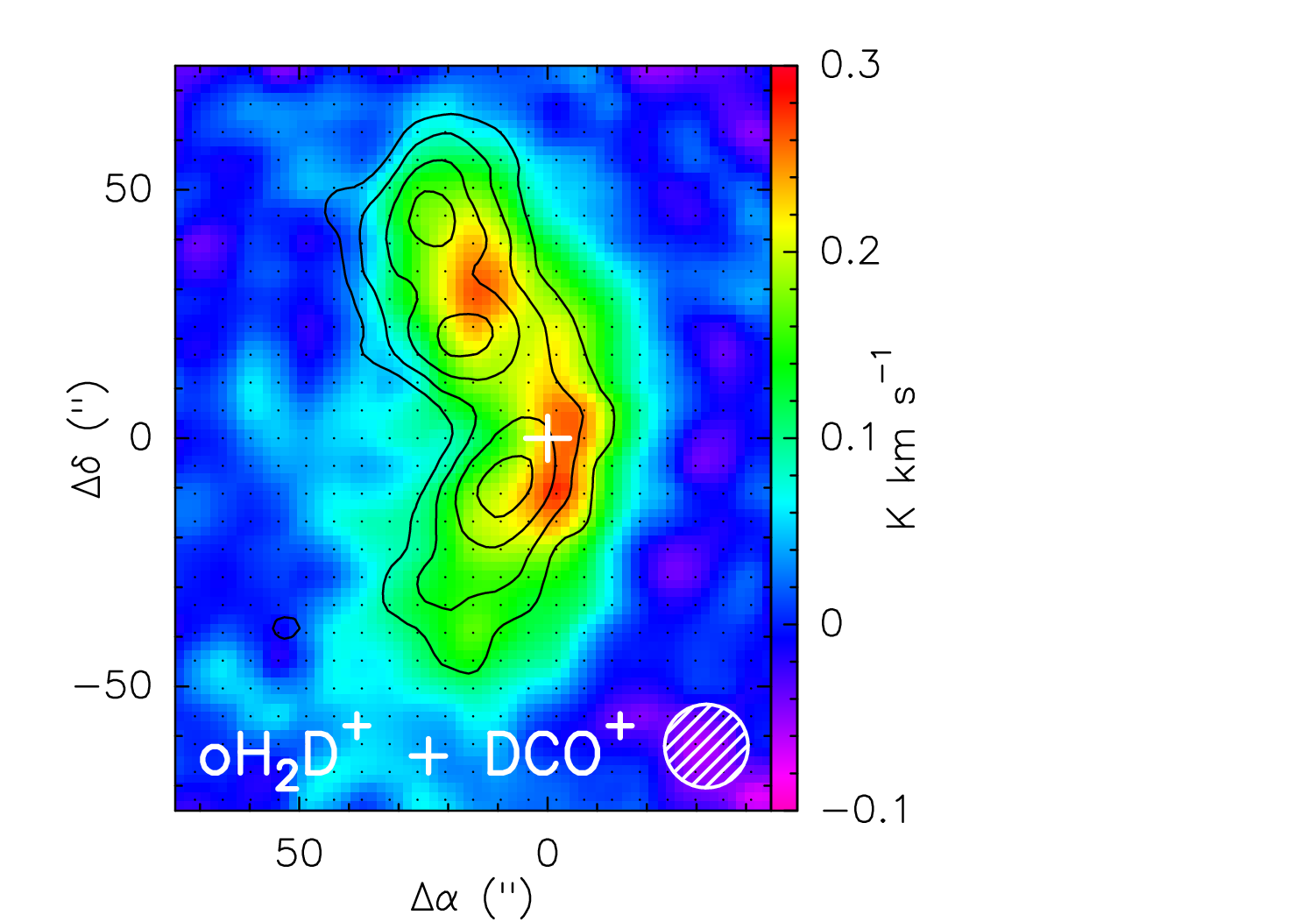}
\end{picture}}

\end{picture}  
\caption{Integrated o$\htwodplusline$ line intensity maps of H-MM1 with the $\dcoplus(5-4)$ contours superposed. The contour levels are the same as in Fig.~\ref{line_maps}.}
   \label{oh2d+_dco+_maps}
\end{figure}

The integrated intensity maps shown in Fig.~\ref{line_maps} reflect the spatial distributions of the molecules and the excitation conditions (the density and temperature) in the cloud. The lines are integrated over the LSR velocity range $3.3-5.1\,\kms$. The emission in o$\htwodplusline$ is the brightest and most extended of the three lines observed. There we see a peanut-shaped maximum centered some $5\arcsec$ southwest of the density peak (our (0,0)), and another maximum around the offset $\sim (+10\arcsec,+30\arcsec)$. The emission in $\diaz(4-3)$ is relatively compact and peaks $\sim 5\arcsec$ east of (0,0). The secondary $\diaz$ peak in the north-east  coincides with the secondary o$\htwodplus$ peak.  The $\dcoplus(5-4)$ map has three peaks, located at the offsets $\sim(+10\arcsec,-10\arcsec)$, $\sim(+15\arcsec,+20\arcsec)$, and $\sim(+20\arcsec,+45\arcsec)$. { The emission is shifted approximately $10\arcsec$ towards east with respect to the o$\htwodplus$ emission.} This shift is illustrated in Fig.~\ref{oh2d+_dco+_maps}, where the $\dcoplus$ contours are superposed on the o$\htwodplus$ pixel map.  

\begin{table*}
\centering
\caption[]{Parameters of lines observed towards the core centre. Error estimates based on the RMS noise of the spectra are shown in brackets.}
\begin{tabular}{lccccccc} \hline
line & $\int T_{\rm MB} dV$ & $T_{\rm MB, peak}$ & $V_{\rm LSR}$ &
$\Delta V$ & $T_{\rm ex}$ & $\tau$ & $N$ \\ 
 & ($\Kkms$) & (K) & ($\kms$) & ($\kms$) & (K) &  & ($10^{13}\,\persqcm$) \\ \hline
o$\htwodplusline$ & 
0.308 (0.012) & 0.718 (0.041) & 4.152 (0.006) & 0.359 (0.016) &
7.1 (0.4)  & 0.62 (0.15) & 1.1 (0.3) \\
$\diaz(4-3)$ &
0.189 (0.013) &  0.609 (0.047)  & 4.127 (0.007) & 0.198 (0.019)
& 6.3 (0.5) &  0.97 (0.41)  & 0.8 (0.6) \\
$\dcoplus(5-4)$ & 
0.117 (0.006) & 0.499 (0.021)  & 4.355 (0.003) & 0.205 (0.008) &
5.8 (0.3) & 0.78 (0.20) & 2.7 (1.3) \\ \hline
\end{tabular}
\label{line_parameters}
\end{table*}

The line parameters derived by Gaussian fits to the hyperfine structure components are presented in Table~\ref{line_parameters}. In this calculation, it is assumed that all hyperfine components have the same intrinsic linewidth, and that the populations of the hyperfine states are proportional to their statistical weights. { The linewidth, $\Delta V$, listed in Table~\ref{line_parameters} is the full width at half maximum (FWHM) of a single hyperfine component obtained from the Gaussian fit.}  The velocity dispersion of the light $\htwodplus$ molecule ($m\sim4$\,amu) is probably dominated by thermal motion. The { linewidth implies} a maximum kinetic temperature of $11.2\pm1.0$\,K. The hyperfine structure fits give reasonable, albeit uncertain, estimates for the total optical thicknesses, $\tau$ (the sum of the peak optical thicknesses of all components), and the excitation temperatures $T_{\rm ex}$, of the transitions.  Assuming line-of-sight homogeneity and that the $T_{\rm ex}$ is the same for all rotational transitions of the molecule, one obtains { column densities given in the last column of Table~\ref{line_parameters}. For $\diaz$ and $\dcoplus$, the uncertainties propagated to the column density estimates are quite large. }Using a broader, Gaussian gridding kernel (with $\sigma=\sqrt{2}\times$ the cell size) that results in a resolution of $\sim25\arcsec$, the relative uncertainties are alleviated somewhat, and we get  $N({\rm o}\htwodplus)=(1.3\pm0.2)\times10^{13}\,\persqcm$, $N(\diaz)=(6.3\pm2.2)\times10^{12}\,\persqcm$, and $N(\dcoplus)=(2.0\pm0.5)\times10^{13}\,\persqcm$ at (0,0). { Combined with the $\htwo$ column density derived from Herschel, $\sim 6\times 10^{22}\,\persqcm$, these column densities imply the following fractional abundances: $X({\rm o}\htwodplus)\sim 2.2\times10^{-10}$, $X(\diaz)\sim1.0\times10^{-10}$, and $X(\dcoplus)\sim3.3\times10^{-10}$. Given the large formal errors, and uncertain assumptions involved in the LTE method, we estimate abundances by modelling line emission maps.} 

The centroid LSR velocity and the velocity dispersion maps derived from hyperfine structure fits to the observed lines are presented in Fig.~\ref{v_and_sigmav}. The fits are made to image cubes created by the GILDAS\footnote{https://www.iram.fr/IRAMFR/GILDAS} task {\tt xy\_map} from the reduced spectra with the {\tt /nogrid} option. Here we assume optically thin emission (by fixing the total optical thickness to 0.1) to avoid eventual degeneracy between the $\tau$ and the linewidth. The pixel size corresponds to the grid step $5\farcs6$. Besides maps from the o$\htwodplusline$, $\diaz(4-3)$, and $\dcoplus(5-4)$ lines, we also include there the corresponding distributions of the $\ammo(1,1)$ lines observed with the Very Large Array (VLA) by \cite{2022AJ....163..294P}. The resolution of the $\ammo$ mapping was $6\arcsec$, but these data have been resampled to the grid used for the present LAsMA maps. The original $V_{\rm LSR}$ and $\sigma_V$ maps from the VLA are shown in Figure 3 of \cite{2022AJ....163..294P}. The line parameters can be derived accurately only for a small portion of each LAsMA map. The VLA $\ammo$ maps are more extended and of better quality. Nevertheless, the LAsMA maps show the same basic features as seen on the VLA map: 1) the most red-shifted gas is found in { an} elongated region pointing towards the core centre in the { centroid LSR velocity} maps (upper panels), and 2) a region with low velocity dispersion can be discerned in the northern part of the core in the $\sigma_V$ maps (lower panels).

\section{Core model}
\label{core_model}

\begin{figure}
\unitlength=1mm
\begin{picture}(80,73)(0,0)
\put(-10,0){
\begin{picture}(0,0) 
\includegraphics[width=10cm,angle=0]{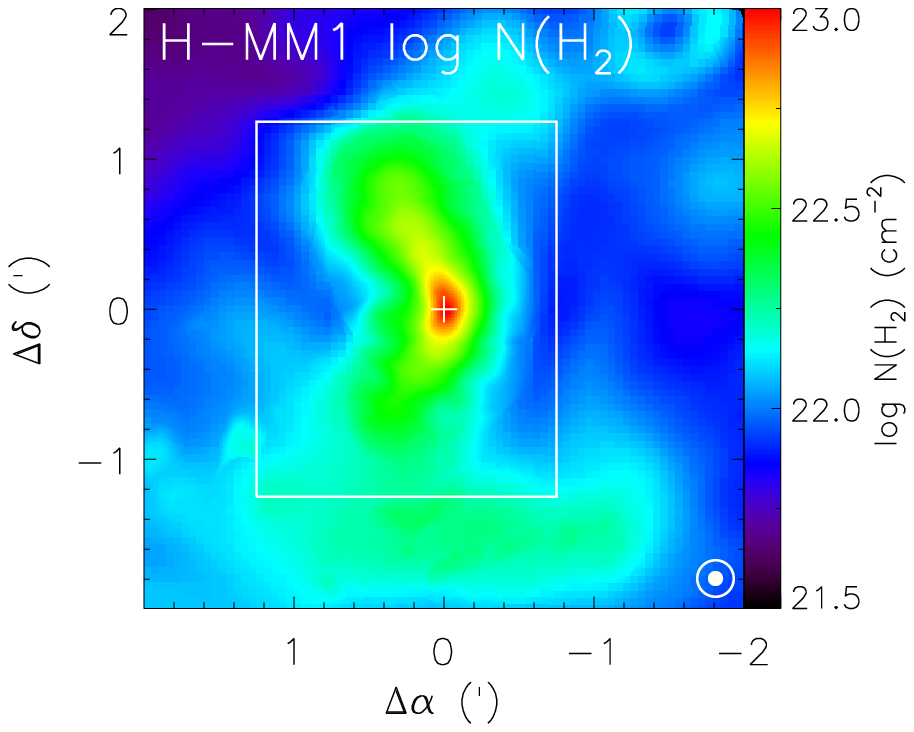}
\end{picture}}
\end{picture}
\caption{$\htwo$ column density map pf H-MM1 derived by combining { the} $8\,\mu$m surface brightness map observed with {\sl Spitzer} with SPIRE HiRes far-infrared maps. The angular resolution is $6\arcsec$ in the central parts of the map, and $15\arcsec$  on the outskirts. The effective beam sizes are shown in the bottom right corner. The box shows the region covered by the molecular line maps.}
\label{h2col_comb}
\end{figure}

We have updated the three-dimensional model for H-MM1 described in \cite{2022AJ....163..294P}. As before, the model is based on a $\htwo$ column density map with a resolution of $6\arcsec$ derived from $8\,\mu$m absorption observed by {\sl Spitzer}, combined with $850\,\mu$m emission map from SCUBA and the dust colour temperature ($T_{\rm C}$) map from {\sl Herschel} (see Appendix A of \citealt{2020ApJ...895..101H}).  Instead of original {\sl Herschel}/SPIRE maps used previously, we have calculated the $T_{\rm C}$ map from deconvolved SPIRE HiRes images, having an angular resolution of approximately $15\arcsec$ (instead of $\sim 38\arcsec$ achieved previously)\footnote{http://herschel.esac.esa.int/Docs/SPIRE/spire\_handbook.pdf}. This resolution corresponds to the SCUBA beam at $850\,\mu$m. Moreover, as the $N(\htwo)$ map from {\sl Spitzer} misses extended structures, we have combined it with the extended  $N(\htwo)$ map derived from SPIRE HiRes at 250, 350, and $500\,\mu$m. The SPIRE data have been downloaded from the Herschel Science Archive\footnote{https://archives.esac.esa.int/hsa}. { The  $N(\htwo)$ and the $T_{\rm C}$ maps of H-MM1 derived from the SPIRE HiRes data with $\sim 15\arcsec$ resolution are shown in Fig.~\ref{spire_maps}. The $N(\htwo)$ map resulting from combining  the {\sl Spitzer} and SPIRE HiRes data is shown in Fig.~\ref{h2col_comb}. There, the angular resolution is $6\arcsec$ in the core region, and $15\arcsec$ around it.}  

\begin{figure}
\unitlength=1mm
\begin{picture}(80,120)(0,0)
\put(0,60){
\begin{picture}(0,0) 
\includegraphics[width=8cm,angle=0]{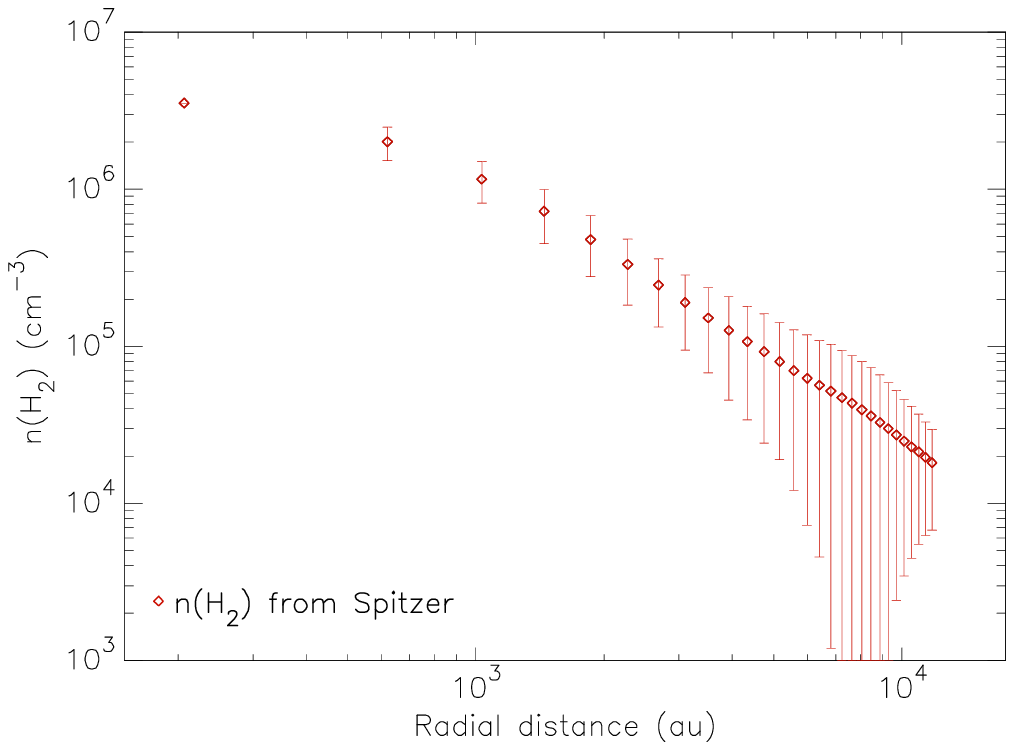}
\end{picture}}
\put(0,0){
\begin{picture}(0,0) 
\includegraphics[width=8cm,angle=0]{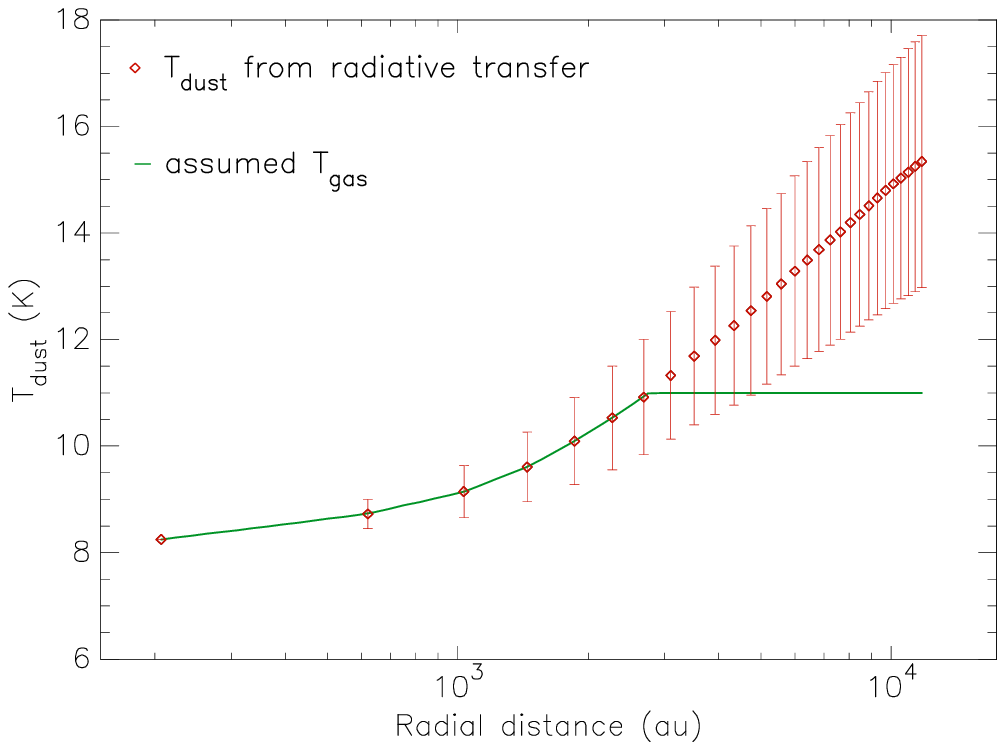}
\end{picture}}
\put(70,107){\Large \bf a}
\put(70,47){\Large \bf b}
\end{picture}
\caption{Density and dust temperature distributions in the core model as functions of the distance from the density peak. The squares and error bars show the mean values and standard deviations in concentric spherical shells. The assumed gas temperature profile is shown with green in the bottom panel.}
\label{nT_radial}
\end{figure}

The density distribution of the core is derived from the $N(\htwo)$ map { shown in Fig.~\ref{h2col_comb}} by fitting Plummer-type profiles to horizontal cuts of the nearly north-south oriented dense ridge of H-MM1. The inversion method is { adopted from \cite{2011A&A...529L...6A}, and described in some detail in Appendix D of \cite{2022AJ....163..294P}.} The density model is then immersed in the interstellar radiation field (ISRF) to compute the dust temperature distribution. The assumptions about radiation field are the same as used previously in \cite{2020ApJ...895..101H} (see their Appendix B), that is, it consists of an isotropic component and radiation from a B star (HD\,147889) located on the western side of the core approximately 1.2\,pc away. The distributions of the average density and dust temperature as functions of the radial distance from the density maximum of the core are shown in Fig.~\ref{nT_radial}. As in \cite{2022AJ....163..294P}, we assume that in the outer parts of the core the gas temperature is lower than the dust temperature because of molecular line cooling. Based on the $T_{\rm kin}$ distribution derived from $\ammo$, we assume the gas temperature does not rise above 11 K. The assumed gas temperature profile is shown with a green curve in Fig.~\ref{nT_radial}.

To predict line intensity maps from the core model requires assumptions about molecular abundances. The simplest assumption is that the fractional abundance of a molecule is constant throughout the cloud. We tested this assumption using the estimates towards our (0,0) { position} from the LTE method described in Sect.~\ref{spectra_maps}. While the predicted line intensities using these constant abundances are comparable to those observed in the core centre, the extent of the emission is clearly more compact than observed. This indicates that the fractional abundances of all three molecules are higher in the outer parts than in the centre of the core. { In what follows, we use a chemical model to predict the abundance profiles of the observed molecules in the core, given the density and temperature distributions derived above.}

\section{Chemistry model}
\label{chemistry_model}

\begin{figure}
\unitlength=1mm
\begin{picture}(80,175)(0,0)
\put(0,120){
\begin{picture}(0,0) 
\includegraphics[width=8cm,angle=0]{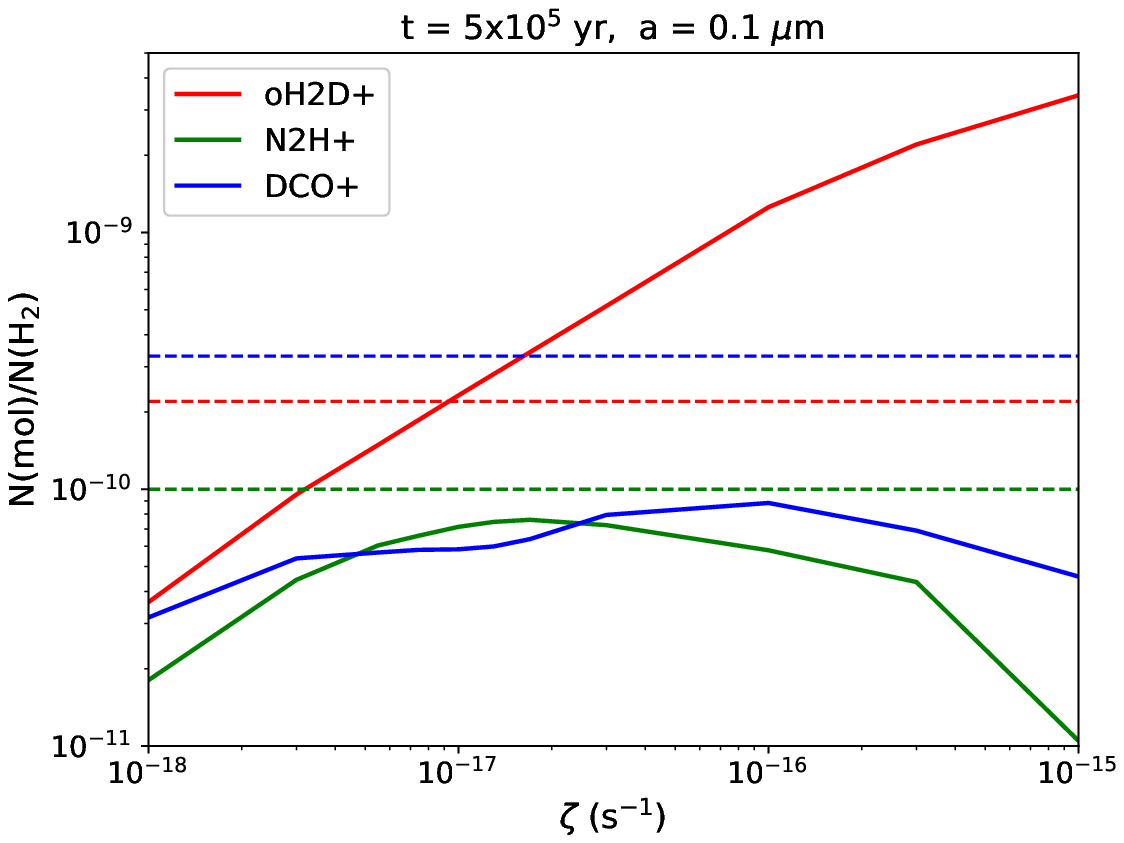}
\end{picture}}
\put(0,60){
\begin{picture}(0,0) 
\includegraphics[width=8cm,angle=0]{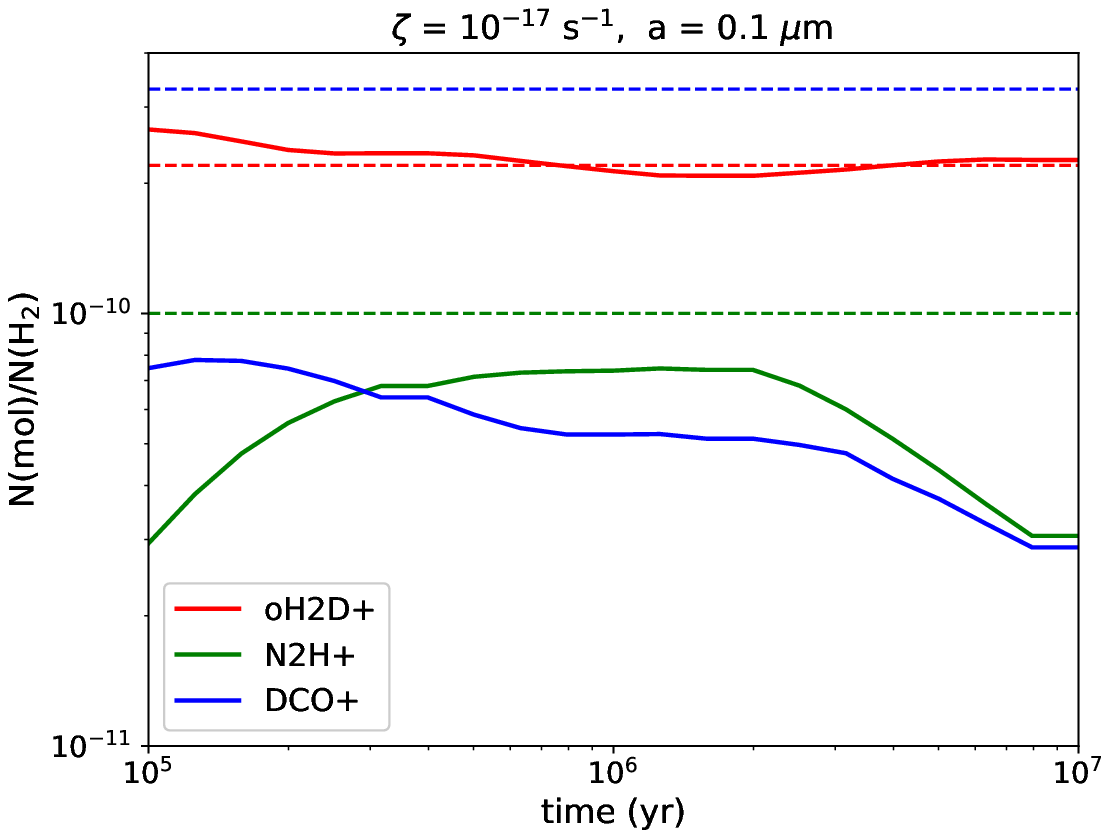}
\end{picture}}
\put(0,0){
\begin{picture}(0,0) 
\includegraphics[width=8cm,angle=0]{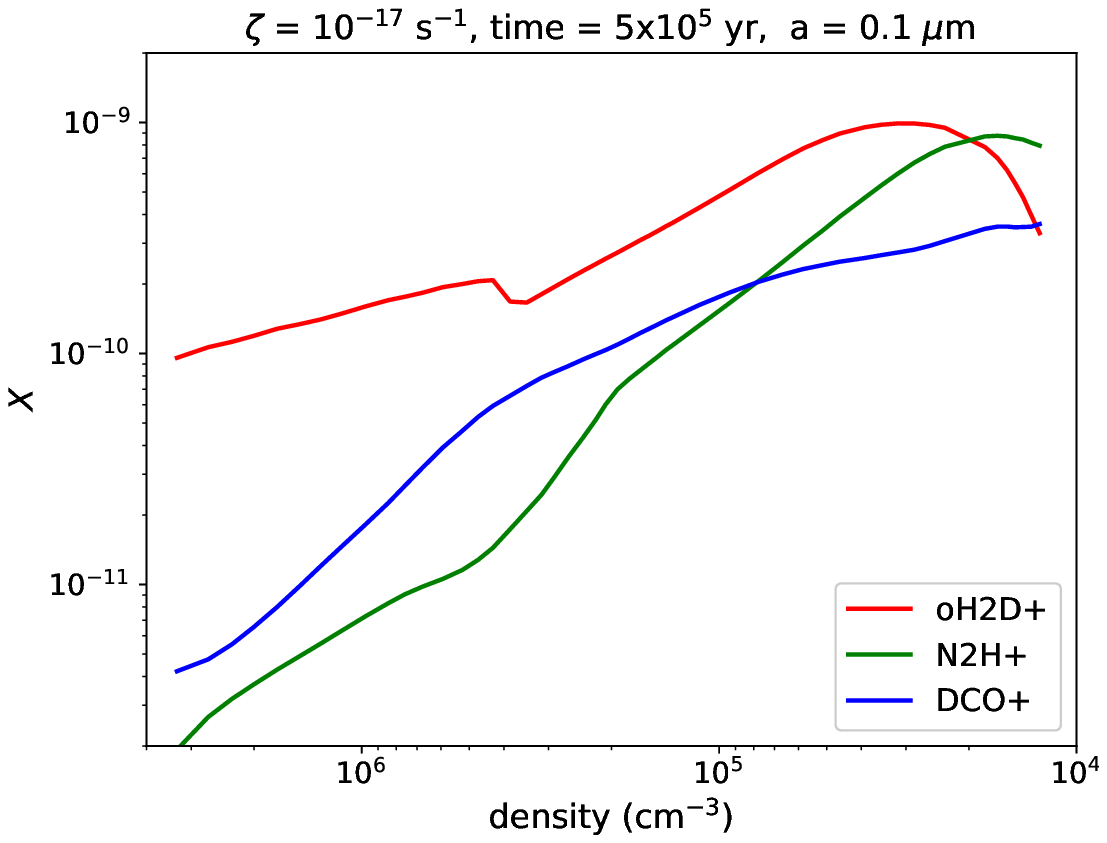}
\end{picture}}
\put(65,167){\Large \bf a}
\put(65,107){\Large \bf b}
\put(65,47){\Large \bf c}
\end{picture}
\caption{Predictions for the fractional o$\htwodplus$, $\diaz$, and $\dcoplus$ abundances as functions of the cosmic-ray ionisation rate $\zeta_{\htwo}$ ({\bf a}), time ({\bf b}), and gas density ({\bf c}) according to our chemistry model. The first two ({\bf a} and {\bf b}) represent the column density ratios $N({\rm mol})/N(\htwo)$ through the centre of the core.  In panels {\bf a} and {\bf c}, the simulation time is $5\times10^5$\,yr. In panels {\bf b} and {\bf c}, $\zeta_{\htwo}=10^{-17}\,\pers$. The dashed lines in panels {\bf a} and {\bf b} show the fractional abundances estimated by the LTE method in Sect.~\ref{line_maps}. { In panel {\bf c}, higher densities are on the left to be consistent with Fig.~\ref{nT_radial}.}}
\label{x_vs_z_t_n}
\end{figure}

 We use for the simulations the gas-grain chemical model pyRate described in \citet{2019A&A...631A..63S}, \citet{2023A&A...680A..87R}, and \citet{2024A&A...682A...8H}. We assume constant chemical desorption with 1\% of exothermic surface reactions leading to desorption, and adopt the proton hop chemical network in the gas phase \citep{2019A&A...631A..63S}. In this model, proton donation and abstraction reactions involving O, C, and N are assumed to proceed via the exchange of one proton (or deuteron) between the reactants, without multiple atom exchanges. On the other hand, full scrambling is assumed for the $\hthreeplus + \htwo$ reaction system and its deuterated variants, and the rate coefficients and branching ratios for these reactions are adopted from \cite{2009JChPh.130p4302H}. In the chemical calculations we employ a one-dimensional physical model for H-MM1, which is constructed using the average densities and temperatures over concentric shells shown in Fig.~\ref{nT_radial}. We adopt the same initial abundances as in \citet{2019A&A...631A..63S}, and the  simulation parameters are also the same as in that paper except for the cosmic-ray ionisation rate of $\htwo$, $\zeta_{\htwo}$, for which we test here different values ranging from $10^{-18}\,\pers$ to $10^{-15}\,\pers$. We assume that $\zeta_{\htwo}$ is constant throughout the core, because the column densities in the core region (at the resolution achieved here) are confined to a narrow range, $N(\htwo)\sim 10^{22}\,\persqcm - 6\times 10^{22}\,\persqcm$, where attenuation models (e.g., \citealt{2018A&A...614A.111P}) do not predict drastic changes in $\zeta_{\htwo}$. Besides, the angular resolution of the observations is not sufficient for studying spatial variations of $\zeta_{\htwo}$.  The adopted abundances of oxygen, carbon, and nitrogen relative to hydrogen are ${\rm O/H}=1.8\times10^{-4}$, ${\rm C/H}=7.3\times10^{-5}$, and ${\rm N/H}=5.3\times10^{-5}$. Previous observations of ammonia and its deuterated forms towards H-MM1 have been reproduced well by these abundances (\citealt{2017A&A...600A..61H}; \citealt{2022AJ....163..294P}; \citealt{2024A&A...682A...8H}). We also tested higher elemental abundances (${\rm O/H}=2.56\times10^{-4}$, ${\rm C/H}=1.2\times10^{-4}$, and ${\rm N/H}=7.6\times10^{-5}$) used previously in chemical models for the prestellar core L1544 (e.g., \citealt{2023A&A...680A..87R}). { The chemistry model assumes a unique grain radius, and this is fiducially set to 0.1\,$\mu$m. We also examined how increasing the grain radius would affect the abundances of the observed molecules. The results of tests with different elemental abundances and grain radii are discussed at the end of Sect.~\ref{simulated_maps}.}

Simulation results for the one-dimensional core model { with our fiducial set of parameters} are shown in Fig.~\ref{x_vs_z_t_n}. There we show the average fractional abundances of o$\htwodplus$, $\diaz$, and $\dcoplus$ as functions of the cosmic-ray ionisation rate $\zeta_{\htwo}$ at a certain time of simulation (Fig~\ref{x_vs_z_t_n}a), and the temporal evolution of these abundances for $\zeta_{\htwo}=10^{-17}\,\pers$ (Fig~\ref{x_vs_z_t_n}b). The abundances derived using the LTE approximation in Sect.~\ref{line_maps} are shown with dashed lines in panels a and b. The bottom panel (Fig~\ref{x_vs_z_t_n}c) shows the fractional o$\htwodplus$, $\diaz$, and $\dcoplus$ abundances as functions of the local density in the core at the simulation time $5\times10^5$\,yr assuming $\zeta_{\htwo}=10^{-17}\,\pers$. In panels a and b, the average fractional abundance is the column density ratio $N({\rm mol})/N(\htwo)$ computed along a pencil beam going through the core centre. The best-fit abundances discussed in Sect.~\ref{simulated_maps} are derived using simulated maps convolved to the resolution of the observations. 

The chemical model predicts that o$\htwodplus$ is clearly more abundant than the $\dcoplus$ and $\diaz$ ions, except for the core boundary, and that its abundance strongly correlates with the cosmic-ray ionisation rate. The abundance derived in Sect.~\ref{spectra_maps} by the LTE method, $X({\rm o}\htwodplus)\sim 2.2\times10^{-10}$ would agree with $\zeta_{\htwo}=10^{-17}\,\pers$. On the other hand, the predicted average $\dcoplus$ abundance is for all values of $\zeta_{\htwo}$ and at all times clearly lower than that derived from the observations (suggesting $X(\dcoplus)\sim 3.3
\times10^{-10}$).  All three ions show a decreasing tendency as a function of density in the interior parts of the core (Fig~\ref{x_vs_z_t_n}c). The decrease in the o$\htwodplus$ abundance is, however, shallower than for $\diaz$ and $\dcoplus$. At early times, we see a rise and fall pattern in the abundances when going from the outer edge to the centre of the core. The peak shifts outwards, towards lower densities, with time. At the time $5\times 10^5$\,yr the pattern is clearly discerned only for o$\htwodplus$. Because the core model does not include densities below $\sim 10^4\, \percc$, the fractional abundances are high at the core boundary at late times. Models used in \cite{2022A&A...668A.131S} for the prestellar core L1544 reach down to low densities, and there we see a drastic drop in the $\diaz$ and $\dcoplus$ abundances in the outer parts of the clouds (their Fig.~5).
 
\section{Simulated line maps}
\label{simulated_maps}

The chemical abundances calculated using the one-dimensional core model at different times and for different values of the cosmic-ray ionisation rate $\zeta_{\htwo}$ were interpolated to the three-dimensional model described in Sect.~\ref{core_model}. In the interpolation it was assumed that the abundance of a molecule depends only on the density. This means that the temperature variations in a certain density regime owing to the anisotropic external radiation field was ignored. The simulations were done using the LOC radiative transfer program \citep{2020A&A...644A.151J}. The goodness of the simulation result was evaluated by comparing the simulated integrated intensity maps with the observed ones. For this purpose, we used the image cubes from GILDAS mentioned in Sect.~\ref{line_maps} that gave a pixel size of $5\farcs6$. The simulated integrated intensity maps were resampled to the same grid, and the $\chi^2$ statistic was evaluated over pixels where the observed integrated intensity exceeded a certain threshold. The threshold was set to $0.1\,\Kkms$ for $\diaz$ and $\dcoplus$, and $0.2\,\Kkms$ for o$\htwodplus$. This selection gave $\sim 70-80$ pixels for each map. For the $\sigma$ we used the RMS integrated intensity on the outskirts of the observed maps, where the surface brightness is dominated by noise.    

The best agreement between the simulated and observed integrated o$\htwodplusline$ intensity map in terms of $\chi^2$ was found with $\zeta_{\htwo}=5.5\times10^{-18}\,\pers$ at the simulation time $5\times10^5$\,yr. Nearly as good agreement is obtained with $\zeta_{\htwo}=7.4\times10^{-18}\,\pers$ at $6.3\times10^5$\,yr. With the latter value of $\zeta_{\htwo}$, the agreement with the observed o$\htwodplus$ map does not depend strongly on the simulation time after $6.3\times 10^5$\,yr. Image-style plots of $\chi^2$ as functions of $\zeta_{\htwo}$ and time are shown in Fig.~\ref{chi2_maps}.  There we show chi-square maps for the two sets of elemental abundances mentioned in Sect.~\ref{chemistry_model}, { and the same plot for a model with the grain radius $a=0.3\,\mu$m (instead of our standard $a=0.1\,\mu$m, see below). One can see that changes in the abundances of O, C, and N, or the assumed grain radius do not cause notable differences in the o$\htwodplus$ emission.}  

The average fractional o$\htwodplus$ abundance of the best-fit models (with $\zeta_{\htwo}= (5.5-7.4)\times10^{-18}\,\pers$) is  $\sim 2.2\times10^{-10}$ towards the centre of the core, derived from column densities convolved to the angular resolution of the observations. The value is close to that estimated in Sect.~\ref{line_maps}. As can be anticipated from Fig.~\ref{x_vs_z_t_n}a, higher values of $\zeta_{\htwo}$ produce too bright o$\htwodplus$ lines, and for lower $\zeta_{\htwo}$ the emission is weaker than observed. The { "baseline" model with the default elemental abundances and grain radius} under-predicts the $\diaz(4-3)$ and the $\dcoplus(5-4)$ intensities for all tested values of $\zeta_{\htwo}$ at all times.  Adopting the abundance profiles for the model with $\zeta_{\htwo}=7.4\times 10^{-18}\,\pers$ at the time $6.3\times10^5$\,yr, but allowing the $\diaz$ and $\dcoplus$ abundances to be scaled up, an approximate agreement with the observed spectra and maps is reached when the $\diaz$ abundances are multiplied by 4.8, and the $\dcoplus$ abundances are multiplied by 4. With these scaling factors, the average fractional abundances towards the centre of the core are $X(\diaz)\sim 4.3\times 10^{-10}$ and $X(\dcoplus)\sim 2.9\times10^{-10}$ at the angular resolution of the observations. Maps and spectra towards a few selected positions calculated from the model with $\zeta_{\htwo}=7.4\times10^{-18}\,\pers$ are shown in Figs.~\ref{obs_model_maps} and \ref{obs_model_spectra}. Here the $\diaz$ and $\dcoplus$ abundances are multiplied by the factors mentioned above. 

We also tested the set of higher O, C, and N abundances mentioned in Sect.~\ref{chemistry_model}. This did not result in significant changes in the o$\htwodplus$ abundance, and also for $\diaz$ and $\dcoplus$ the effect is small for low cosmic-ray ionisation rates. For $\zeta_{\htwo}$ values higher than $\sim5\times10^{-17}\,\pers$, the fractional $\dcoplus$ and $\diaz$ abundances increase using these elemental abundances, but also then the simulated intensities remain { below the observed level.} 

Another possible way to increase the $\diaz$ and $\dcoplus$ abundances is to decrease the destruction of their precursors, N$_2$ and CO, through accretion onto grains. { A unique grain radius of 0.1\,$\mu$m assumed standardly in the present chemistry models, together with the assumed grain material density ($2.5\,{\rm g}\,\percc$) and dust-to-gas mass ratio (0.01) imply a total surface area of dust grains of $7\times10^{-22}\,{\rm cm^2}$ per H atom. Yet another series of simulations was made varying the grain radius. The tested values were 0.2\,$\mu$m, 0.3\,$\mu$m, and 0.4\,$\mu$m. Keeping the grain material density and the dust-to-gas mass ratio constant, the total surface area is inversely proportional to $a$, so with the value $a=0.4\,\mu$m the grain surface area is one quarter of that in our "standard" model. The $\diaz$ abundance increased noticeably with the grain radii $a=0.2\,\mu$m and $a=0.3\,\mu$m, and decreased again with $a=0.4\,\mu$m. The observed intensity levels in the $\diaz(4-3)$ map were best reproduced with the parameter values $\zeta_{\htwo}=7.4\times10^{-18}\,\pers$, $a=0.3\,\mu$m at time $t=7.9\times10^{5}$\,yr. For these values of $\zeta_{\htwo}$ and $a$, the best-fit time for o$\htwodplus$ is $5\times10^5$\,yr.  The o$\htwodplus$ abundances decrease gradually with an increasing grain radius, which must be compensated by increasing the cosmic-ray ionisation rate in order to reproduce the observed line intensities. For the grain radii $a=0.1\,\mu$m and 0.2\,$\mu$m, the minimum $\chi^2$ was found at $\zeta_{\htwo}=5.5\times10^{-18}\,\pers$, whereas with $a=0.3\,\mu$m and 0.4\,$\mu$m, the best agreement with the o$\htwodplus$ maps was achieved at $\zeta_{\htwo}=7.4\times10^{-18}\,\pers$. The $\dcoplus$ abundance increases when grains are made larger, and finally with $a=0.4\,\mu$m, and with a high cosmic-ray ionisation rate of $\zeta_{\htwo}\sim 10^{-16}\,\pers$, the modelled peak intensities attain the observed values. However, cosmic-ray ionisation rates above $10^{-17}\,\pers$ are inconsistent with the o$\htwodplus$ and $\diaz$ data. The observed displacement of the $\dcoplus$ emission peaks from the highest densities is not reproduced by any model tested here.}
  
{ Simulated o$\htwodplus$, $\diaz$, and $\dcoplus$ maps from the model with $\zeta_{\htwo}=7.4\times10^{-18}\,\pers$ and $a = 0.3\,\mu$m are shown in Fig.~\ref{model_maps_a3}. Again, the $\dcoplus$ abundance is scaled by a factor of 4 to attain the observed peak intensity, but the o$\htwodplus$ and $\diaz$ maps are as predicted by the model, although the $\diaz$ map represents a later simulation time ($7.9\times10^5$ yr) than the o$\htwodplus$ and $\dcoplus$ maps ($5\times10^5$\,yr).}

\section{Discussion}
\label{discussion}

The observed o$\htwodplus$ map follows the general structure of the H-MM1 core seen in the far-infrared and sub-millimetre continuum (Figs.~\ref{spire_maps} and \ref{line_maps}). The core model misses the secondary density peak near the offset $+10\arcsec,+30\arcsec$ from the core centre (Fig.~\ref{obs_model_maps}). Besides the o$\htwodplus$ and $\diaz$ maps presented here, the presence of the secondary peak is also evident on the previous high-resolution $\dammo$ and $\ammo$ maps of the core (\citealt{2020ApJ...895..101H}; \citealt{2022AJ....163..294P}). According to the chemistry model, the fractional o$\htwodplus$ abundance decreases typically by one order of magnitude from the outer parts to the centre of the core (Fig.~\ref{x_vs_z_t_n}c). For $\dcoplus$ and $\diaz$, the decrease in the fractional abundance is about two orders of magnitude.  Combined with the density model of the core (Fig.~\ref{nT_radial}), this means that while the number densities of $\dcoplus$ and $\diaz$ ($\percc$) remain approximately constant throughout the core, the number density of o$\htwodplus$ increases by a factor of ten towards the centre. Provided that it is not optically thick, or absorbed in an ambient cloud, the o$\htwodplusline$ line is therefore a useful probe of the interior parts of dense cores. In H-MM1, we do not see signs of self-absorption in this line. 

The centroid velocity and velocity dispersion maps of o$\htwodplus$ shown in Fig.~\ref{v_and_sigmav} do not bring new information about the gas kinematics in the centre of the core. The previous maps derived from the $\ammo(1,1)$ line observed with the VLA are superior thanks to their high spatial resolution ($6\arcsec$ versus $18\arcsec$ achieved here) and sensitivity, despite the heavy depletion of ammonia in this region (\citealt{2022AJ....163..294P}). The sudden drop in the velocity dispersion seen at a declination offset of about $+30\arcsec$ in $\ammo$, $\dcoplus$, and $\diaz$ is { less pronounced} in o$\htwodplus$ because the linewidth of this molecule is dominated by thermal motions. All three maps observed here confirm the presence of red-shifted gas in a narrow region pointing to the centre of the core. This probably represents a stream of gas falling into the centre of gravity, resembling thus larger-scale accretion streams discovered previously in association with protostars (\citealt{2020NatAs...4.1158P}; \citealt{2022A&A...667A..12V}; \citealt{2023A&A...677A..92V}). The putative stream, best discernible on the $\ammo(1,1)$ velocity map (Fig.~\ref{v_and_sigmav}, top right panel), may have a role in the conception of a protostar in this core.   

The shift between the $\dcoplus$ and $\htwodplus$ maps, illustrated in Fig.~\ref{oh2d+_dco+_maps}, is reminiscent of what was seen in high-resolution maps of $\meth$ and $\dammo$ towards this core (\citealt{2020ApJ...895..101H}). The bright methanol emission at its shaded, eastern boundary was suggested to be caused by vigorous desorption, possibly { induced by grain-grain collisions in a layer of strong velocity shear.} { Methanol formation requires CO-rich ice (e.g., \citealt{2017ApJ...842...33V}), and carbon monoxide eventually released in this process could also have led to an intensified production of $\hcoplus$ and $\dcoplus$. In low-mass star-forming regions, bright $\dcoplus$ emission sometimes marks interaction with a low-velocity outflow \citep{2016ApJ...827..133L}. To examine the possible connection of the methanol feature to the $\dcoplus$ "displacement", we have overlaid in Fig.~\ref{line_maps+meth} contours of strong methanol emission on the integrated intensity maps of the three lines observed here. The figure shows that the $\dcoplus$ peaks are not found in the direction of the brightest methanol emission, but lie west of it, slightly inward from the core edge. Also the $\diaz$ emission peak seems to reside, at least in projection, next to methanol contours. On the whole, the differences between the distributions of $\dcoplus$, $\diaz$ and o$\htwodplus$ relative to $\meth$ are small. Nevertheless, the fact that $\dcoplus$ emission peaks on the same side of the core where also most methanol is found can probably be traced to their common precursor, CO. It is possible that the distributions have been shaped by the asymmetric radiation field owing to nearby B-type stars, which has led to a measurable difference in the dust temperature between the western and eastern boundaries of the core (\citealt{2020ApJ...895..101H}, Sect.~7.1). The effects of uneven illumination on carbon chemistry and the formation of methanol in prestellar and starless cores have been discussed  in \cite{2016A&A...592L..11S} and \cite{2020A&A...643A..60S}; the latter paper also presents a single-dish methanol map of H-MM1. The large-scale distribution of methanol may be understood in terms favourable conditions for the formation of CO, and CO-rich ice, on the sheltered eastern side of the core, and relative scarcity of CO on the western side side which is exposed to strong radiation. How these conditions influence the $\dcoplus$ distribution is not clear to us. The present chemistry model does not include local variations of the desorption efficiency or anisotropies of the external radiation field. This may partly explain the discrepancy between the observed and modelled $\dcoplus$ abundances.} 

The decrease in the fractional abundances of electrons and molecular ions towards higher densities {predicted by the chemistry model (Fig.~\ref{x_vs_z_t_n}c)} is a natural consequence of more frequent recombination reactions; the recombination rate ($\percc\,\pers$) between molecular ions and electrons is approximately proportional to the cosmic-ray ionisation rate (here assumed to be constant) and the number density of $\htwo$ (\citealt{1989ApJ...345..782M}, Appendix B). For $\dcoplus$ and $\diaz$ the decrease is accentuated by the disappearance of their precursors CO and N$_2$ owing to freezing onto dust grains. In contrast, o$\htwodplus$ and the other forms of $\hthreeplus$ benefit from the depletion of these molecules, and the density dependence of their fractional abundance is in our model slightly shallower than $\sim n^{-0.5}$ predicted by \cite{1989ApJ...345..782M} (see also Fig.~13 of \citealt{2022AJ....163..294P}). 

\begin{figure}
\unitlength=1mm
\begin{picture}(80,65)(0,0)
\put(0,0){
\begin{picture}(0,0) 
\includegraphics[width=9cm,angle=0]{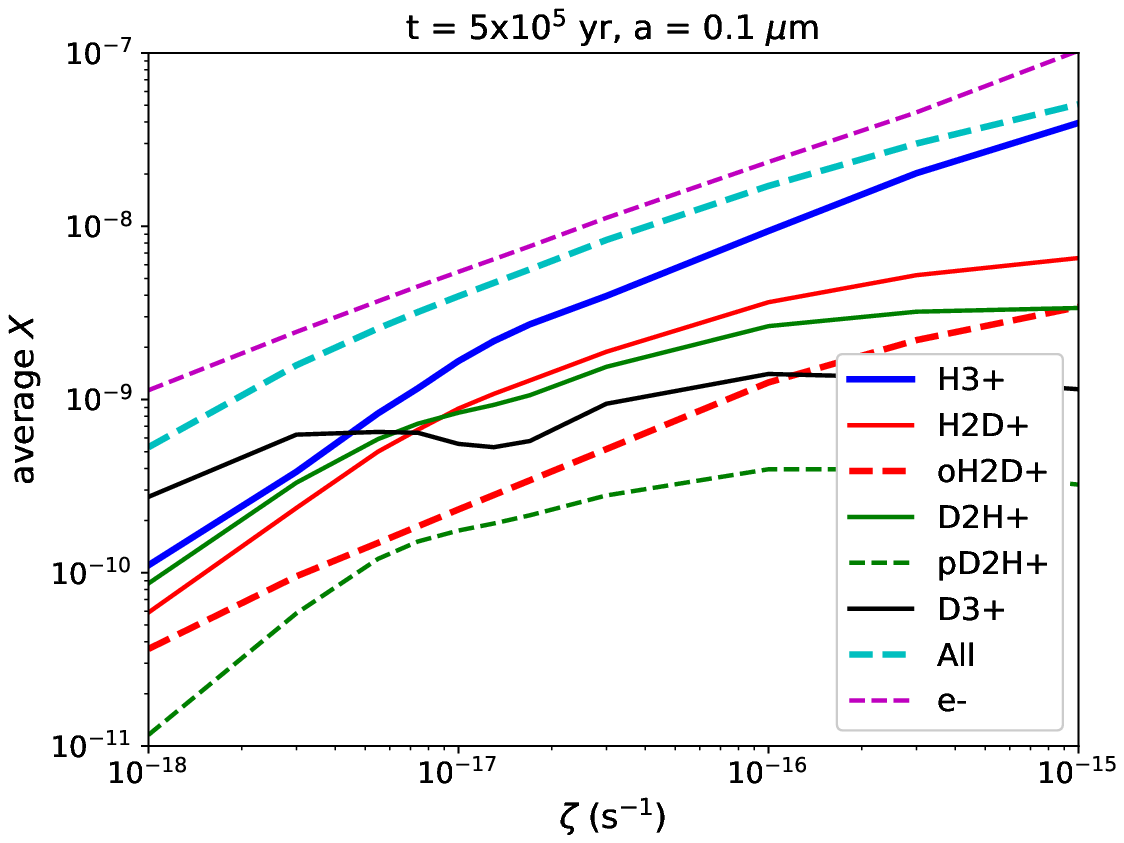}
\end{picture}}
\end{picture}
\caption{Average fractional abundances of $\hthreeplus$, its deuterated isotopologues, and electrons in the core model as functions of the cosmic-ray ionisation rate $\zeta_{\htwo}$. The solid red curve shows the total (ortho + para) abundance of $\htwodplus$, whereas the dashed red curve shows separately the fractional o$\htwodplus$ abundance. Likewise, the dashed green curve shows the fractional p$\dtwohplus$ abundance. The combined fractional abundance of all modifications of $\hthreeplus$, denoted $\toththree$ in the text, is shown with the dashed cyan curve.}
\label{xh3+_combined}
\end{figure}

The average o$\htwodplus$ abundance is predicted to be nearly constant in time and to correlate strongly with the cosmic-ray ionisation rate (Fig.~\ref{x_vs_z_t_n} a and b). The same is true for $\hthreeplus$ and for the combined abundance of all its modifications. In Fig.~\ref{xh3+_combined}, we show the average fractional abundances of $\hthreeplus$, $\htwodplus$, $\dtwohplus$, $\dthreeplus$ as functions of $\zeta_{\htwo}$. The fractional abundances of the observable o$\htwodplus$ and p$\dtwohplus$ cations are shown there with dashed red and green curves, respectively. Also shown is the total abundance of all $\hthreeplus$ isotopologues (cyan dashed curve; denoted $\toththree$). The fractional o$\htwodplus$ abundance follows remarkably well the total abundance { of} $\toththree$, at about one order of magnitude lower level. This is true also for other simulation times. 

The constancy and the correlation of the average o$\htwodplus$ abundance with the cosmic-ray ionisation rate is a consequence of the fact that the formation rate of $\hthreeplus$ is practically equal to $\zeta_{\htwo}\times n(\htwo)$. The $\hthreeplus$ ion is deuterated through reactions with HD, starting with $\hthreeplus + {\rm HD} \leftrightharpoons{} \htwodplus + \htwo$.  { In dense gas, the fractional HD abundance is reduced slowly from the initial value $\sim 3\times10^{-5}$} (\citealt{2013A&A...554A..92S}), and deuteration is boosted when the fractional CO  abundance decreases below that of HD. At high cosmic-ray ionisation rates, deuterium fractionation is, however, suppressed as $\hthreeplus$ is destroyed in electron recombination reactions rather than reactions with HD. In this respect, the situation resembles that in diffuse clouds. On the other hand, at low values of $\zeta_{\htwo}$, $\dthreeplus$ can become the dominant ion. The isotopologues of $\hthreeplus$ together form by far the most abundant group of cations in our core model, and the total $\hthreeplus$ abundance, $X(\toththree)$, is nearly equal to the electron fraction, $X({\rm e^-})$. Therefore, it seems reasonable to apply the approximation presented by \cite{1976RvMP...48..513W} for the minimum fractional ionisation (his Eq.\,4.50) to the total $\hthreeplus$ abundance, and write $X(\toththree) \approx  X({\rm e^-}) \approx \sqrt{\zeta_{\htwo}/(\alpha \bar{n}(\htwo))}$, where $\alpha$ is the rate coefficient of dissociative electron recombination, and  $\bar{n}(\htwo)$ is the average $\htwo$ density. In our core model $\bar{n}(\htwo)\approx 2\times10^5\,\percc$. The formula approximates the dependency of $\toththree$ on $\zeta_{\htwo}$ very well in our models if we set $\alpha=1.9\times10^{-6}\,{\rm cm^3}\,\pers$, giving $X(\toththree) \approx 1.6 \sqrt{\zeta_{\htwo}/\pers}$. The $\alpha$ value quoted is, however, about ten times larger than typical recombination rate coefficients for $\hthreeplus$ and its variants at 10\,K (\citealt{2009A&A...494..623P}; their Table\,B.1), and it can be thought to represent the removal of molecules from the sequence $\hthreeplus \leftrightharpoons{} \htwodplus \leftrightharpoons{} \dtwohplus \leftrightharpoons{} \dthreeplus$ by both dissociative electron recombination (including those on grains) and proton transfer reactions, primarily to CO and N$_2$. Inspection of the main production and destruction pathways during the simulation reveals that at $\zeta_{\htwo}=10^{-17}\,\pers$, the share of electron recombination reactions is approximately 60\% of the total destruction rate of the $\hthreeplus$ family, whereas CO and N$_2$ together are responsible for about 40\% of it. The relative importance of these reaction types change with $\zeta_{\htwo}$, but our simulations suggest that their combined effect on $X(\toththree)$ remains approximately constant. A great majority of reactions involving the altogether nine isotopologues and spin modifications of $\hthreeplus$ is, however, conversions between these forms induced by HD and ortho- or para-$\htwo$.   

The cosmic-ray ionisation rate derived in H-MM1, $5\times10^{-18}\,\pers \lesssim \zeta_{\htwo} \lesssim 10^{-17}\,\pers$, is somewhat lower than previous estimates for dense clouds ($1-5\times10^{-17}\,\pers$; e.g., \citealt{2006PNAS..10312269D}; \citealt{2024A&A...682A.131P}). Direct comparison with previous results is difficult, though, because different molecular tracers, and a number of assumptions about their abundances, have been employed (see below). Cosmic ray propagation models predict that the flux of particles responsible for $\htwo$ ionisation is attenuated by large columns of gas (\citealt{2009A&A...501..619P}; \citealt{2015ApJ...812..135I}; \citealt{2018A&A...614A.111P}; \citealt{2021ApJ...909..107I}; \citealt{2024A&A...682A.131P}). According to attenuation models, $\zeta_{\htwo}$ should stay above $10^{-17}\,\pers$ at column densities typical for starless dense cores ($N(\htwo)\sim 10^{23}\,\persqcm$, $\Sigma \sim 0.5\,{\rm g\,cm^{-2}}$) (e.g., \citealt{2018A&A...614A.111P}, their Appendix F). { The upper limits for the cosmic-ray ionisation
rates derived by \cite{2012ApJ...745...91I} towards diffuse sight lines
in the Ophiuchus-Scorpius region are some of the lowest in their
sample, suggesting that Ophiuchus clouds may be exposed to a weaker
than average cosmic ray flux (see their Section 6.2.4). The results presented here
only pertain the local value of $\zeta_{\htwo}$ in the dense core H-MM1.} Our simulations with two different sets of elemental O, C, and N abundances, and { different} grain sizes (affecting the total surface area of grains, and thus the depletion of CO  and N$_2$ and other molecules with heavy elements) gave similar best-fit $\zeta_{\htwo}$ values, based on the agreement with the observed o$\htwodplus$ abundance, which correlates strongly with $\zeta_{\htwo}$. There is no such simple relationship between the cosmic-ray ionisation rate and the abundances of the two other molecules observed here. { The failure to fit simultaneously the o$\htwodplus$ and $\diaz$ maps, and the underestimation of the $\dcoplus$ abundance are disturbing, but do not undermine the  $\zeta_{\htwo}$ estimate.} 

{ The problem that} our chemical model under-predicts the $\diaz$ abundance has been previously encountered by \cite{2019A&A...629A..15R} and \cite{2021A&A...656A.109R} in the analysis of observations towards the prestellar core L1544. The improvement gained by reducing the grain surface area (end of Sect.~\ref{simulated_maps}) suggests that N$_2$ is excessively depleted in the model. { Effective grain radii exceeding 0.1\,$\mu$m, which were tested in Sect.~\ref{simulated_maps}, would imply that the dust grain size distribution is shifted towards larger grains from the classic dust model of \cite{1977ApJ...217..425M}, with $n(a) \propto a^{-3.5} da$ in the range $a_{\rm min}=50\AA$, $a_{\rm max}=0.25\,\mu$m. Growth of grains to micrometer sizes owing to condensation of icy mantles and coagulation, leading to a decrease in the total surface area of the particles and hence to a lower degree of depletion, is predicted by theoretical models of gas-dust dynamics (e.g., \citealt{1994A&A...291..943O}; \citealt{2009A&A...502..845O}; \citealt{2020EPJWC.22800013L}; \citealt{2021MNRAS.502...15H}). Observational evidence for the presence of large dust particles in dense cores comes from measurements of mid-infrared extinction, the discovery of scattered mid-infrared radiation, variation in the dust opacity at millimetre wavelengths, and changes in the ice band profiles in mid-infrared spectra  (e.g., \citealt{2010A&A...511A...9S}; \citealt{2019A&A...623A.118C};
\citealt{2024NatAs...8..359D}).}

To test if the derived low cosmic-ray ionisation rate can possibly be true, we make a quick calculation of the gas temperature, $T_{\rm gas}$, in the centre of the core and at its outer edge, using the density and the dust temperature distributions of the core model shown in Fig.~\ref{nT_radial}. Assuming that the line cooling rate is very low in the core centre, the cosmic-ray heating rate, $\Gamma_{\rm cr}$, is approximately equal to the gas-dust energy transfer rate, $\Lambda_{\rm gd}$, and adopting the expressions for these quantities from \cite{2001ApJ...557..736G} and \cite{2002A&A...394..275G} (Eqs.\,8 and 10 in the latter paper), the equilibrium $T_{\rm gas}$ corresponding to $\zeta_{\htwo}=7.4\times10^{-18}\,\pers$ is 7.2\,K. Because of the strong gas-dust coupling at densities prevailing in the centre of core, $T_{\rm gas}$ would not rise much above $T_{\rm dust}$ even for substantially higher cosmic-ray ionisation rates. The value measured at the outer edge of the core is, therefore, critical. There, line cooling is important, and $T_{\rm gas}$ is obtained from the equation $\Gamma_{\rm cr} - \Lambda_{\rm g} - \Lambda_{\rm gd} = 0$, where $\Lambda_{\rm g}$ is the gas cooling rate by molecular and atomic lines. Again, we adopt the parametrization of \cite{2001ApJ...557..736G} for this quantity, $\Lambda_{\rm g}=\alpha (T_{\rm gas}/10\,{\rm K})^\beta$. According to our chemistry model, the fractional CO abundance at the outer edge of the core is $\sim 6\times10^{-6}$, which corresponds to a depletion factor (DF) of 10. Consequently, we use the values of $\alpha$ and $\beta$ given in Table\,4 of \cite{2001ApJ...557..736G} for model DF10, interpolated to the density $n(\htwo) =2\times10^4\,\percc$, characteristic of the outer edge of the core (giving $\alpha=6\times10^{-24}\,{\rm ergs\,cm^{-3}\,s^{-1}}$, $\beta=2.7$). The dust temperature there is $\sim 15$\,K. Substituting these numbers into the thermal balance equation one finds that $T_{\rm gas}=11.2$\,K for $\zeta_{\htwo}=7.4\times10^{-18}$. This agrees with the gas temperature model shown in Fig.~\ref{nT_radial}, which is based on the VLA ammonia map of \cite{2022AJ....163..294P}. At the outer edge, the gas is heated by dust grains ($\Lambda_{\rm gd}<0$), otherwise the adopted $\zeta_{\htwo}$ value would imply a lower gas temperature ($\sim 9.3$\,K, if gas-dust energy transfer would be neglected).

Finally, we briefly discuss assumptions made in analytical methods of deriving the electron abundance and the cosmic-ray ionisation from observed abundances. For a thorough analysis of these methods, we refer to \cite{2024A&A...685A..67R}. The analytic methods commonly assume steady state, and derive the abundance of the dominant ion, $\hthreeplus$, from that of an abundant observable ion, such as $\hcoplus$ or o$\htwodplus$.  Here, one has to consider the importance of proton transfer reactions (for example, $\hthreeplus + {\rm CO} \rightarrow \hcoplus + \htwo$) relative to the dissociative electron recombination ($\hthreeplus +{\rm e^-} \rightarrow \htwo + {\rm H}$) in the destruction of $\hthreeplus$ (or its deuterated forms). When high densities are concerned, a proper judgement on this issue would require modelling. If o$\htwodplus$ or p$\dtwohplus$ are used as substitutes of $\hthreeplus$, the difficulty is that neither the ortho/para ratios of these molecules nor the fractionation ratios $\htwodplus/\hthreeplus$ and $\dtwohplus/\hthreeplus$ are known a priori (Fig.~\ref{xh3+_combined}).  A common assumption is that $\dcoplus/\hcoplus$ and $\ddiaz/\diaz$ are $\sim1/3\times\htwodplus/\hthreeplus$. This is based on the idea that $\htwodplus$ is the principal deuterated form of $\hthreeplus$, and that one third of the reactions ${\rm CO}+\htwodplus$ and ${\rm N_2} + \htwodplus$ leads to $\dcoplus$ and $\ddiaz$, respectively. The fact that also $\dtwohplus$ and $\dthreeplus$ contribute to the formation of $\dcoplus$ and $\ddiaz$, are taken into account in the steady-state formula for the fractionation ratios $\ddiaz/\diaz$ and  $\dcoplus/\hcoplus$ presented by \cite{2008A&A...492..703C} and \cite{2009A&A...494..623P} (their Eqs.\,13 and 10, respectively):
$$
\frac{\ddiaz}{\diaz} \approx \frac{\dcoplus}{\hcoplus} \approx \frac{\htwodplus + 2 \dtwohplus + 3\dthreeplus}{3\hthreeplus + 2 \htwodplus + \dtwohplus} \; .
\label{pagani}
$$

In Fig.~\ref{RD} we show the ratios $\ddiaz/\diaz$ and $\dcoplus/\hcoplus$ from our models as functions of the cosmic-ray ionisation rate and time, along with the prediction from the equation above. Also shown is the ratio $1/3 \times \htwodplus/\hthreeplus$, which often is assumed to be equal to $\dcoplus/\hcoplus$ or $\ddiaz/\diaz$. { Besides the model with the fiducial set of parameters, diagrams are shown for the model with the grain radius $a=0.3\,\mu$m, which can approximately reproduce both o$\htwodplus$ and $\diaz$ intensities.} While the steady-state formula shown above gives a nearly perfect match, one can see that the $\dcoplus/\hcoplus$ and $\ddiaz/\diaz$ ratios do not serve as good templates of the quantity $1/3\times \htwodplus/\hthreeplus$ in this case. The discrepancy is a factor of 2-3, and thus still within the estimated accuracy of the common approximation according to \cite{2024A&A...685A..67R}. For targets with lower average densities the agreement is probably better. Unfortunately, the present observations do not include $\hcoplus$ or $\ddiaz$ lines to test the predictions for the fractionation ratios. 

\section{Conclusions}
\label{conclusions}

Each of the three common molecular ions, o$\htwodplus$, $\diaz$, and $\dcoplus$, presents a different picture of the prestellar core H-MM1. The o$\htwodplusline$ emission follows rather closely the general structure of the core seen in the far-infrared and in ammonia maps. The line is excited already at densities around $10^5\,\percc$, but its optical thickness remains moderate also towards the densest regions, partly because of the broad thermal linewidth and the hyperfine structure. The $\diaz(4-3)$ emission is more compact, concentrating on the density maxima. Like $\diaz(4-3)$, also the $\dcoplus(5-4)$ line has a high critical density, but its emission maxima are shifted eastward from the density peaks, towards the more sheltered side of the core.

The chemistry model used to interpret the observed map brings out two advantages of o$\htwodplus$ observations: 1) the fractional o$\htwodplus$ abundance remains relatively high also in the interiors of the core, where $\diaz$ and $\dcoplus$ are depleted by a factor of hundred or more; 2) the average o$\htwodplus$ abundance is nearly constant in time, and correlates strongly with the cosmic-ray ionisation rate owing to its direct relationship to the primary ionisation product $\hthreeplus$. The observed o$\htwodplus$ line emission constrains the cosmic-ray ionisation rate to be $\zeta_{\htwo}\sim5\times10^{-18}\,\pers - 1\times10^{-17}\,\pers$. A substantial increase in the assumed elemental O, C, and N abundances, { or a decrease in the total surface area of the dust grains} in the chemistry model, do not change the best-fit $\zeta_{\htwo}$ value { markedly}, and the estimate $\zeta_{\htwo}\lesssim10^{-17}\,\pers$ in H-MM1 seems robust. The estimate is further supported by the low gas temperature measured on the outskirts of the core. The $\htwo$ column density of the ambient cloud around the core is of the order { of} $10^{22}\,\persqcm$, and the low $\zeta_{\htwo}$ value derived in H-MM1 compared with estimates in diffuse clouds can probably be understood in terms of attenuation of the cosmic ray flux in dense gas (e.g., \citealt{2018A&A...614A.111P}), although propagation models predict higher ionisation rate at this column density. Our modelling results corroborate the conclusion from the analysis presented in \cite{2020MNRAS.495L...7B} and \cite{2024A&A...685A..67R} that o$\htwodplus$ observations are useful for deriving the cosmic-ray ionisation rate in dense dark clouds where $\hthreeplus$ is not observable. However, the o$\htwodplus$ abundance relative to that of $\hthreeplus$ cannot be reliably obtained from the observable fractionation ratios $\dcoplus/\hcoplus$ and $\ddiaz/\diaz$. Modelling { the} o$\htwodplus$ line emission using $\zeta_{\htwo}$ as a parameter provides a more straightforward method.     

\begin{acknowledgements}
We thank the APEX staff for performing the observations presented here, and the referee for helpful suggestions to improve the manuscript. The authors gratefully acknowledge financial support from the Max Planck Society. 
\end{acknowledgements}

\bibliographystyle{aa} 
   \bibliography{bibliography.bib} 

\begin{thebibliography}{71}
\expandafter\ifx\csname natexlab\endcsname\relax\def\natexlab#1{#1}\fi

\bibitem[{{Arzoumanian} {et~al.}(2011){Arzoumanian}, {Andr{\'e}}, {Didelon},
  {K{\"o}nyves}, {Schneider}, {Men'shchikov}, {Sousbie}, {Zavagno}, {Bontemps},
  {di Francesco}, {Griffin}, {Hennemann}, {Hill}, {Kirk}, {Martin}, {Minier},
  {Molinari}, {Motte}, {Peretto}, {Pezzuto}, {Spinoglio}, {Ward-Thompson},
  {White}, \& {Wilson}}]{2011A&A...529L...6A}
{Arzoumanian}, D., {Andr{\'e}}, P., {Didelon}, P., {et~al.} 2011, \aap, 529, L6

\bibitem[{{Bovino} {et~al.}(2020){Bovino}, {Ferrada-Chamorro}, {Lupi},
  {Schleicher}, \& {Caselli}}]{2020MNRAS.495L...7B}
{Bovino}, S., {Ferrada-Chamorro}, S., {Lupi}, A., {Schleicher}, D.~R.~G., \&
  {Caselli}, P. 2020, \mnras, 495, L7

\bibitem[{{Burke} \& {Hollenbach}(1983)}]{1983ApJ...265..223B}
{Burke}, J.~R. \& {Hollenbach}, D.~J. 1983, \apj, 265, 223

\bibitem[{{Caselli} \& {Dore}(2005)}]{2005A&A...433.1145C}
{Caselli}, P. \& {Dore}, L. 2005, \aap, 433, 1145

\bibitem[{{Caselli} {et~al.}(2003){Caselli}, {van der Tak}, {Ceccarelli}, \&
  {Bacmann}}]{2003A&A...403L..37C}
{Caselli}, P., {van der Tak}, F.~F.~S., {Ceccarelli}, C., \& {Bacmann}, A.
  2003, \aap, 403, L37

\bibitem[{{Caselli} {et~al.}(2008){Caselli}, {Vastel}, {Ceccarelli}, {van der
  Tak}, {Crapsi}, \& {Bacmann}}]{2008A&A...492..703C}
{Caselli}, P., {Vastel}, C., {Ceccarelli}, C., {et~al.} 2008, \aap, 492, 703

\bibitem[{{Caselli} {et~al.}(1998){Caselli}, {Walmsley}, {Terzieva}, \&
  {Herbst}}]{1998ApJ...499..234C}
{Caselli}, P., {Walmsley}, C.~M., {Terzieva}, R., \& {Herbst}, E. 1998, \apj,
  499, 234

\bibitem[{{Caselli} {et~al.}(2002){Caselli}, {Walmsley}, {Zucconi}, {Tafalla},
  {Dore}, \& {Myers}}]{2002ApJ...565..344C}
{Caselli}, P., {Walmsley}, C.~M., {Zucconi}, A., {et~al.} 2002, \apj, 565, 344

\bibitem[{{Chac{\'o}n-Tanarro} {et~al.}(2019){Chac{\'o}n-Tanarro}, {Pineda},
  {Caselli}, {Bizzocchi}, {Gutermuth}, {Mason}, {G{\'o}mez-Ruiz}, {Harju},
  {Devlin}, {Dicker}, {Mroczkowski}, {Romero}, {Sievers}, {Stanchfield},
  {Offner}, \& {S{\'a}nchez-Arg{\"u}elles}}]{2019A&A...623A.118C}
{Chac{\'o}n-Tanarro}, A., {Pineda}, J.~E., {Caselli}, P., {et~al.} 2019, \aap,
  623, A118

\bibitem[{{Dalgarno}(2006)}]{2006PNAS..10312269D}
{Dalgarno}, A. 2006, Proceedings of the National Academy of Science, 103, 12269

\bibitem[{{Dartois} {et~al.}(2024){Dartois}, {Noble}, {Caselli}, {Fraser},
  {Jim{\'e}nez-Serra}, {Mat{\'e}}, {McClure}, {Melnick}, {Pendleton},
  {Shimonishi}, {Smith}, {Sturm}, {Taillard}, {Wakelam}, {Boogert},
  {Drozdovskaya}, {Erkal}, {Harsono}, {Herrero}, {Ioppolo}, {Linnartz},
  {McGuire}, {Perotti}, {Qasim}, \& {Rocha}}]{2024NatAs...8..359D}
{Dartois}, E., {Noble}, J.~A., {Caselli}, P., {et~al.} 2024, Nature Astronomy,
  8, 359

\bibitem[{{Denis-Alpizar} {et~al.}(2020){Denis-Alpizar}, {Stoecklin}, {Dutrey},
  \& {Guilloteau}}]{2020MNRAS.497.4276D}
{Denis-Alpizar}, O., {Stoecklin}, T., {Dutrey}, A., \& {Guilloteau}, S. 2020,
  \mnras, 497, 4276

\bibitem[{{Galli} \& {Padovani}(2015)}]{2015arXiv150203380G}
{Galli}, D. \& {Padovani}, M. 2015, Cosmic Rays and the InterStellar Medium,
  24-27 June 2014, Montpellier, France (Proceedins of Science,
  https://pos.sissa.it/221/), arXiv:1502.03380

\bibitem[{{Galli} {et~al.}(2002){Galli}, {Walmsley}, \&
  {Gon{\c{c}}alves}}]{2002A&A...394..275G}
{Galli}, D., {Walmsley}, M., \& {Gon{\c{c}}alves}, J. 2002, \aap, 394, 275

\bibitem[{{Goldsmith}(2001)}]{2001ApJ...557..736G}
{Goldsmith}, P.~F. 2001, \apj, 557, 736

\bibitem[{{Guelin} {et~al.}(1977){Guelin}, {Langer}, {Snell}, \&
  {Wootten}}]{1977ApJ...217L.165G}
{Guelin}, M., {Langer}, W.~D., {Snell}, R.~L., \& {Wootten}, H.~A. 1977, \apjl,
  217, L165

\bibitem[{{Harju} {et~al.}(2017){Harju}, {Daniel}, {Sipil{\"a}}, {Caselli},
  {Pineda}, {Friesen}, {Punanova}, {G{\"u}sten}, {Wiesenfeld}, {Myers},
  {Faure}, {Hily-Blant}, {Rist}, {Rosolowsky}, {Schlemmer}, \&
  {Shirley}}]{2017A&A...600A..61H}
{Harju}, J., {Daniel}, F., {Sipil{\"a}}, O., {et~al.} 2017, \aap, 600, A61

\bibitem[{{Harju} {et~al.}(2024){Harju}, {Pineda}, {Sipil{\"a}}, {Caselli},
  {Belloche}, {Wyrowski}, {Riedel}, {Redaelli}, \&
  {Vasyunin}}]{2024A&A...682A...8H}
{Harju}, J., {Pineda}, J.~E., {Sipil{\"a}}, O., {et~al.} 2024, \aap, 682, A8

\bibitem[{{Harju} {et~al.}(2020){Harju}, {Pineda}, {Vasyunin}, {Caselli},
  {Offner}, {Goodman}, {Juvela}, {Sipil{\"a}}, {Faure}, {Le Gal}, {Hily-Blant},
  {Alves}, {Bizzocchi}, {Burkert}, {Chen}, {Friesen}, {G{\"u}sten}, {Myers},
  {Punanova}, {Rist}, {Rosolowsky}, {Schlemmer}, {Shirley}, {Spezzano},
  {Vastel}, \& {Wiesenfeld}}]{2020ApJ...895..101H}
{Harju}, J., {Pineda}, J.~E., {Vasyunin}, A.~I., {et~al.} 2020, \apj, 895, 101

\bibitem[{{Hirashita} {et~al.}(2021){Hirashita}, {Il'in}, {Pagani}, \&
  {Lef{\`e}vre}}]{2021MNRAS.502...15H}
{Hirashita}, H., {Il'in}, V.~B., {Pagani}, L., \& {Lef{\`e}vre}, C. 2021,
  \mnras, 502, 15

\bibitem[{{Hugo} {et~al.}(2009){Hugo}, {Asvany}, \&
  {Schlemmer}}]{2009JChPh.130p4302H}
{Hugo}, E., {Asvany}, O., \& {Schlemmer}, S. 2009, \jcp, 130, 164302

\bibitem[{{Indriolo} \& {McCall}(2012)}]{2012ApJ...745...91I}
{Indriolo}, N. \& {McCall}, B.~J. 2012, \apj, 745, 91

\bibitem[{{Ivlev} {et~al.}(2015){Ivlev}, {Padovani}, {Galli}, \&
  {Caselli}}]{2015ApJ...812..135I}
{Ivlev}, A.~V., {Padovani}, M., {Galli}, D., \& {Caselli}, P. 2015, \apj, 812,
  135

\bibitem[{{Ivlev} {et~al.}(2021){Ivlev}, {Silsbee}, {Padovani}, \&
  {Galli}}]{2021ApJ...909..107I}
{Ivlev}, A.~V., {Silsbee}, K., {Padovani}, M., \& {Galli}, D. 2021, \apj, 909,
  107

\bibitem[{{Ivlev} {et~al.}(2019){Ivlev}, {Silsbee}, {Sipil{\"a}}, \&
  {Caselli}}]{2019ApJ...884..176I}
{Ivlev}, A.~V., {Silsbee}, K., {Sipil{\"a}}, O., \& {Caselli}, P. 2019, \apj,
  884, 176

\bibitem[{{Jensen} {et~al.}(1997){Jensen}, {Paidarova}, {Spirko}, \&
  {Sauer}}]{1997MolPh..91..319J}
{Jensen}, P., {Paidarova}, I., {Spirko}, V., \& {Sauer}, S. P.~A. 1997,
  Molecular Physics, 91, 319

\bibitem[{{Jusko} {et~al.}(2017){Jusko}, {T{\"o}pfer}, {M{\"u}ller}, {Ghosh},
  {Schlemmer}, \& {Asvany}}]{2017JMoSp.332...33J}
{Jusko}, P., {T{\"o}pfer}, M., {M{\"u}ller}, H. S.~P., {et~al.} 2017, Journal
  of Molecular Spectroscopy, 332, 33

\bibitem[{{Juvela}(2020)}]{2020A&A...644A.151J}
{Juvela}, M. 2020, \aap, 644, A151

\bibitem[{{Lattanzi} {et~al.}(2007){Lattanzi}, {Walters}, {Drouin}, \&
  {Pearson}}]{2007ApJ...662..771L}
{Lattanzi}, V., {Walters}, A., {Drouin}, B.~J., \& {Pearson}, J.~C. 2007, \apj,
  662, 771

\bibitem[{{Lef{\`e}vre} {et~al.}(2020){Lef{\`e}vre}, {Pagani}, {Ladjelate},
  {Min}, {Hirashita}, \& {Zylka}}]{2020EPJWC.22800013L}
{Lef{\`e}vre}, C., {Pagani}, L., {Ladjelate}, B., {et~al.} 2020, in European
  Physical Journal Web of Conferences, Vol. 228, mm Universe @ NIKA2 -
  Observing the mm Universe with the NIKA2 Camera, 00013

\bibitem[{{Lis} {et~al.}(2016){Lis}, {Wootten}, {Gerin}, {Pagani}, {Roueff},
  {van der Tak}, {Vastel}, \& {Walmsley}}]{2016ApJ...827..133L}
{Lis}, D.~C., {Wootten}, H.~A., {Gerin}, M., {et~al.} 2016, \apj, 827, 133

\bibitem[{{Mangum} {et~al.}(2007){Mangum}, {Emerson}, \&
  {Greisen}}]{2007A&A...474..679M}
{Mangum}, J.~G., {Emerson}, D.~T., \& {Greisen}, E.~W. 2007, \aap, 474, 679

\bibitem[{{Mathis} {et~al.}(1977){Mathis}, {Rumpl}, \&
  {Nordsieck}}]{1977ApJ...217..425M}
{Mathis}, J.~S., {Rumpl}, W., \& {Nordsieck}, K.~H. 1977, \apj, 217, 425

\bibitem[{{McKee}(1989)}]{1989ApJ...345..782M}
{McKee}, C.~F. 1989, \apj, 345, 782

\bibitem[{{Neufeld} \& {Wolfire}(2017)}]{2017ApJ...845..163N}
{Neufeld}, D.~A. \& {Wolfire}, M.~G. 2017, \apj, 845, 163

\bibitem[{{Ormel} {et~al.}(2009){Ormel}, {Paszun}, {Dominik}, \&
  {Tielens}}]{2009A&A...502..845O}
{Ormel}, C.~W., {Paszun}, D., {Dominik}, C., \& {Tielens}, A.~G.~G.~M. 2009,
  \aap, 502, 845

\bibitem[{{Ossenkopf} \& {Henning}(1994)}]{1994A&A...291..943O}
{Ossenkopf}, V. \& {Henning}, T. 1994, \aap, 291, 943

\bibitem[{{Padovani} {et~al.}(2009){Padovani}, {Galli}, \&
  {Glassgold}}]{2009A&A...501..619P}
{Padovani}, M., {Galli}, D., \& {Glassgold}, A.~E. 2009, \aap, 501, 619

\bibitem[{{Padovani} {et~al.}(2024){Padovani}, {Galli}, {Scarlett}, {Grassi},
  {Rehill}, {Zammit}, {Bray}, \& {Fursa}}]{2024A&A...682A.131P}
{Padovani}, M., {Galli}, D., {Scarlett}, L.~H., {et~al.} 2024, \aap, 682, A131

\bibitem[{{Padovani} {et~al.}(2018){Padovani}, {Ivlev}, {Galli}, \&
  {Caselli}}]{2018A&A...614A.111P}
{Padovani}, M., {Ivlev}, A.~V., {Galli}, D., \& {Caselli}, P. 2018, \aap, 614,
  A111

\bibitem[{{Padovani} {et~al.}(2020){Padovani}, {Ivlev}, {Galli}, {Offner},
  {Indriolo}, {Rodgers-Lee}, {Marcowith}, {Girichidis}, {Bykov}, \&
  {Kruijssen}}]{2020SSRv..216...29P}
{Padovani}, M., {Ivlev}, A.~V., {Galli}, D., {et~al.} 2020, \ssr, 216, 29

\bibitem[{{Pagani} {et~al.}(2009{\natexlab{a}}){Pagani}, {Daniel}, \&
  {Dubernet}}]{2009A&A...494..719P}
{Pagani}, L., {Daniel}, F., \& {Dubernet}, M.~L. 2009{\natexlab{a}}, \aap, 494,
  719

\bibitem[{{Pagani} {et~al.}(2009{\natexlab{b}}){Pagani}, {Vastel}, {Hugo},
  {Kokoouline}, {Greene}, {Bacmann}, {Bayet}, {Ceccarelli}, {Peng}, \&
  {Schlemmer}}]{2009A&A...494..623P}
{Pagani}, L., {Vastel}, C., {Hugo}, E., {et~al.} 2009{\natexlab{b}}, \aap, 494,
  623

\bibitem[{{Parise} {et~al.}(2011){Parise}, {Belloche}, {Du}, {G{\"u}sten}, \&
  {Menten}}]{2011A&A...526A..31P}
{Parise}, B., {Belloche}, A., {Du}, F., {G{\"u}sten}, R., \& {Menten}, K.~M.
  2011, \aap, 526, A31

\bibitem[{{Pineda} {et~al.}(2022){Pineda}, {Harju}, {Caselli}, {Sipil{\"a}},
  {Juvela}, {Vastel}, {Rosolowsky}, {Burkert}, {Friesen}, {Shirley},
  {Maureira}, {Choudhury}, {Segura-Cox}, {G{\"u}sten}, {Punanova}, {Bizzocchi},
  \& {Goodman}}]{2022AJ....163..294P}
{Pineda}, J.~E., {Harju}, J., {Caselli}, P., {et~al.} 2022, \aj, 163, 294

\bibitem[{{Pineda} {et~al.}(2020){Pineda}, {Segura-Cox}, {Caselli},
  {Cunningham}, {Zhao}, {Schmiedeke}, {Maureira}, \&
  {Neri}}]{2020NatAs...4.1158P}
{Pineda}, J.~E., {Segura-Cox}, D., {Caselli}, P., {et~al.} 2020, Nature
  Astronomy, 4, 1158

\bibitem[{{Redaelli} {et~al.}(2019){Redaelli}, {Bizzocchi}, {Caselli},
  {Sipil{\"a}}, {Lattanzi}, {Giuliano}, \& {Spezzano}}]{2019A&A...629A..15R}
{Redaelli}, E., {Bizzocchi}, L., {Caselli}, P., {et~al.} 2019, \aap, 629, A15

\bibitem[{{Redaelli} {et~al.}(2024){Redaelli}, {Bovino}, {Lupi}, {Grassi},
  {Gaete-Espinoza}, {Sabatini}, \& {Caselli}}]{2024A&A...685A..67R}
{Redaelli}, E., {Bovino}, S., {Lupi}, A., {et~al.} 2024, \aap, 685, A67

\bibitem[{{Redaelli} {et~al.}(2021){Redaelli}, {Sipil{\"a}}, {Padovani},
  {Caselli}, {Galli}, \& {Ivlev}}]{2021A&A...656A.109R}
{Redaelli}, E., {Sipil{\"a}}, O., {Padovani}, M., {et~al.} 2021, \aap, 656,
  A109

\bibitem[{{Riedel} {et~al.}(2023){Riedel}, {Sipil{\"a}}, {Redaelli}, {Caselli},
  {Vasyunin}, {Dulieu}, \& {Watanabe}}]{2023A&A...680A..87R}
{Riedel}, W., {Sipil{\"a}}, O., {Redaelli}, E., {et~al.} 2023, \aap, 680, A87

\bibitem[{{Roberts} {et~al.}(2003){Roberts}, {Herbst}, \&
  {Millar}}]{2003ApJ...591L..41R}
{Roberts}, H., {Herbst}, E., \& {Millar}, T.~J. 2003, \apjl, 591, L41

\bibitem[{{Sabatini} {et~al.}(2023){Sabatini}, {Bovino}, \&
  {Redaelli}}]{2023ApJ...947L..18S}
{Sabatini}, G., {Bovino}, S., \& {Redaelli}, E. 2023, \apjl, 947, L18

\bibitem[{{Sawada} {et~al.}(2008){Sawada}, {Ikeda}, {Sunada}, {Kuno},
  {Kamazaki}, {Morita}, {Kurono}, {Koura}, {Abe}, {Kawase}, {Maekawa},
  {Horigome}, \& {Yanagisawa}}]{2008PASJ...60..445S}
{Sawada}, T., {Ikeda}, N., {Sunada}, K., {et~al.} 2008, \pasj, 60, 445

\bibitem[{{Sch{\"o}ier} {et~al.}(2005){Sch{\"o}ier}, {van der Tak}, {van
  Dishoeck}, \& {Black}}]{2005A&A...432..369S}
{Sch{\"o}ier}, F.~L., {van der Tak}, F.~F.~S., {van Dishoeck}, E.~F., \&
  {Black}, J.~H. 2005, \aap, 432, 369

\bibitem[{{Shirley}(2015)}]{2015PASP..127..299S}
{Shirley}, Y.~L. 2015, \pasp, 127, 299

\bibitem[{{Sipil{\"a}} {et~al.}(2013){Sipil{\"a}}, {Caselli}, \&
  {Harju}}]{2013A&A...554A..92S}
{Sipil{\"a}}, O., {Caselli}, P., \& {Harju}, J. 2013, \aap, 554, A92

\bibitem[{{Sipil{\"a}} {et~al.}(2019){Sipil{\"a}}, {Caselli}, \&
  {Harju}}]{2019A&A...631A..63S}
{Sipil{\"a}}, O., {Caselli}, P., \& {Harju}, J. 2019, \aap, 631, A63

\bibitem[{{Sipil{\"a}} {et~al.}(2022){Sipil{\"a}}, {Caselli}, {Redaelli}, \&
  {Spezzano}}]{2022A&A...668A.131S}
{Sipil{\"a}}, O., {Caselli}, P., {Redaelli}, E., \& {Spezzano}, S. 2022, \aap,
  668, A131

\bibitem[{{Spezzano} {et~al.}(2016){Spezzano}, {Bizzocchi}, {Caselli}, {Harju},
  \& {Br{\"u}nken}}]{2016A&A...592L..11S}
{Spezzano}, S., {Bizzocchi}, L., {Caselli}, P., {Harju}, J., \& {Br{\"u}nken},
  S. 2016, \aap, 592, L11

\bibitem[{{Spezzano} {et~al.}(2020){Spezzano}, {Caselli}, {Pineda},
  {Bizzocchi}, {Prudenzano}, \& {Nagy}}]{2020A&A...643A..60S}
{Spezzano}, S., {Caselli}, P., {Pineda}, J.~E., {et~al.} 2020, \aap, 643, A60

\bibitem[{{Steinacker} {et~al.}(2010){Steinacker}, {Pagani}, {Bacmann}, \&
  {Guieu}}]{2010A&A...511A...9S}
{Steinacker}, J., {Pagani}, L., {Bacmann}, A., \& {Guieu}, S. 2010, \aap, 511,
  A9

\bibitem[{{Valdivia-Mena} {et~al.}(2022){Valdivia-Mena}, {Pineda},
  {Segura-Cox}, {Caselli}, {Neri}, {L{\'o}pez-Sepulcre}, {Cunningham},
  {Bouscasse}, {Semenov}, {Henning}, {Pi{\'e}tu}, {Chapillon}, {Dutrey},
  {Fuente}, {Guilloteau}, {Hsieh}, {Jim{\'e}nez-Serra}, {Marino}, {Maureira},
  {Smirnov-Pinchukov}, {Tafalla}, \& {Zhao}}]{2022A&A...667A..12V}
{Valdivia-Mena}, M.~T., {Pineda}, J.~E., {Segura-Cox}, D.~M., {et~al.} 2022,
  \aap, 667, A12

\bibitem[{{Valdivia-Mena} {et~al.}(2023){Valdivia-Mena}, {Pineda},
  {Segura-Cox}, {Caselli}, {Schmiedeke}, {Choudhury}, {Offner}, {Neri},
  {Goodman}, \& {Fuller}}]{2023A&A...677A..92V}
{Valdivia-Mena}, M.~T., {Pineda}, J.~E., {Segura-Cox}, D.~M., {et~al.} 2023,
  \aap, 677, A92

\bibitem[{{van der Tak} \& {van Dishoeck}(2000)}]{2000A&A...358L..79V}
{van der Tak}, F.~F.~S. \& {van Dishoeck}, E.~F. 2000, \aap, 358, L79

\bibitem[{{Vastel} {et~al.}(2006{\natexlab{a}}){Vastel}, {Caselli},
  {Ceccarelli}, {Phillips}, {Wiedner}, {Peng}, {Houde}, \&
  {Dominik}}]{2006ApJ...645.1198V}
{Vastel}, C., {Caselli}, P., {Ceccarelli}, C., {et~al.} 2006{\natexlab{a}},
  \apj, 645, 1198

\bibitem[{{Vastel} {et~al.}(2006{\natexlab{b}}){Vastel}, {Phillips}, {Caselli},
  {Ceccarelli}, \& {Pagani}}]{2006RSPTA.364.3081V}
{Vastel}, C., {Phillips}, T.~G., {Caselli}, P., {Ceccarelli}, C., \& {Pagani},
  L. 2006{\natexlab{b}}, Philosophical Transactions of the Royal Society of
  London Series A, 364, 3081

\bibitem[{{Vastel} {et~al.}(2004){Vastel}, {Phillips}, \&
  {Yoshida}}]{2004ApJ...606L.127V}
{Vastel}, C., {Phillips}, T.~G., \& {Yoshida}, H. 2004, \apjl, 606, L127

\bibitem[{{Vasyunin} {et~al.}(2017){Vasyunin}, {Caselli}, {Dulieu}, \&
  {Jim{\'e}nez-Serra}}]{2017ApJ...842...33V}
{Vasyunin}, A.~I., {Caselli}, P., {Dulieu}, F., \& {Jim{\'e}nez-Serra}, I.
  2017, \apj, 842, 33

\bibitem[{{Walmsley} {et~al.}(2004){Walmsley}, {Flower}, \& {Pineau des
  For{\^e}ts}}]{2004A&A...418.1035W}
{Walmsley}, C.~M., {Flower}, D.~R., \& {Pineau des For{\^e}ts}, G. 2004, \aap,
  418, 1035

\bibitem[{{Watson}(1976)}]{1976RvMP...48..513W}
{Watson}, W.~D. 1976, Reviews of Modern Physics, 48, 513

\bibitem[{{Wootten} {et~al.}(1979){Wootten}, {Snell}, \&
  {Glassgold}}]{1979ApJ...234..876W}
{Wootten}, A., {Snell}, R., \& {Glassgold}, A.~E. 1979, \apj, 234, 876

\end{thebibliography}

\begin{appendix}

\section{Radial velocity and velocity dispersion maps}

\begin{figure*}
\unitlength=1mm
\begin{picture}(160,90)(0,0)

\put(0,38){
\begin{picture}(0,0) 
\includegraphics[width=5cm,angle=0]{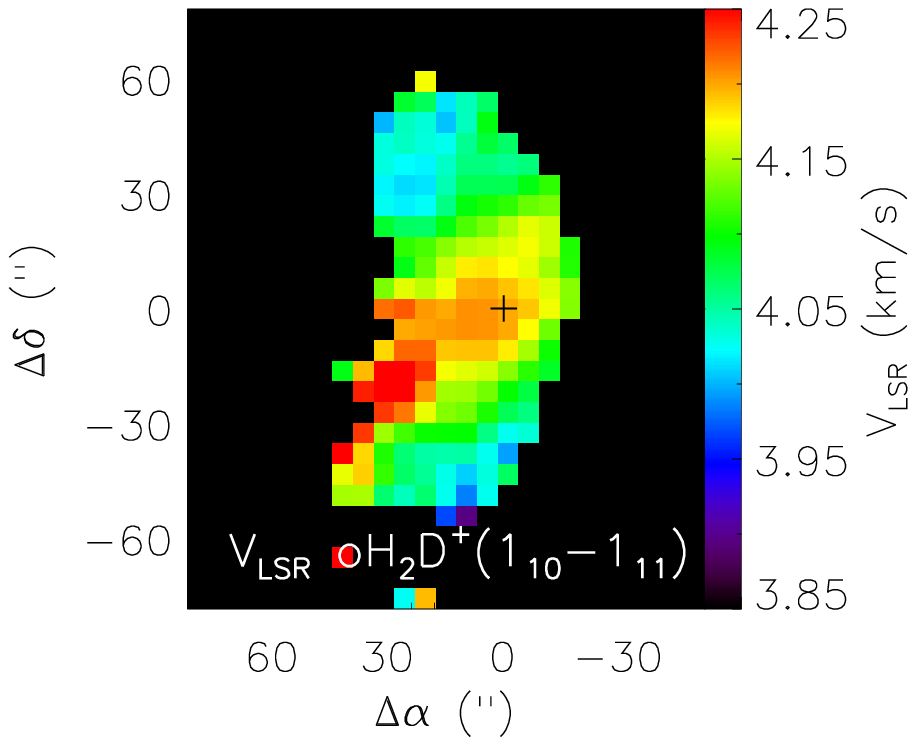}\end{picture}}
\put(45,38){
\begin{picture}(0,0) 
\includegraphics[width=5cm,angle=0]{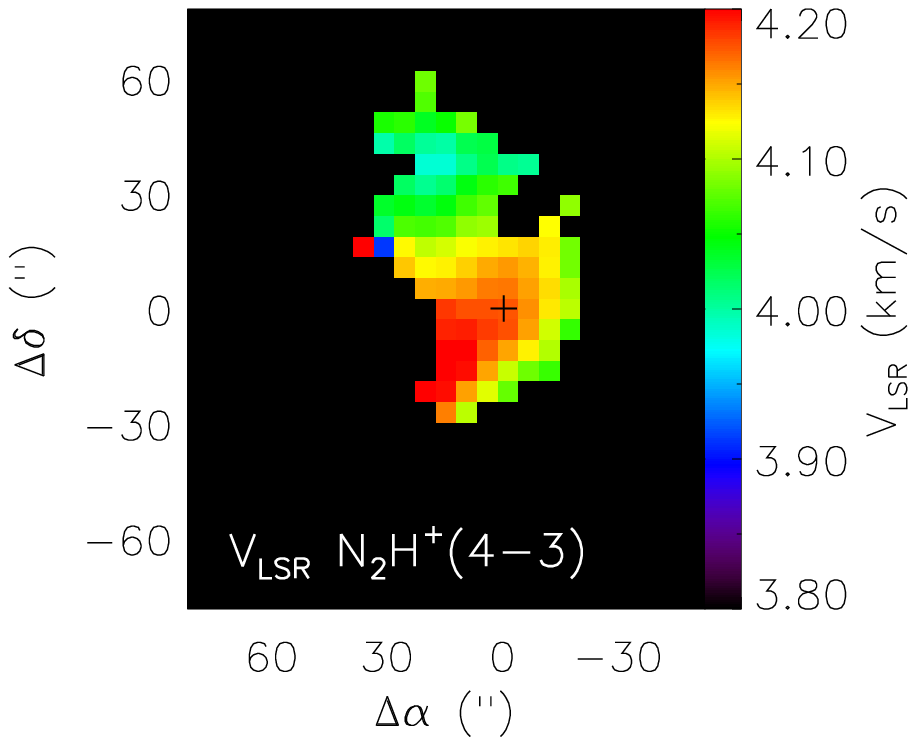}\end{picture}}
\put(90,38){
\begin{picture}(0,0) 
\includegraphics[width=5cm,angle=0]{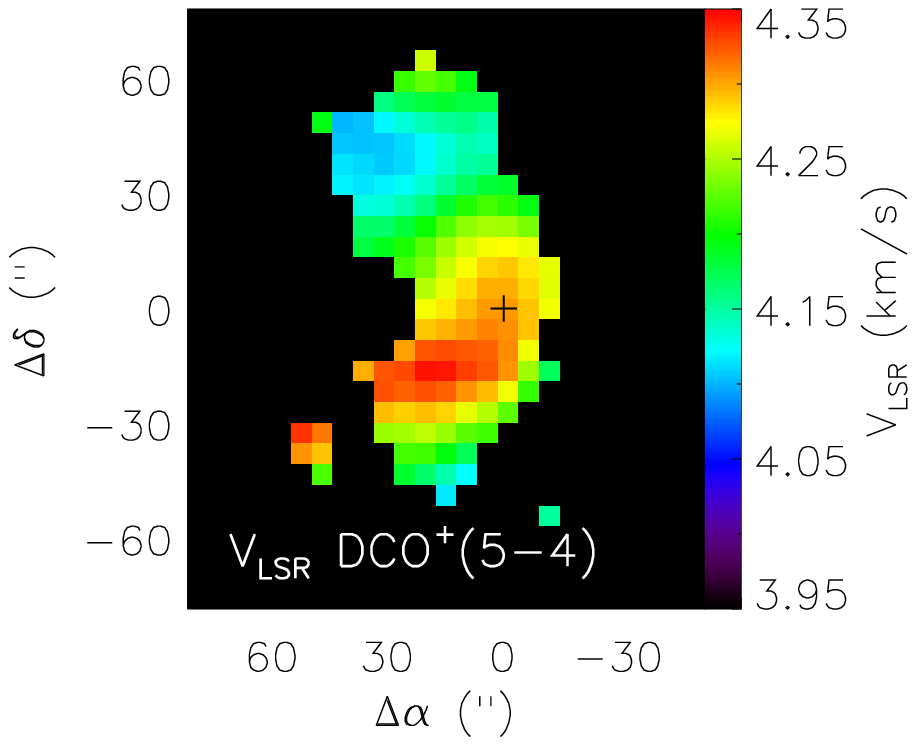}\end{picture}}
\put(135,38){
\begin{picture}(0,0) 
\includegraphics[width=5cm,angle=0]{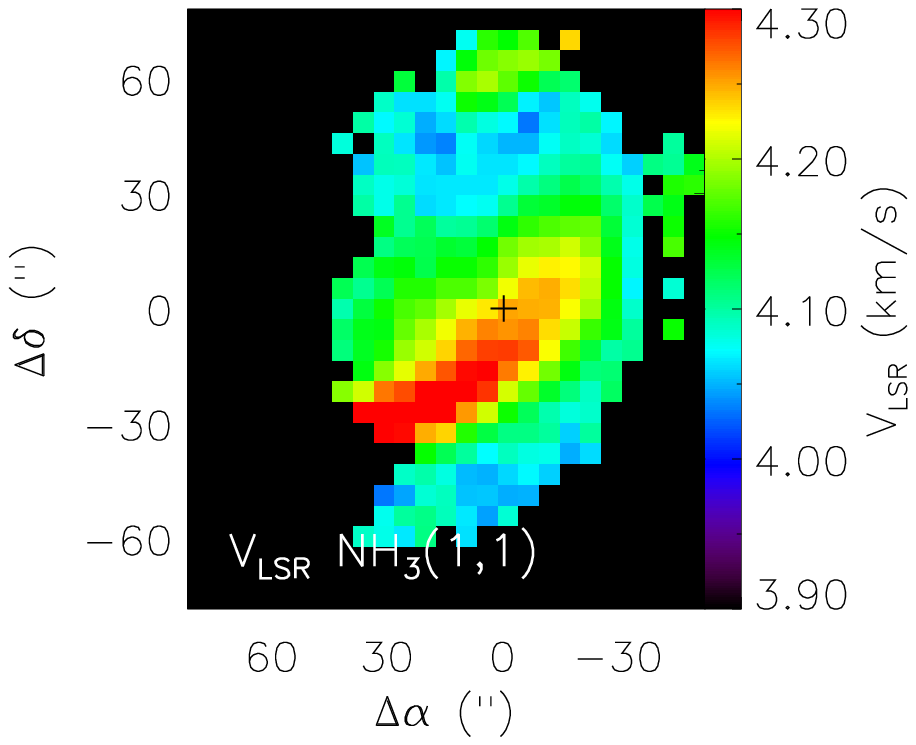}\end{picture}}
\put(0,0){
\begin{picture}(0,0) 
\includegraphics[width=5cm,angle=0]{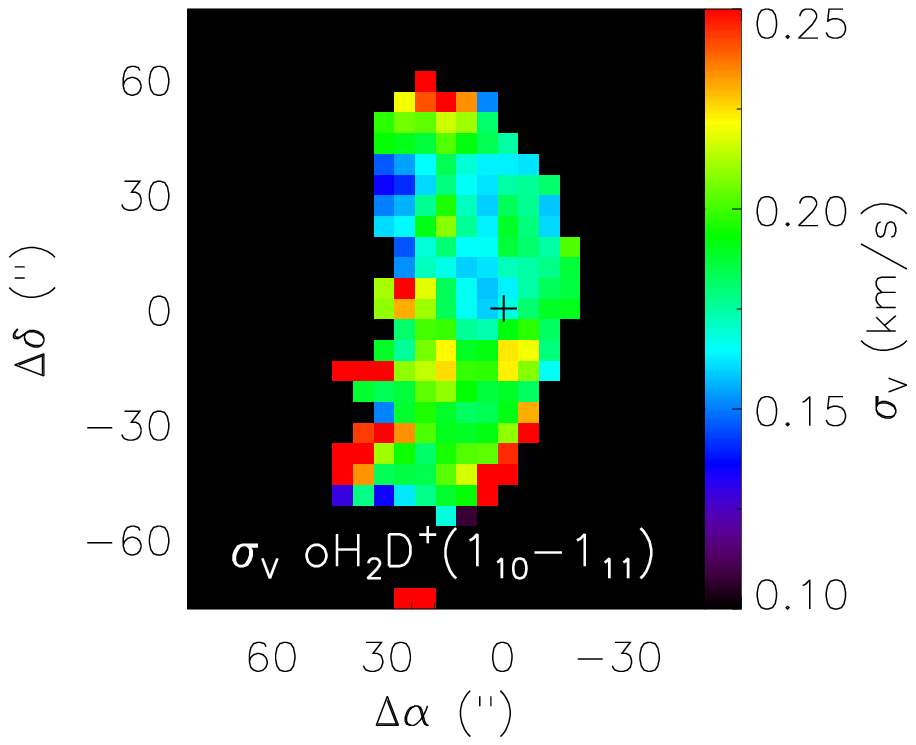}\end{picture}}
\put(45,0){
\begin{picture}(0,0) 
\includegraphics[width=5cm,angle=0]{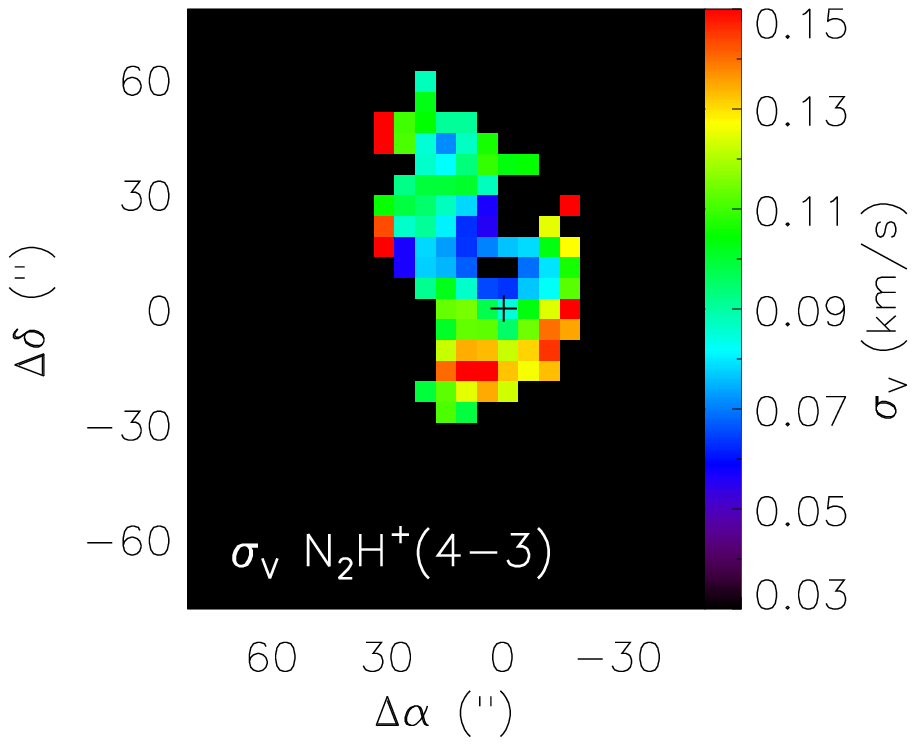}\end{picture}}
\put(90,0){
\begin{picture}(0,0) 
\includegraphics[width=5cm,angle=0]{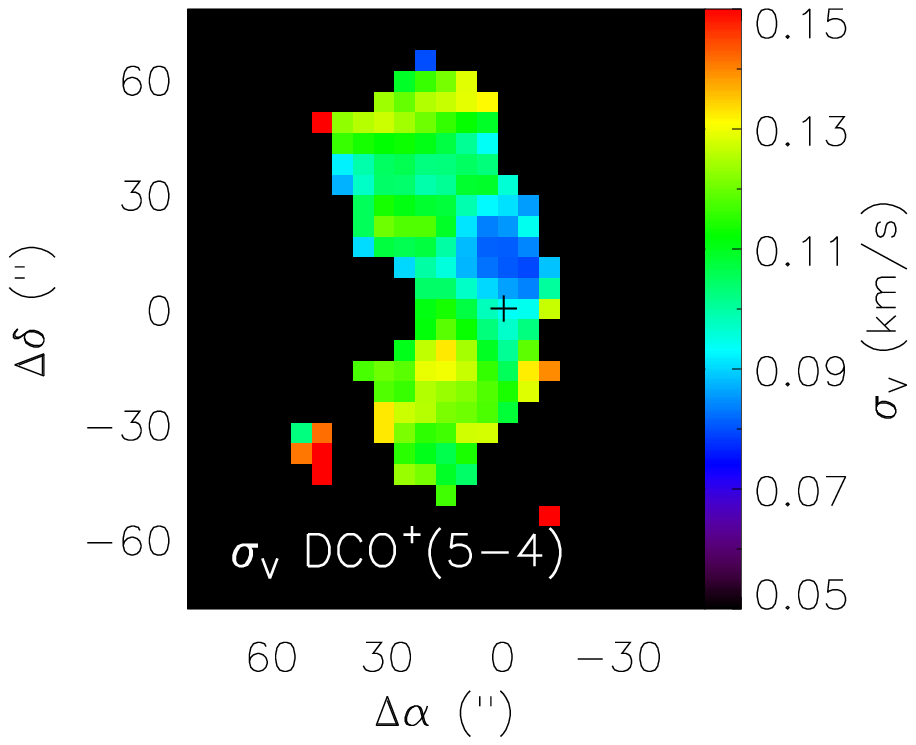}\end{picture}}
\put(135,0){
\begin{picture}(0,0) 
\includegraphics[width=5cm,angle=0]{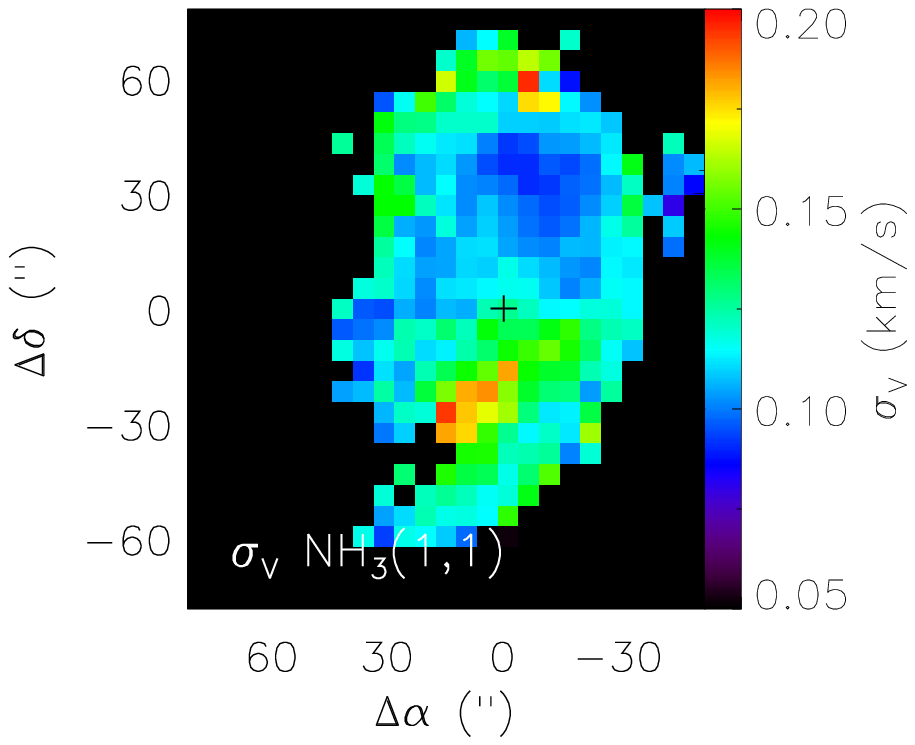}\end{picture}}
\end{picture}
\caption{Distributions of the centroid LSR velocities and velocity dispersions of the lines observed towards H-MM1. Points with low signal-to-noise ratios are excluded. The right hand panel shows results from the VLA $\ammo$ mapping  by \cite{2022AJ....163..294P}, resampled to the grid used for the LAsMA maps.}
\label{v_and_sigmav}
\end{figure*}

The distributions of the { centroid} LSR velocities and the line-of-sight velocity dispersions of the observed lines in H-MM1 are shown in Fig.~\ref{v_and_sigmav}. The points with high signal-to-noise ratios are selected to these images. Besides maps derived from the o$\htwodplus$, $\diaz$, and $\dcoplus$ lines, also shown are the { centroid} $V_{\rm LSR}$ and $\sigma_V$ distributions of the $\ammo(1,1)$ line from the VLA mapping of \cite{2022AJ....163..294P}, resampled to the grid used for the present LAsMA observations. 

\section{Fitting the cosmic-ray ionisation rate}

\begin{figure*}
\unitlength=1mm

\begin{picture}(160,60)(0,0)
\put(-3,0){
\begin{picture}(0,0) 
\includegraphics[width=6cm,angle=0]{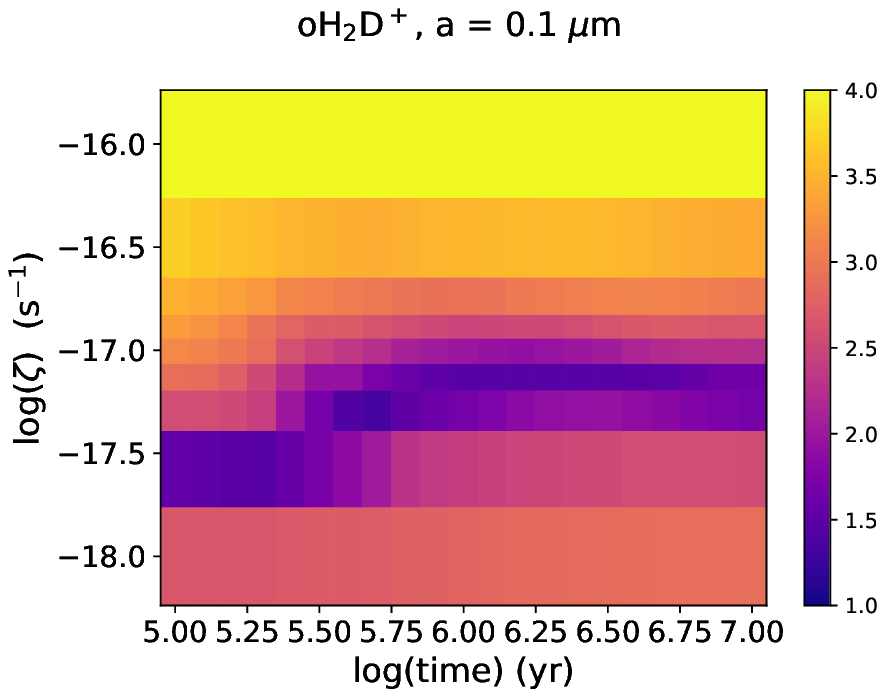}
\end{picture}}

\put(60,0){
\begin{picture}(0,0) 
\includegraphics[width=6cm,angle=0]{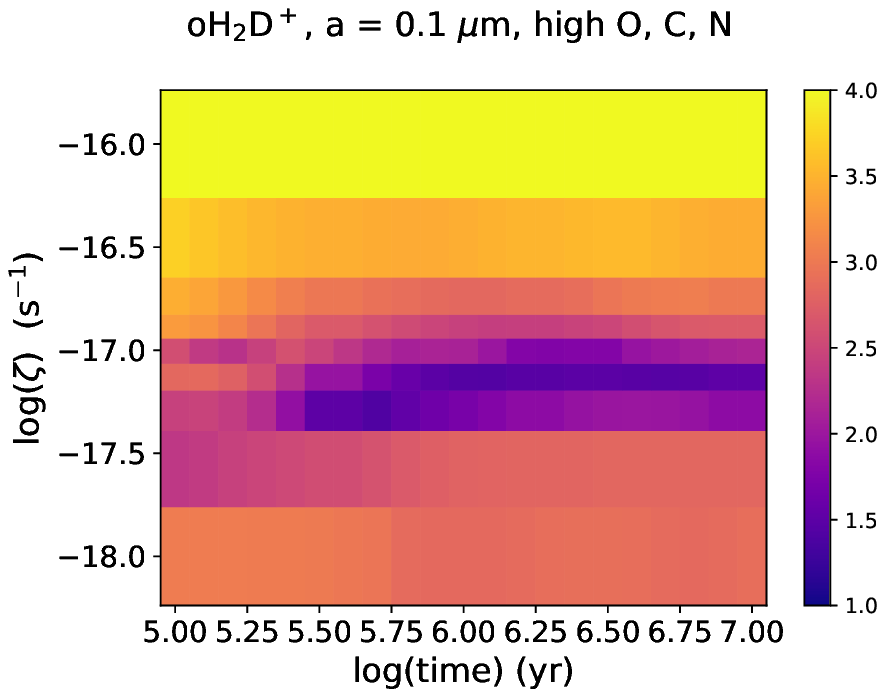}
\end{picture}}

\put(123,0){
\begin{picture}(0,0) 
\includegraphics[width=6cm,angle=0]{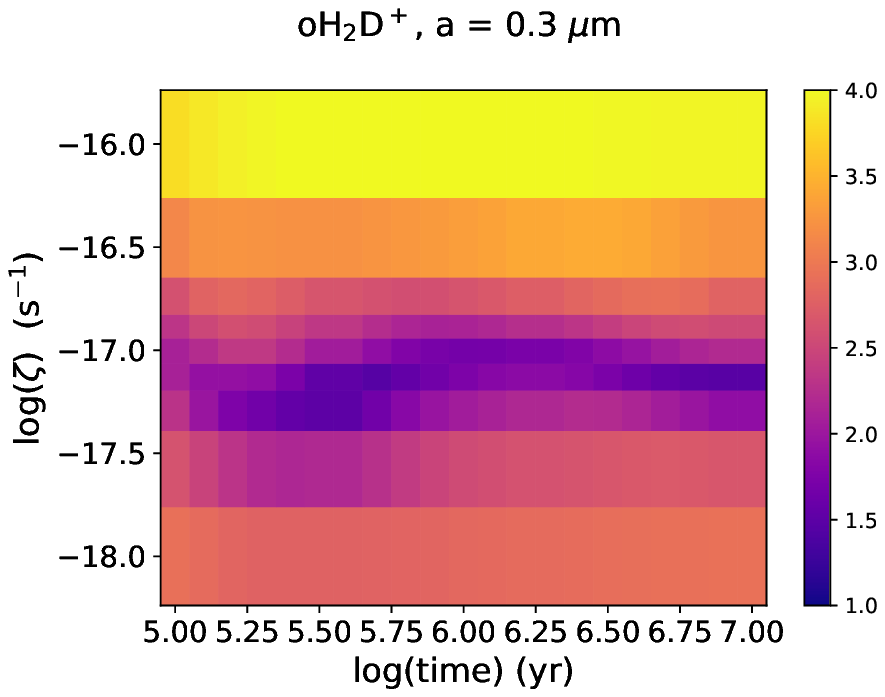}
\end{picture}}

\put(11,34){\bf \Large a}
\put(75,34){\bf \Large b}
\put(137,34){\bf \Large c}
\end{picture}

\caption{Chi-square maps from comparison between the observed and simulated o$\htwodplus$ integrated intensity maps described in Sect.~\ref{simulated_maps}. The $y$-axis in the plots is the logarithm of the cosmic-ray ionisation rate, $\zeta_{\htwo}$, and the $x$-axis is logarithm of time. Also the $\chi^2$ values are presented on a logarithmic scale: a darker shade means a better fit. The plot on the left ({\bf a}) represents simulations with { the grain radius $a=0.1\,\mu$m and} low elemental abundances of O, C, and N, assumed in several previous models of H-MM1, and that { in the middle (b)} simulations with some 40\%-60\% higher elemental O, C, and N abundances. { The chi-square map on the right ({\bf c}), is from a simulation with the grain radius $a=0.3\,\mu$m and low O, C, and N abundances}  (see Sect.~\ref{chemistry_model}).}
\label{chi2_maps}
\end{figure*}

The cosmic-ray ionisation rate in the core is estimated by comparing the simulated integrated intensity maps to those observed as described in Sect.~\ref{simulated_maps}. Fig.~\ref{chi2_maps} shows the chi-square distributions as functions of time and $\zeta_{\htwo}$ for { chemistry models assuming either different elemental abundances or dust grain radii}. The pixelation reflects the discrete sets of simulation times and $\zeta_{\htwo}$ values tested.

\section{Modelled maps and spectra}
\label{simulated_lines}

\begin{figure*}
\unitlength=1mm
\begin{picture}(160,210)(0,0)

\put(90,140){
\begin{picture}(0,0) 
\includegraphics[width=9cm,angle=0]{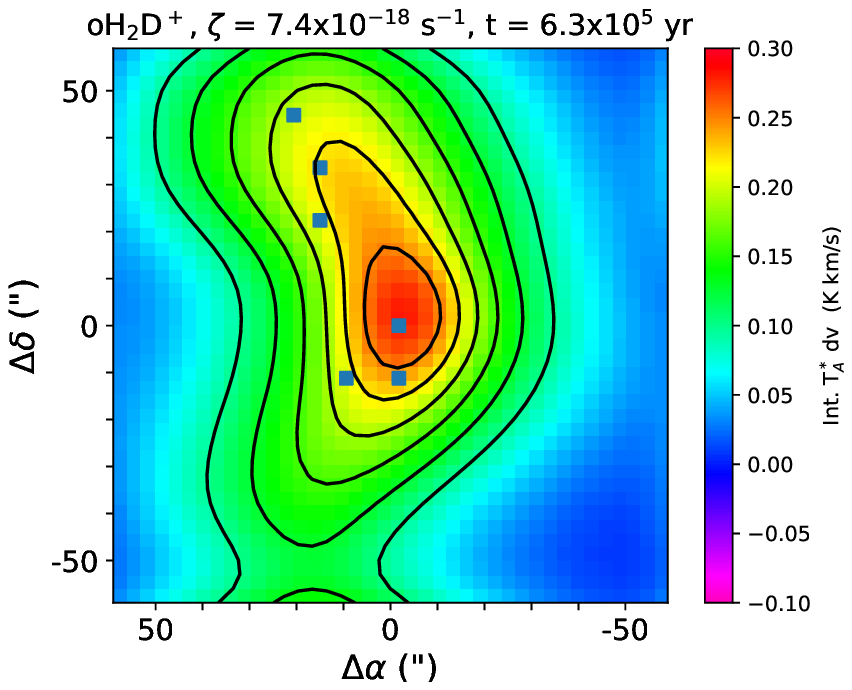}
\end{picture}}

\put(10,205){
\begin{picture}(0,0) 
\includegraphics[width=6.5cm,angle=270]{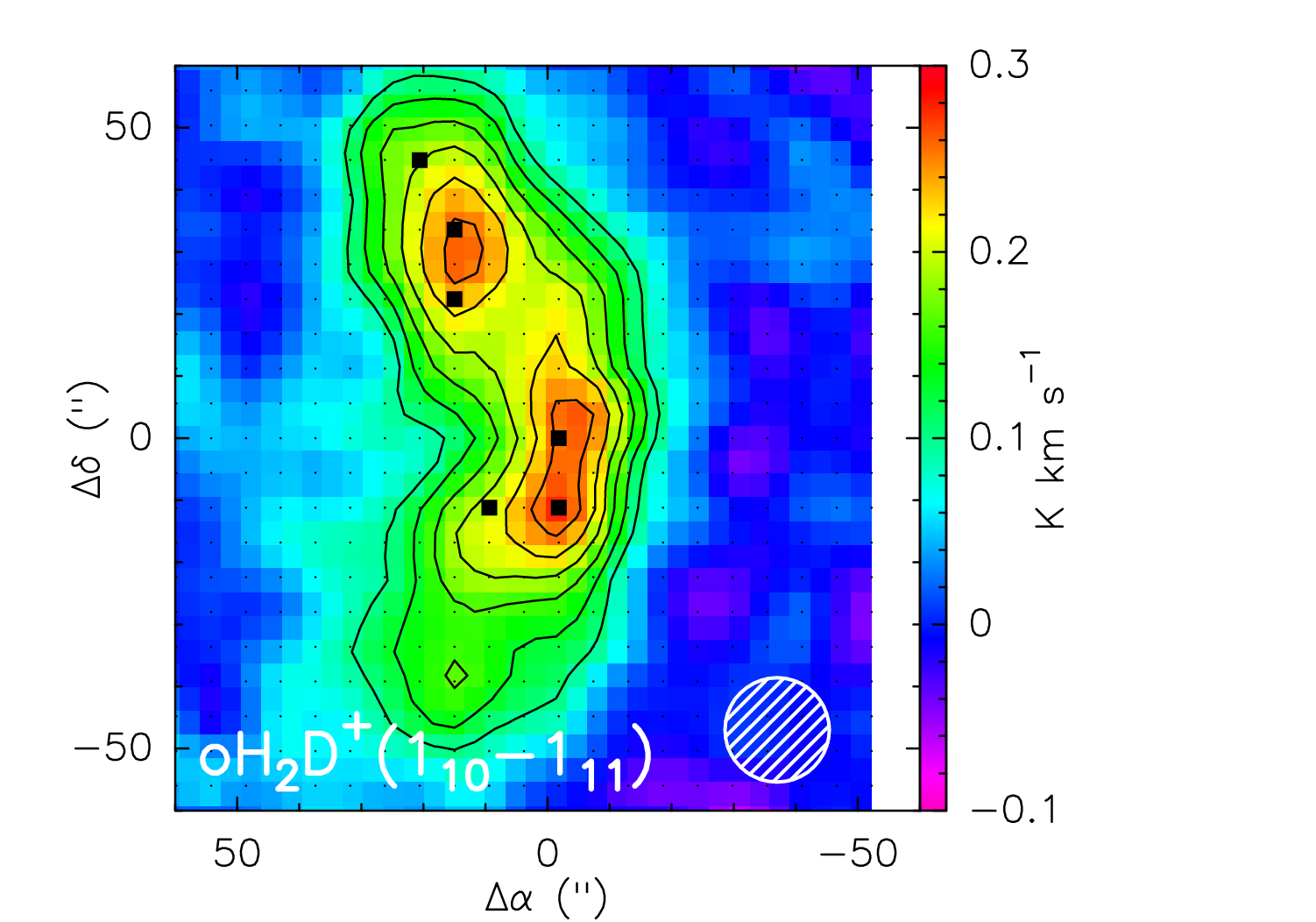}
\end{picture}}

\put(90,70){
\begin{picture}(0,0) 
\includegraphics[width=9cm,angle=0]{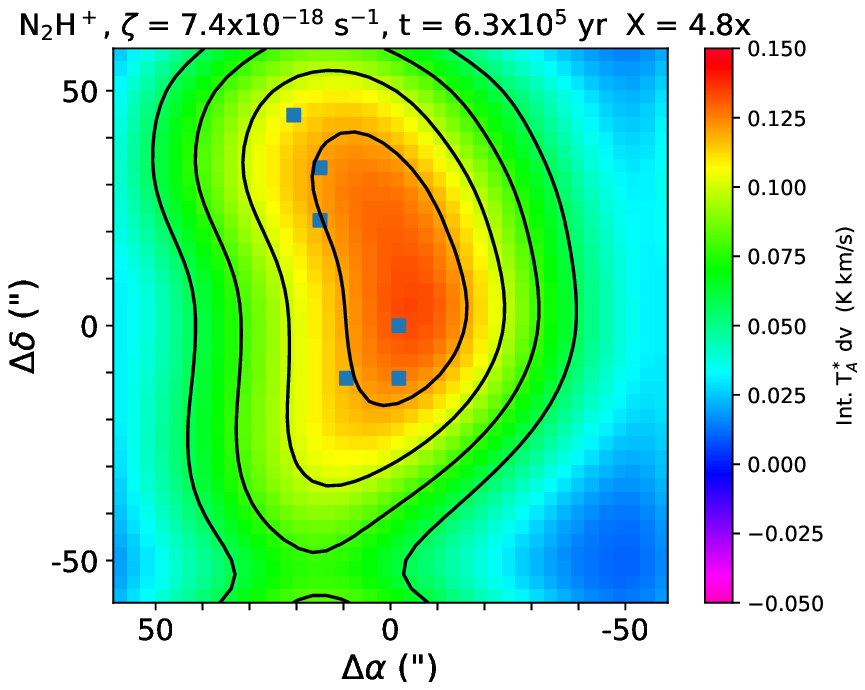}
\end{picture}}

\put(10,135){
\begin{picture}(0,0) 
\includegraphics[width=6.5cm,angle=270]{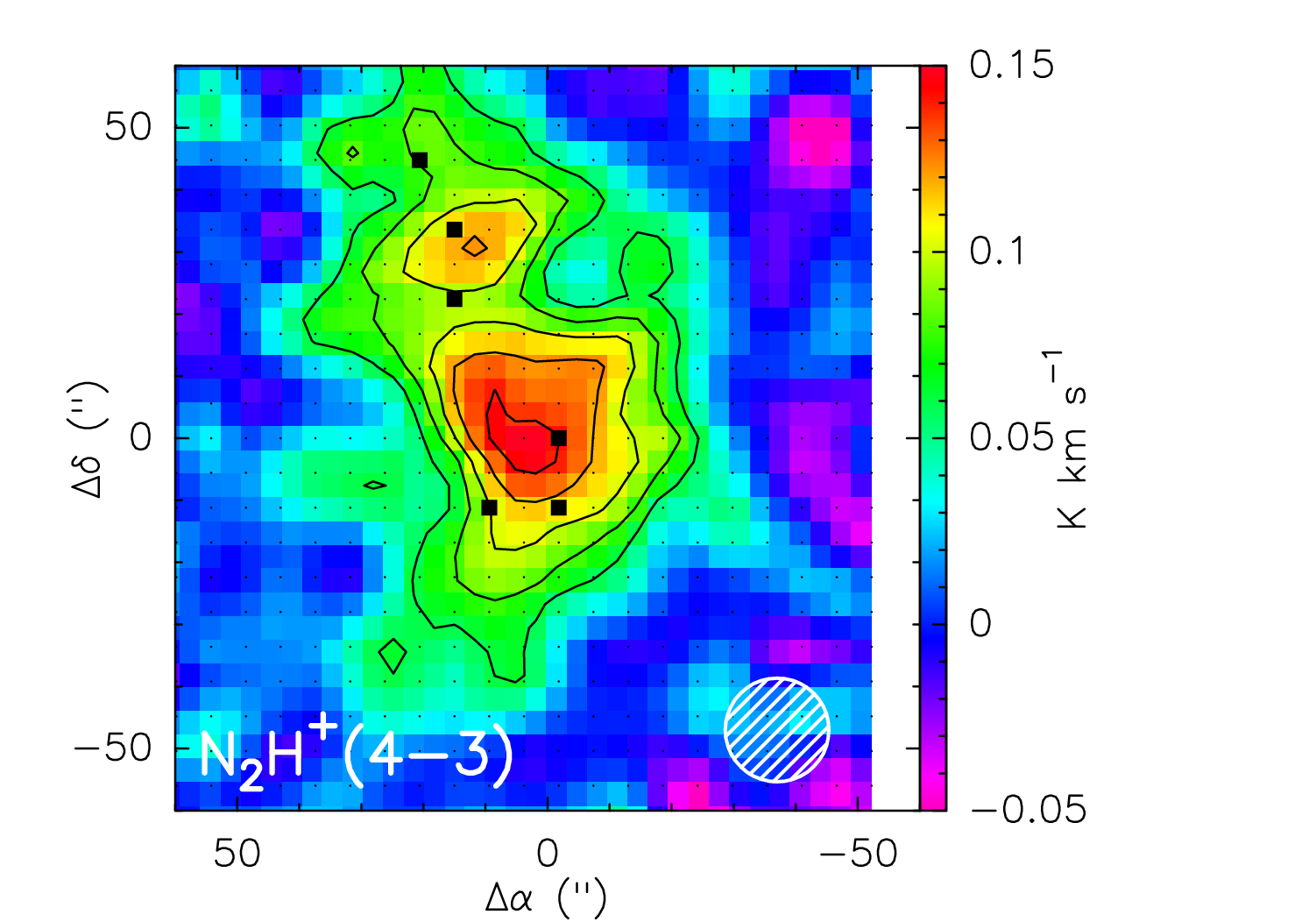}
\end{picture}}

\put(90,0){
\begin{picture}(0,0) 
\includegraphics[width=9cm,angle=0]{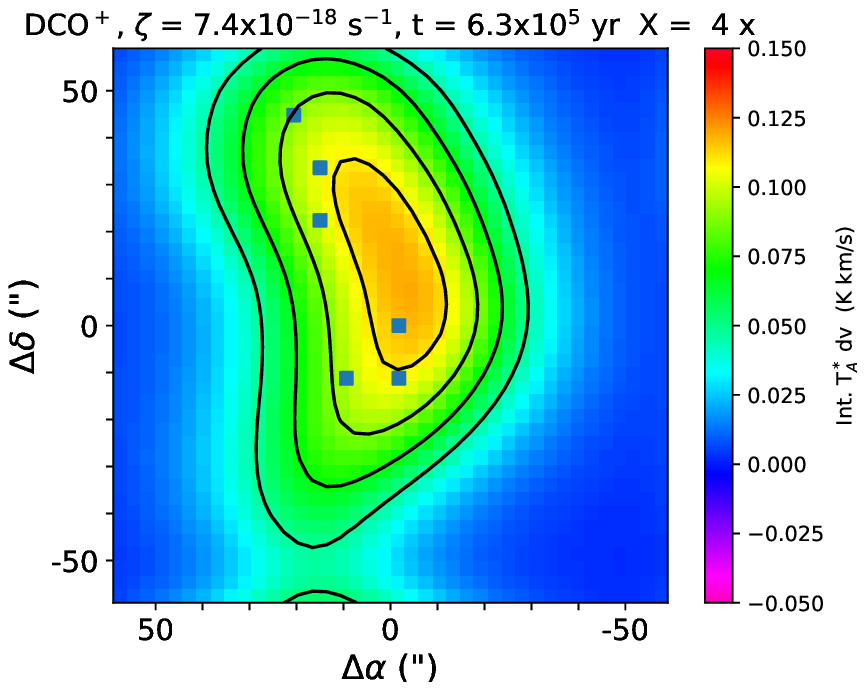}
\end{picture}}

\put(10,65){
\begin{picture}(0,0) 
\includegraphics[width=6.5cm,angle=270]{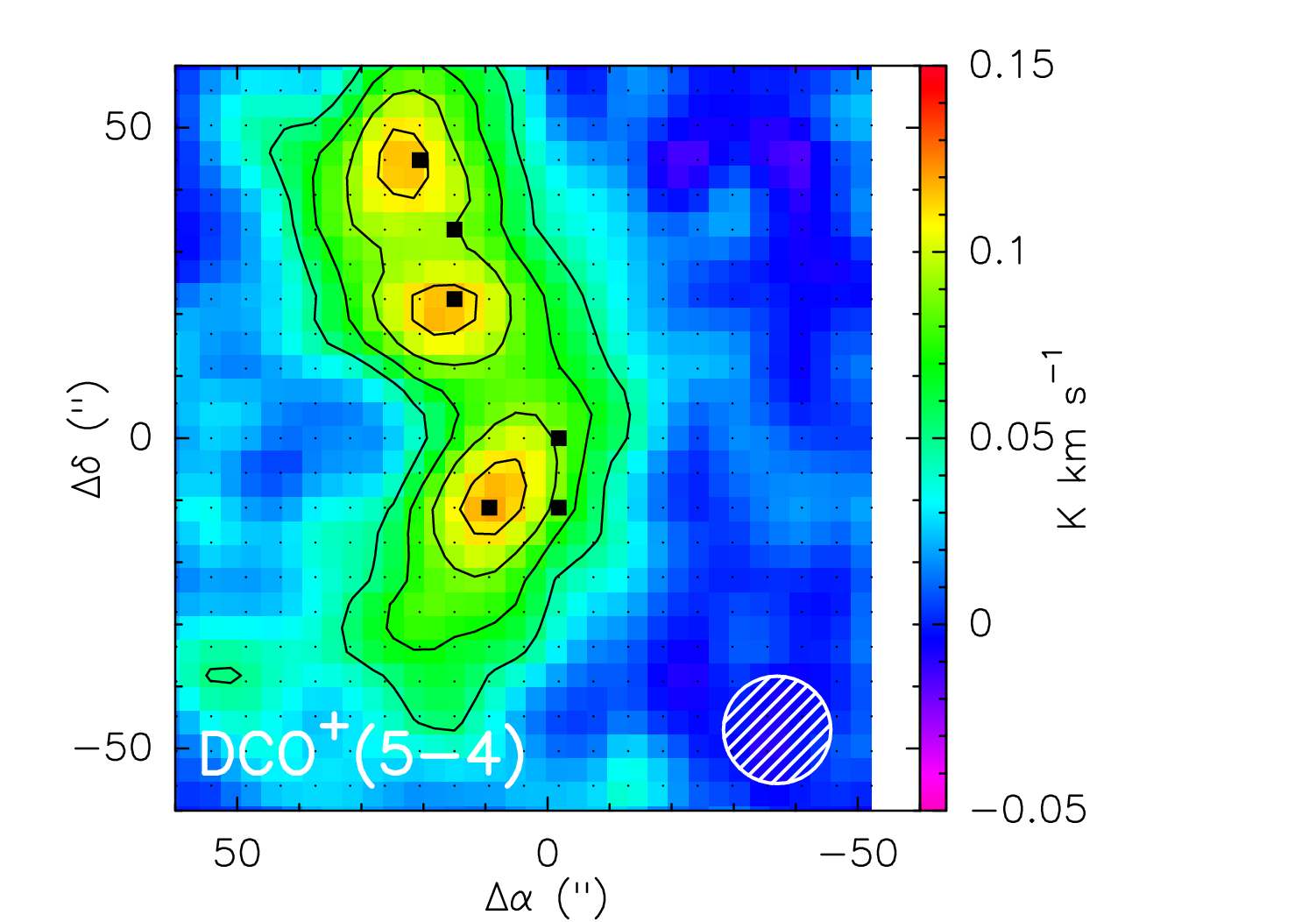}
\end{picture}}

\end{picture}
\caption{Observed (left) and modelled (right) integrated intensity maps of H-MM1 in o$\htwodplusline$ (top), $\diaz(4-3)$ (middle), and $\dcoplus(5-4)$ (bottom). The model assumes a cosmic-ray ionisation rate per $\htwo$ molecule of $\zeta_{\htwo}=7.4\times10^{-18}\,\pers$. The simulation time is $6.3\times10^5$\,yr. The o$\htwodplus$ map is as predicted by the model, whereas the modelled abundances of $\diaz$ and $\dcoplus$ are multiplied by factors 4.8 and 4, respectively, to reproduce the observed intensity levels.  { The intensity scale is $T_{\rm A}^*$.} The colour scales and the contours are the same as in Fig.~\ref{line_maps}. The squares indicate positions from which the spectra shown in Fig.~\ref{obs_model_spectra} are extracted.}
\label{obs_model_maps}
\end{figure*}

\begin{figure*}
\unitlength=1mm
\begin{picture}(160,230)(0,0)

\put(5,0){
\begin{picture}(0,0) 
\includegraphics[width=5cm,angle=0]{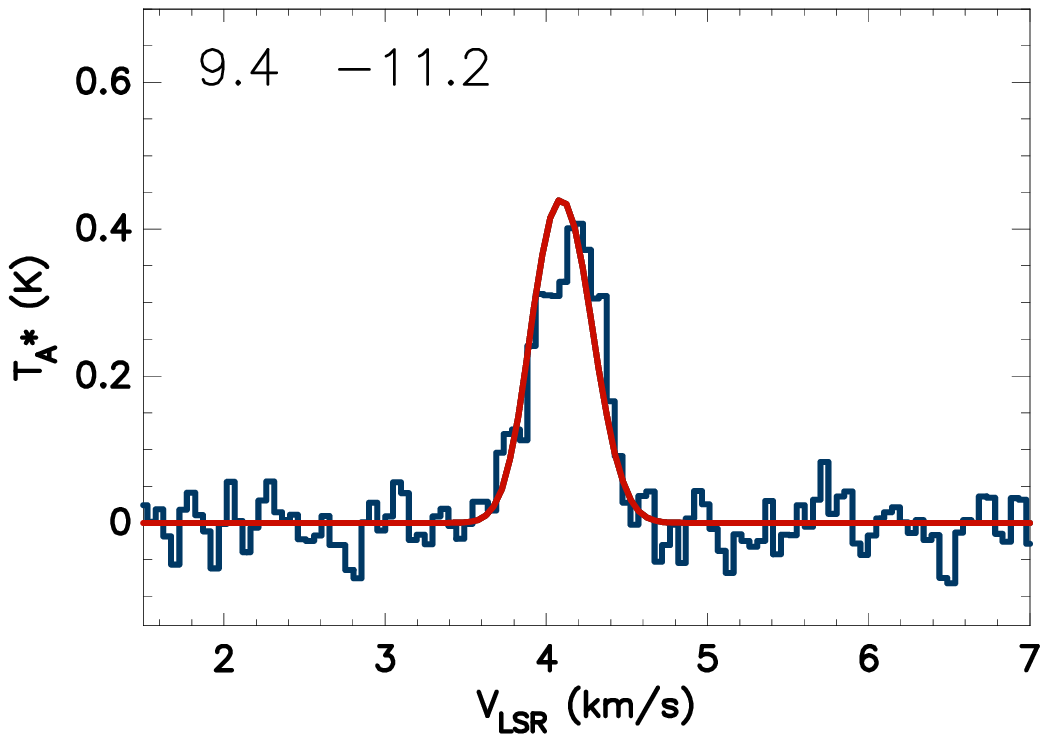}
\end{picture}}

\put(5,37){
\begin{picture}(0,0) 
\includegraphics[width=5cm,angle=0]{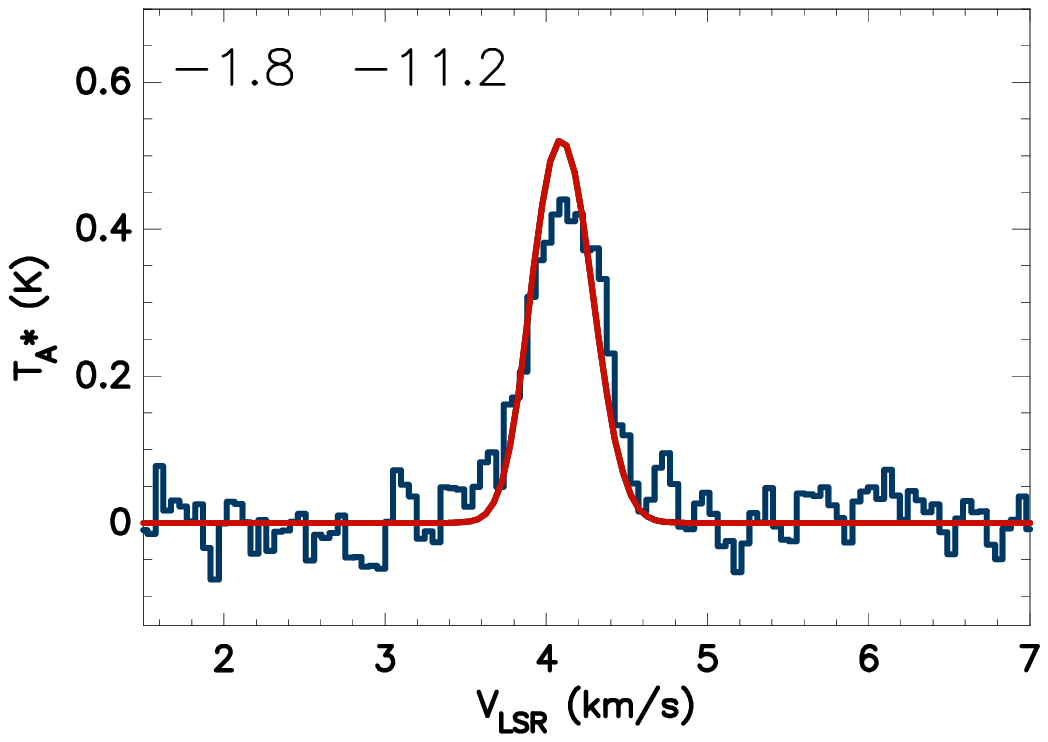}
\end{picture}}

\put(5,74){
\begin{picture}(0,0) 
\includegraphics[width=5cm,angle=0]{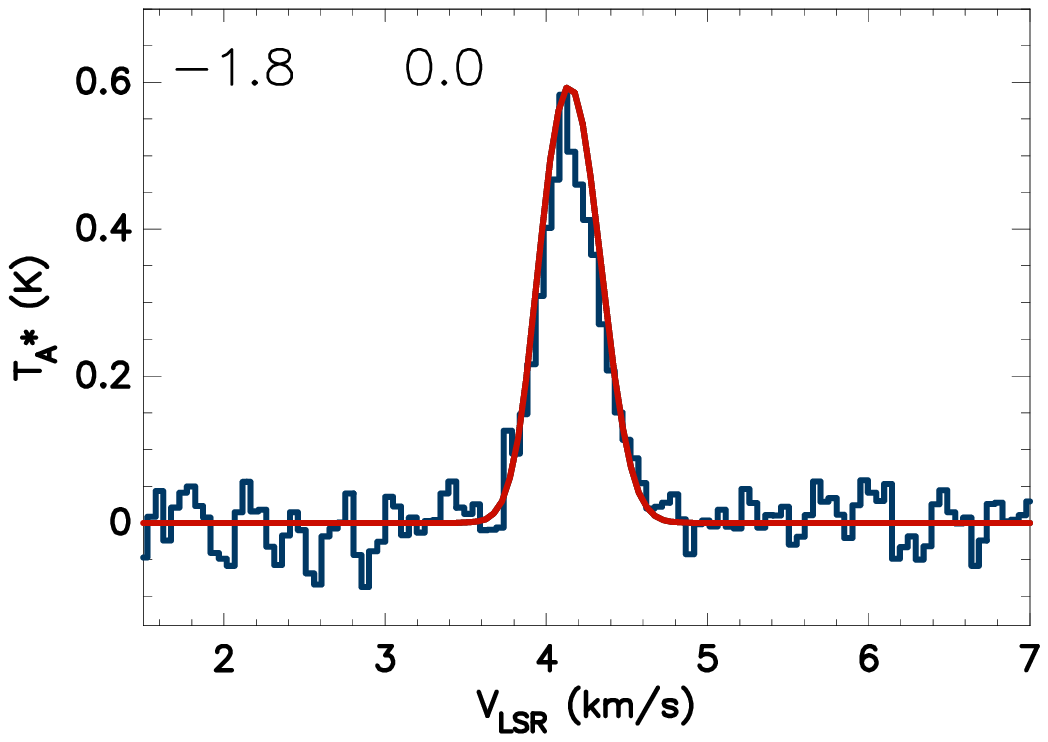}
\end{picture}}

\put(5,111){
\begin{picture}(0,0) 
\includegraphics[width=5cm,angle=0]{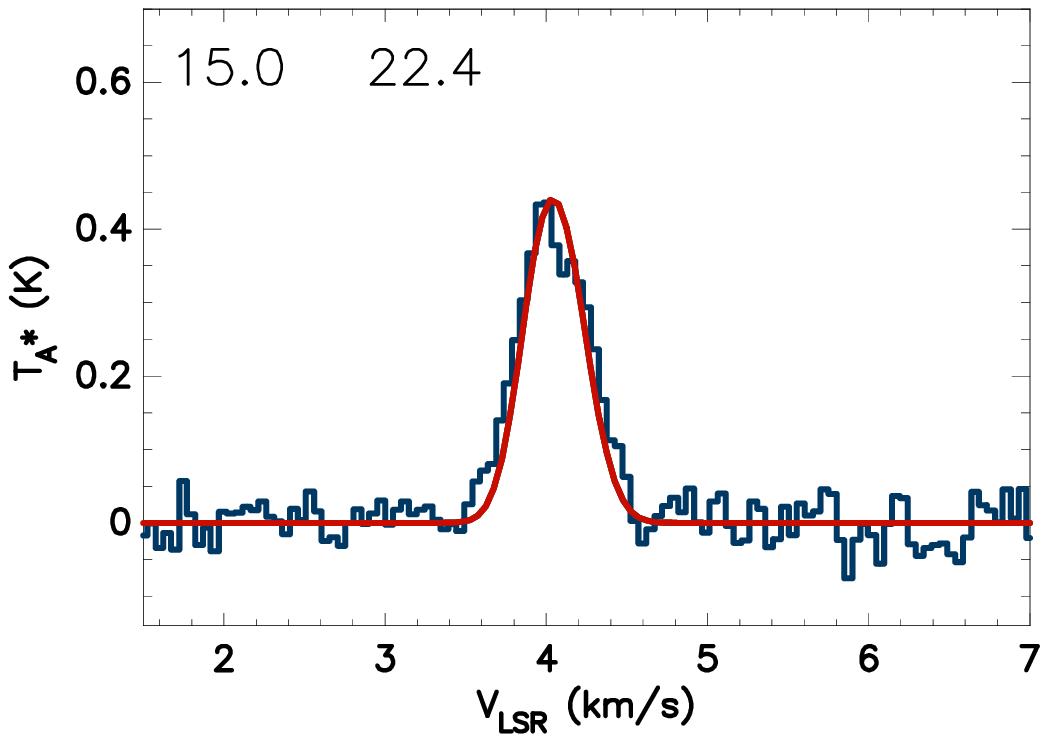}
\end{picture}}

\put(5,148){
\begin{picture}(0,0) 
\includegraphics[width=5cm,angle=0]{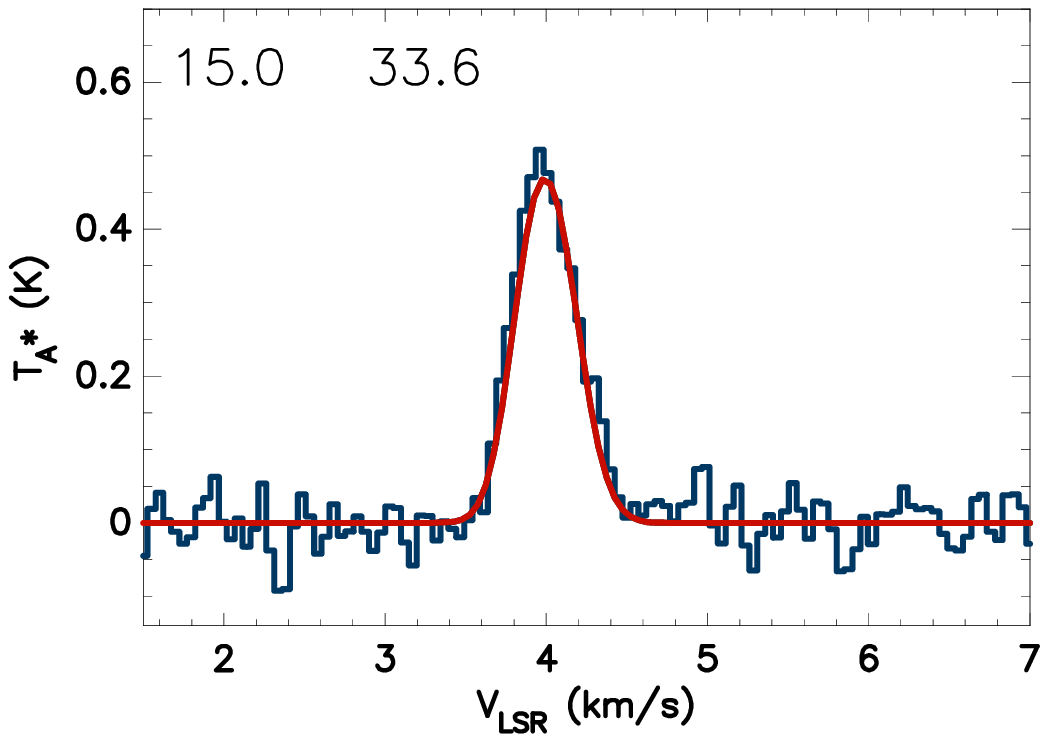}
\end{picture}}

\put(5,185){
\begin{picture}(0,0) 
\includegraphics[width=5cm,angle=0]{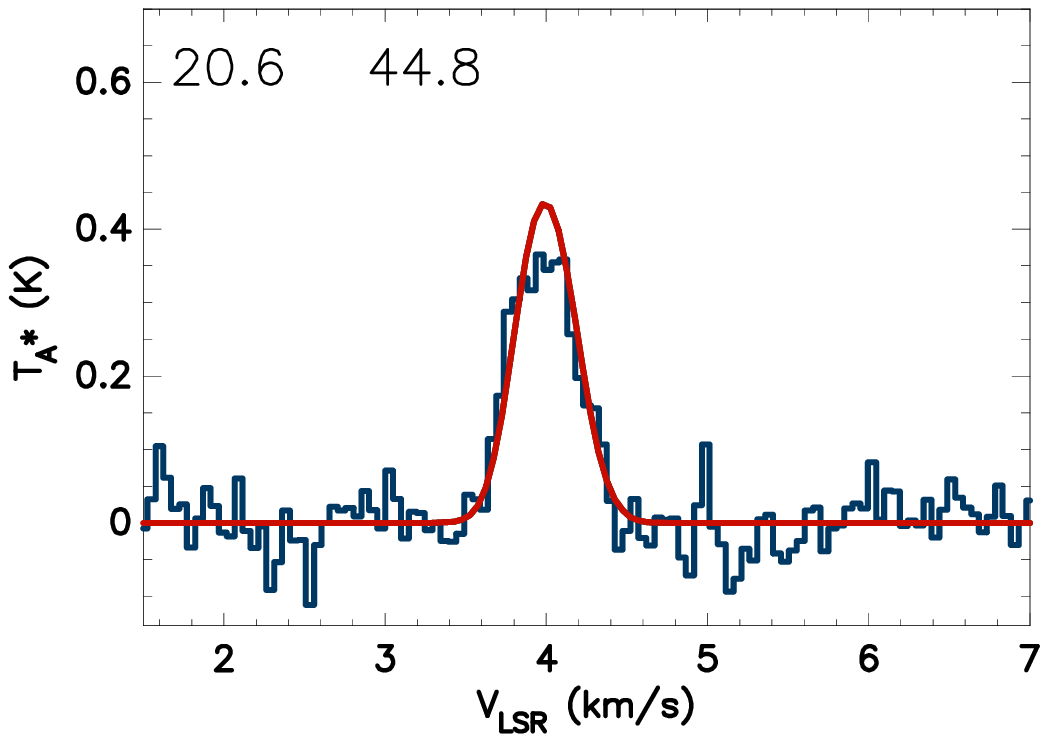}
\end{picture}}

\put(65,0){
\begin{picture}(0,0) 
\includegraphics[width=5cm,angle=0]{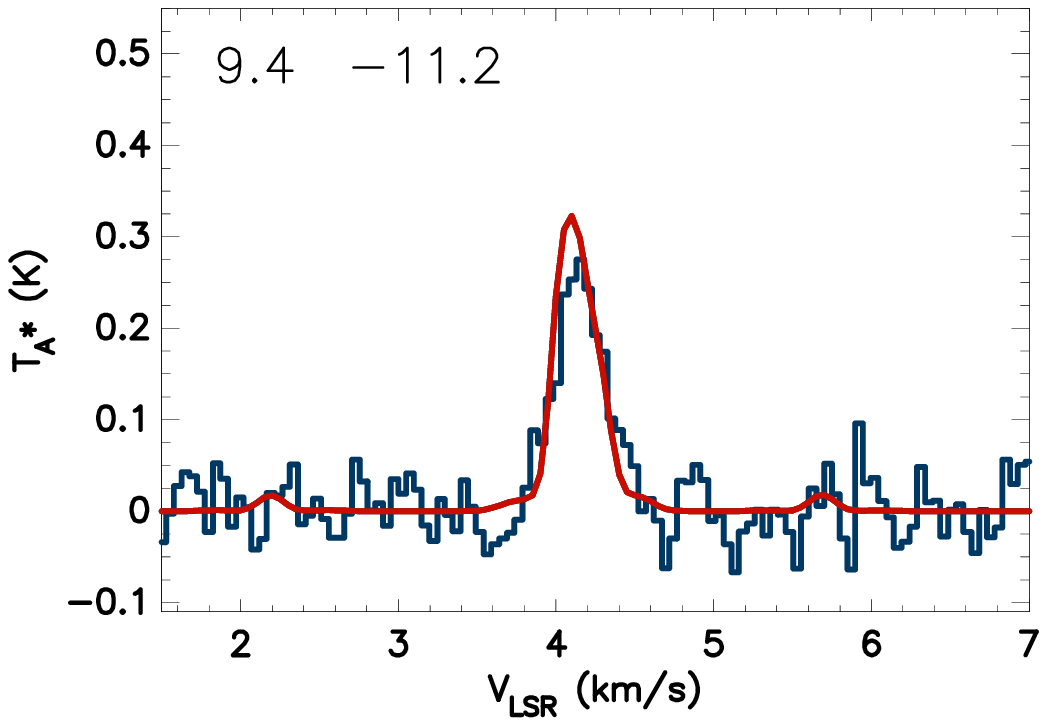}
\end{picture}}

\put(65,37){
\begin{picture}(0,0) 
\includegraphics[width=5cm,angle=0]{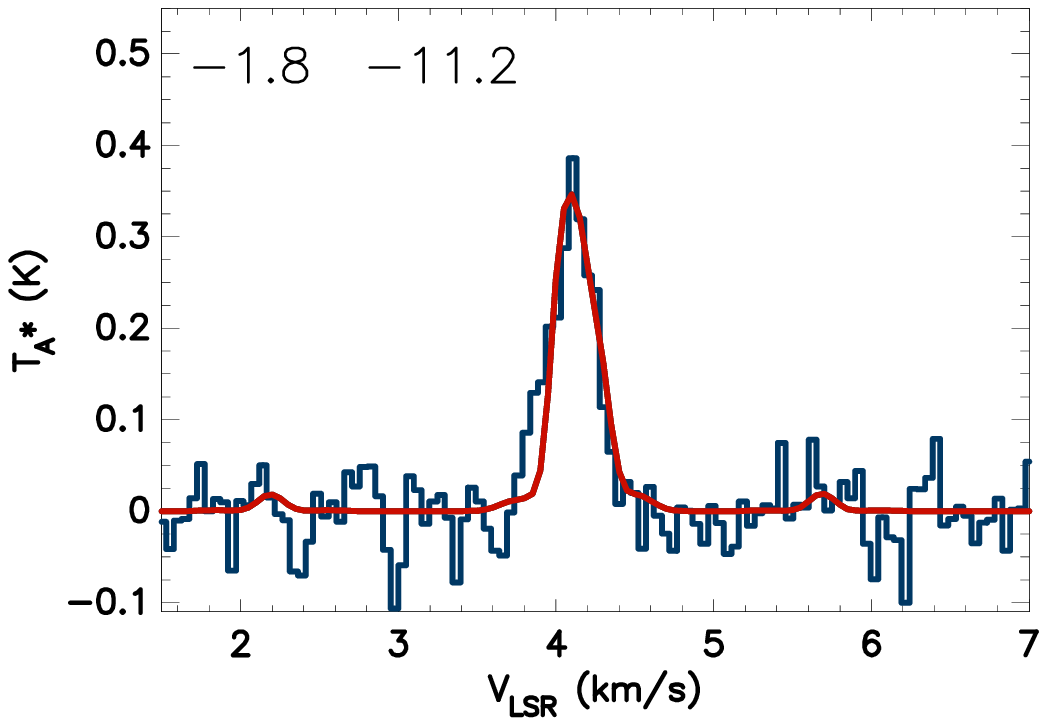}
\end{picture}}

\put(65,74){
\begin{picture}(0,0) 
\includegraphics[width=5cm,angle=0]{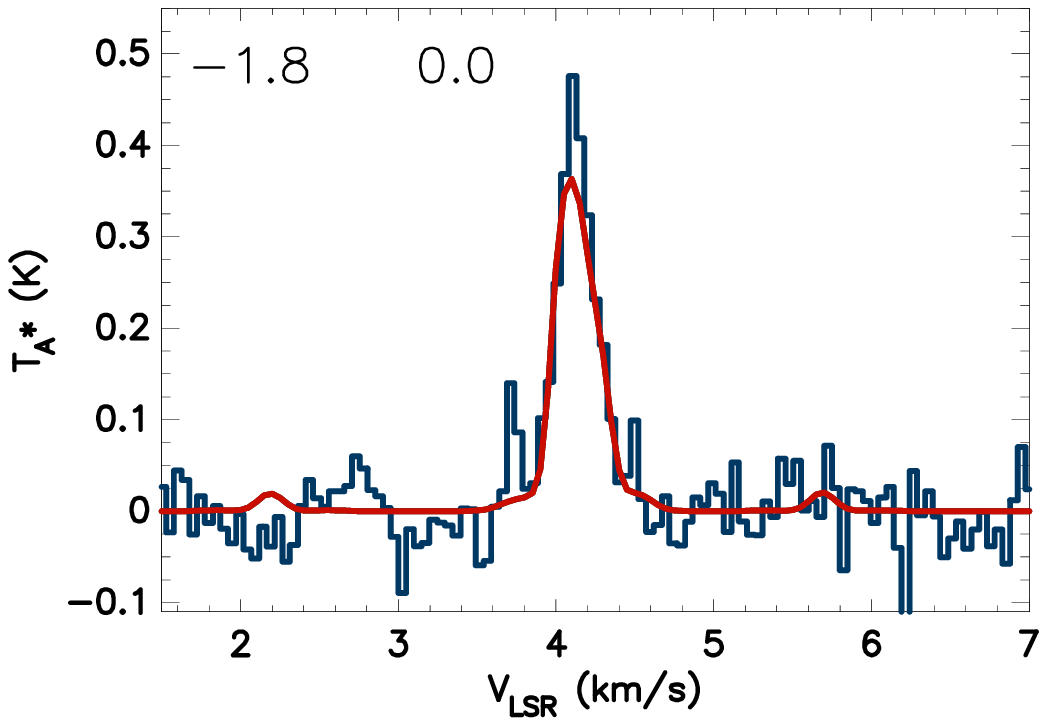}
\end{picture}}

\put(65,111){
\begin{picture}(0,0) 
\includegraphics[width=5cm,angle=0]{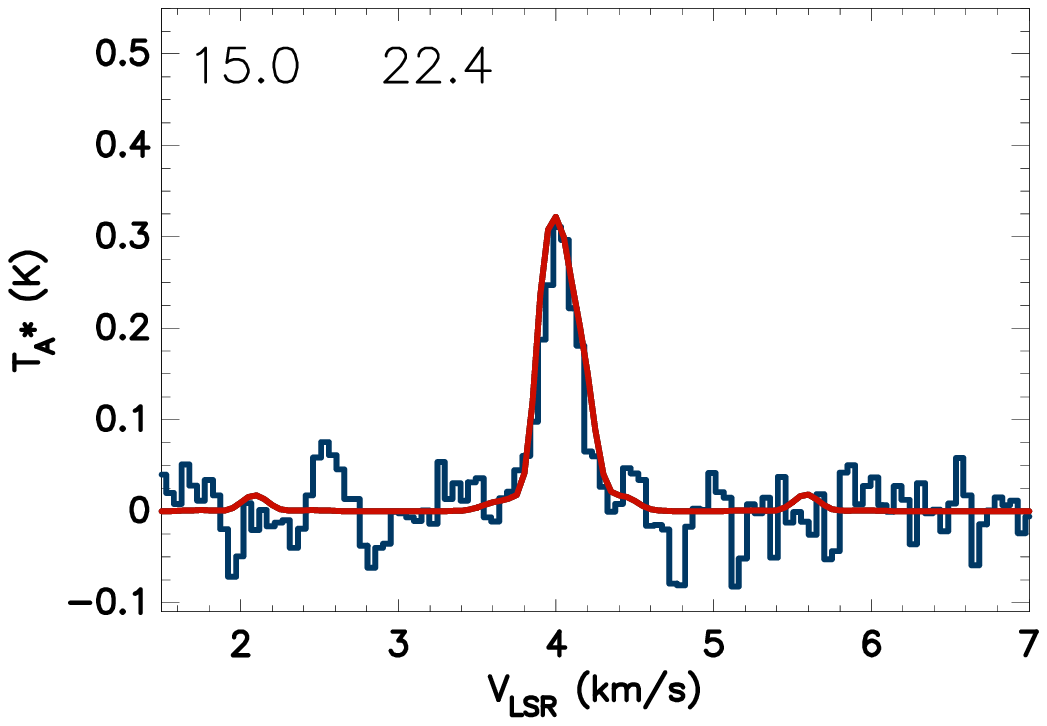}
\end{picture}}

\put(65,148){
\begin{picture}(0,0) 
\includegraphics[width=5cm,angle=0]{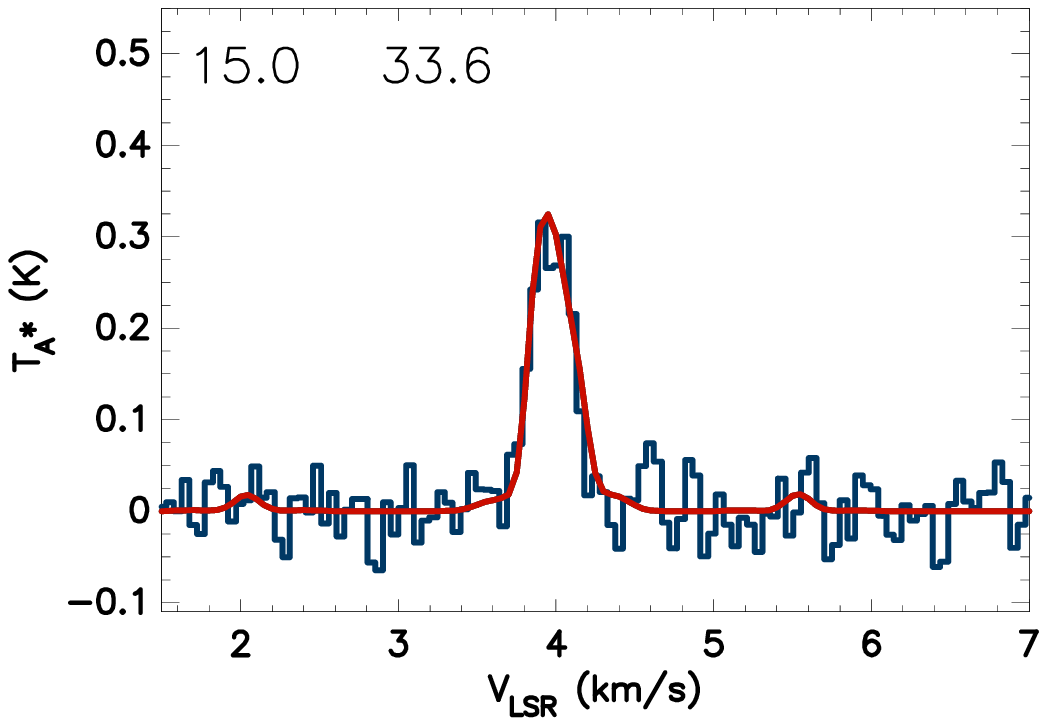}
\end{picture}}

\put(65,185){
\begin{picture}(0,0) 
\includegraphics[width=5cm,angle=0]{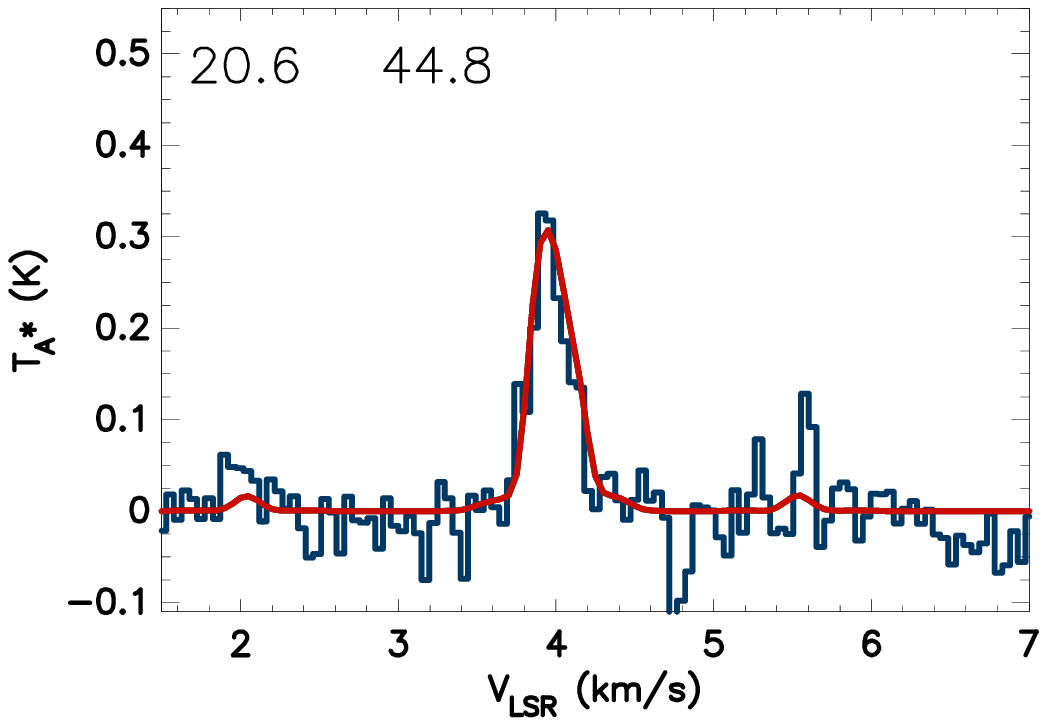}
\end{picture}}

\put(125,0){
\begin{picture}(0,0) 
\includegraphics[width=5cm,angle=0]{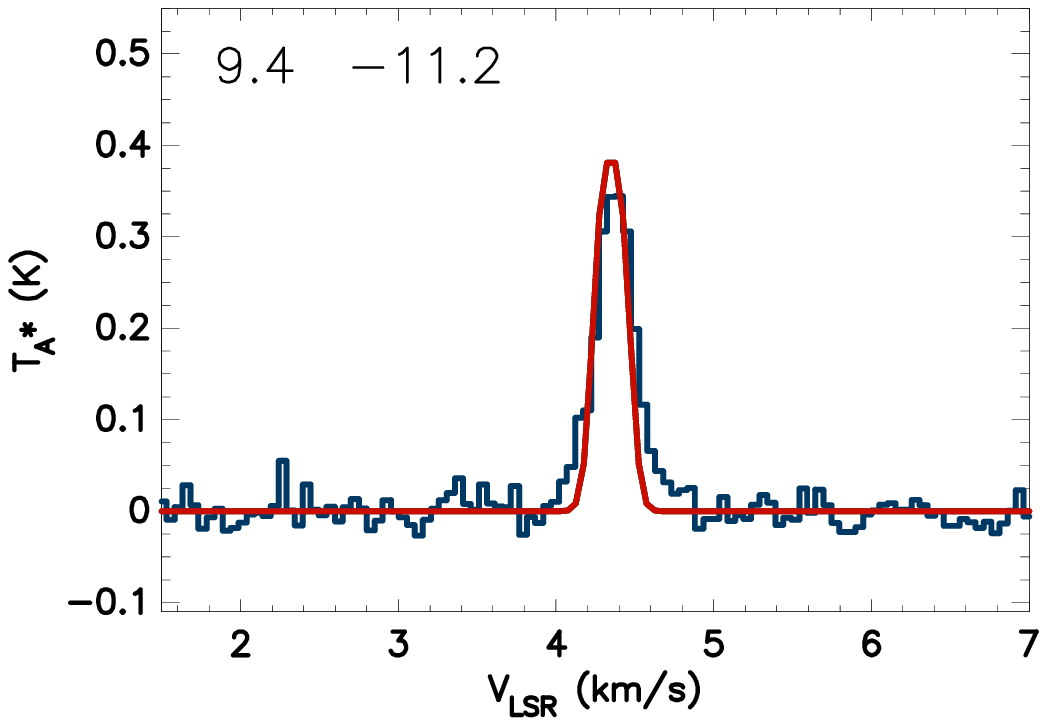}
\end{picture}}

\put(125,37){
\begin{picture}(0,0) 
\includegraphics[width=5cm,angle=0]{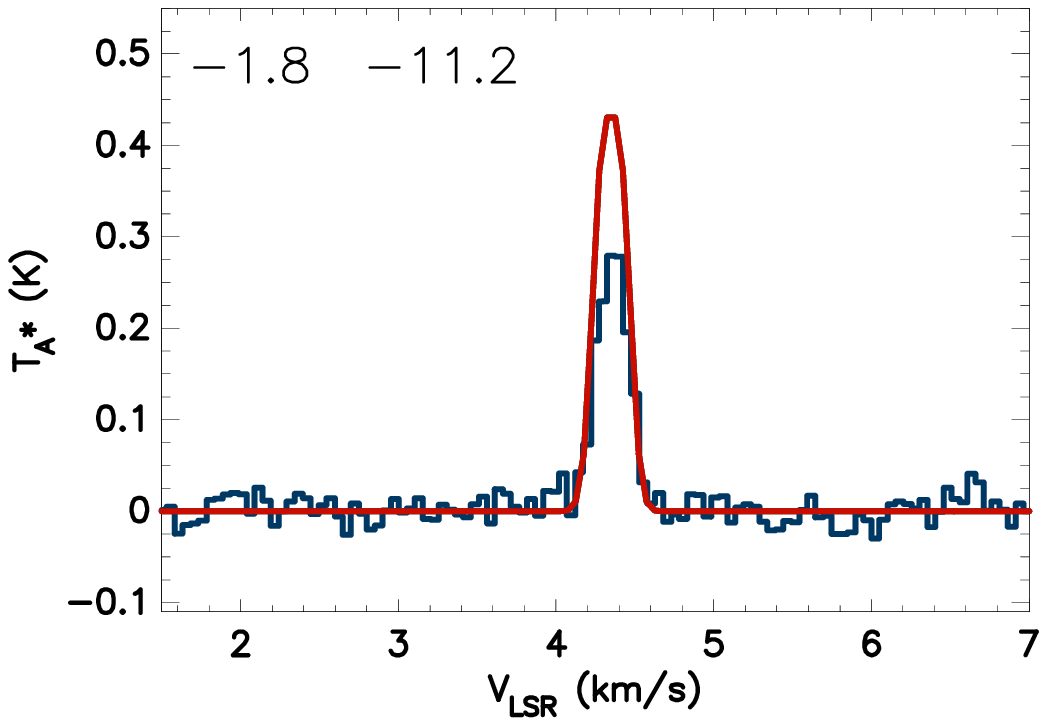}
\end{picture}}

\put(125,74){
\begin{picture}(0,0) 
\includegraphics[width=5cm,angle=0]{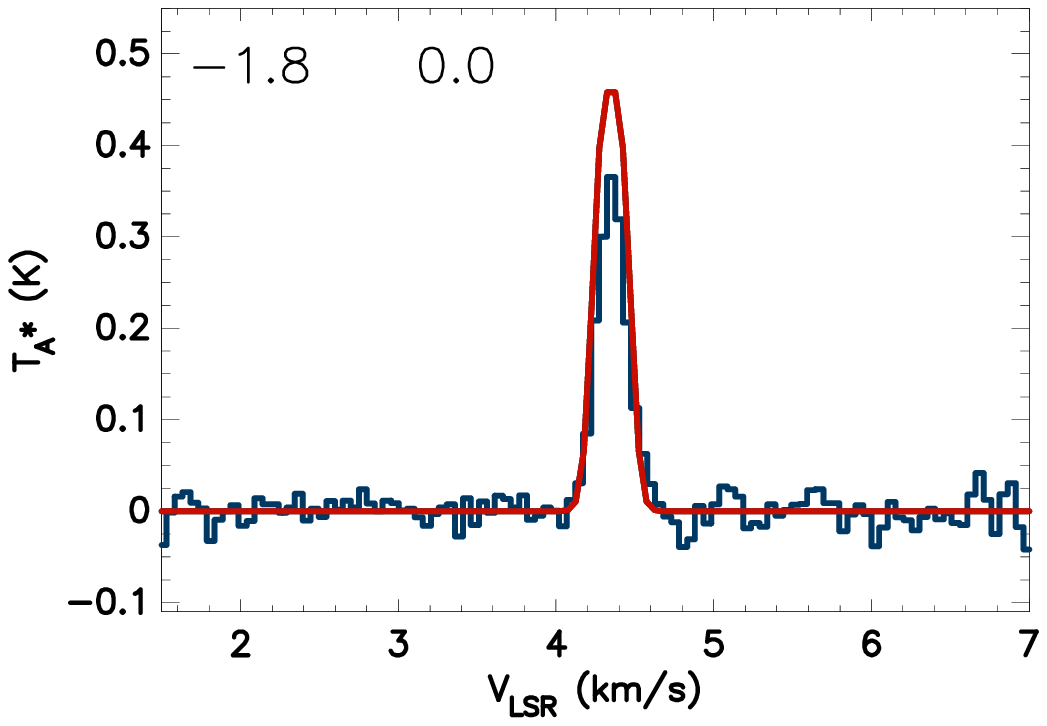}
\end{picture}}

\put(125,111){
\begin{picture}(0,0) 
\includegraphics[width=5cm,angle=0]{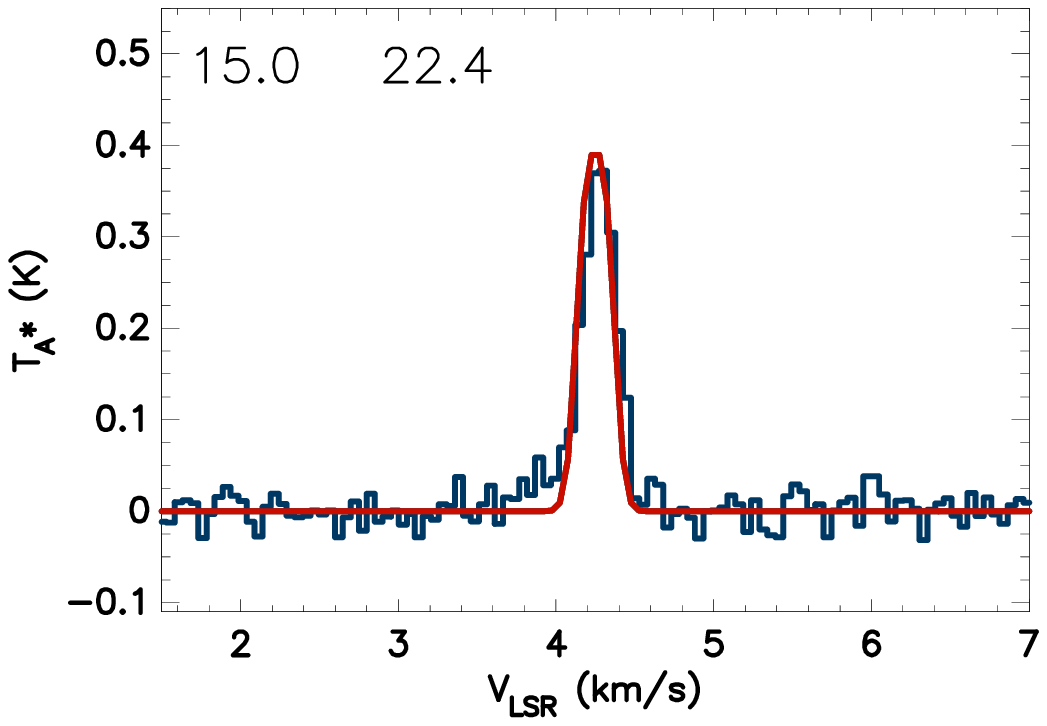}
\end{picture}}

\put(125,148){
\begin{picture}(0,0) 
\includegraphics[width=5cm,angle=0]{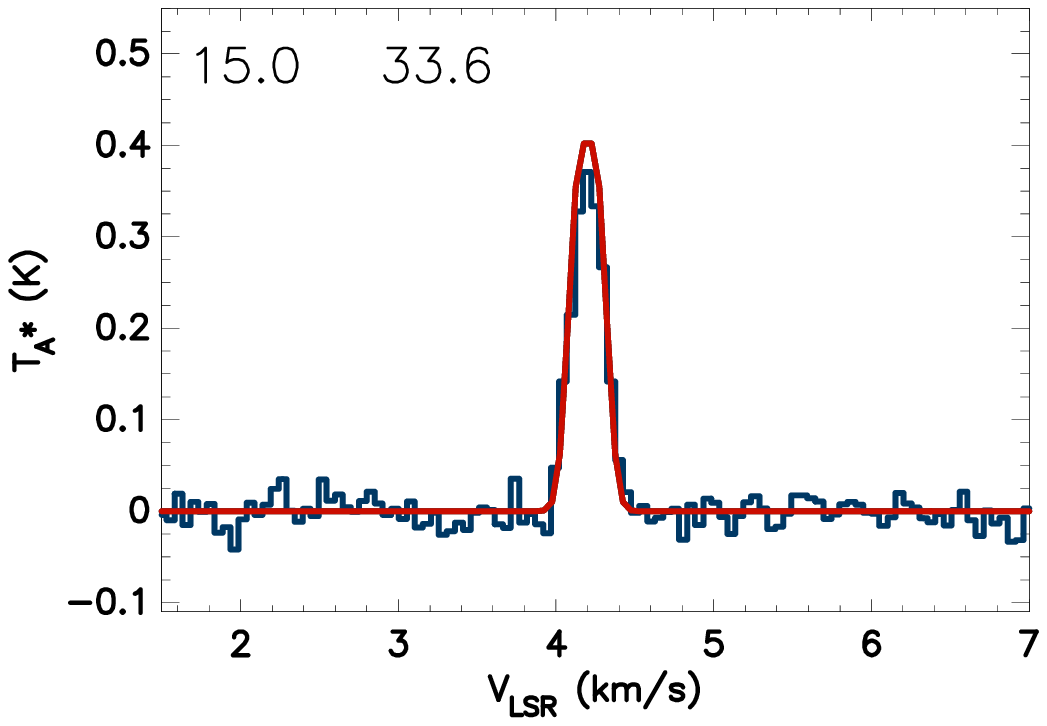}
\end{picture}}

\put(125,185){
\begin{picture}(0,0) 
\includegraphics[width=5cm,angle=0]{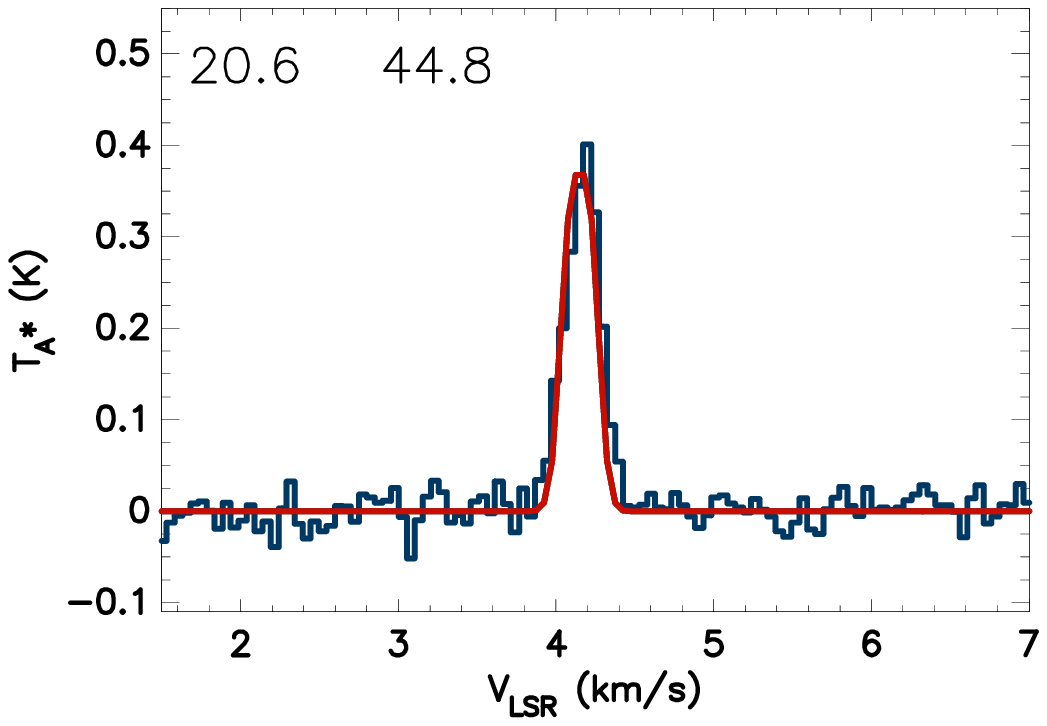}
\end{picture}}

\put(12,225){\large ${\rm o}\htwodplusline$}
\put(72,225){\large $\diaz(4-3)$}
\put(133,225){\large $\dcoplus(5-4)$}
\end{picture}
\caption{Observed and modelled o$\htwodplusline$ (left), $\diaz(4-3)$ (middle), and $\dcoplus(5-4)$ (right) spectra towards selected positions in H-MM1. The positions are indicated with square markers on the maps in Fig.~\ref{obs_model_maps}. The model spectra are shown with red. The offsets are with respect to the column density peak of the core. The model is the same as in Fig~\ref{obs_model_maps}, and also the scaling factors of the $\diaz$ and $\dcoplus$ abundances are the same. The intensity scale is $T_{\rm A}^*$.}
\label{obs_model_spectra}
\end{figure*}

The integrated intensity maps of the o$\htwodplusline$, $\diaz(4-3)$, and $\dcoplus(5-4)$ lines towards H-MM1 and the corresponding simulated maps from a three-dimensional model of the core are shown in Fig.~\ref{obs_model_maps}. Observed and simulated spectra in selected core positions are shown in Fig.~\ref{obs_model_spectra}. 

\section{Simulated maps from the model with a larger grain size}

\begin{figure*}
\unitlength=1mm

\begin{picture}(160,60)(0,0)
\put(-3,0){
\begin{picture}(0,0) 
\includegraphics[width=6cm,angle=0]{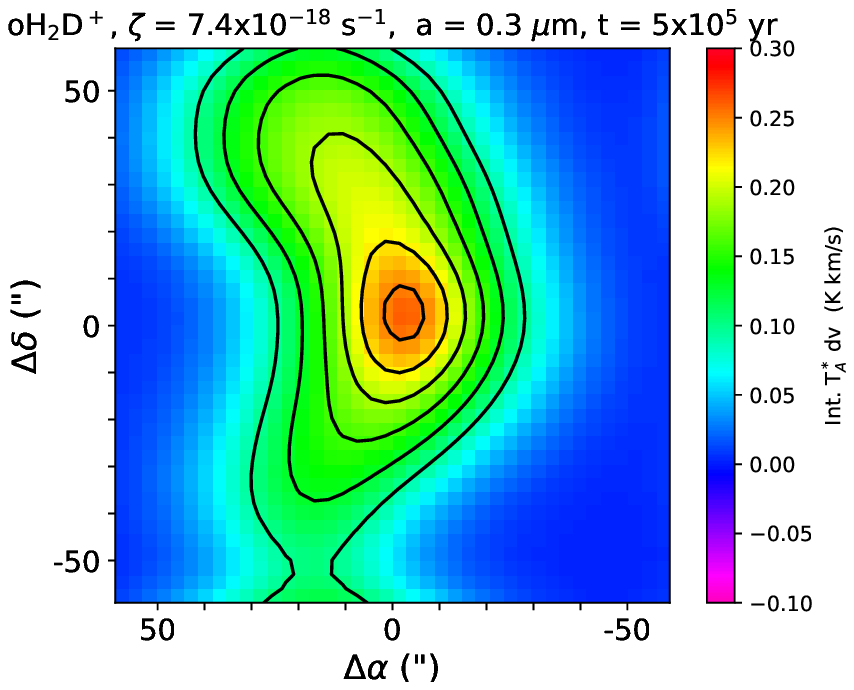}
\end{picture}}

\put(60,0){
\begin{picture}(0,0) 
\includegraphics[width=6cm,angle=0]{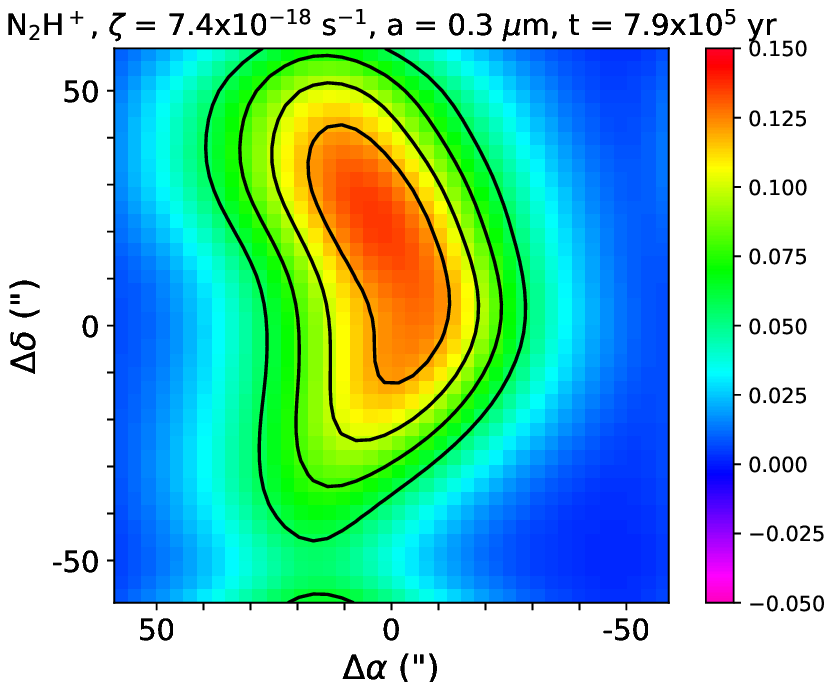}
\end{picture}}

\put(123,0){
\begin{picture}(0,0) 
\includegraphics[width=6cm,angle=0]{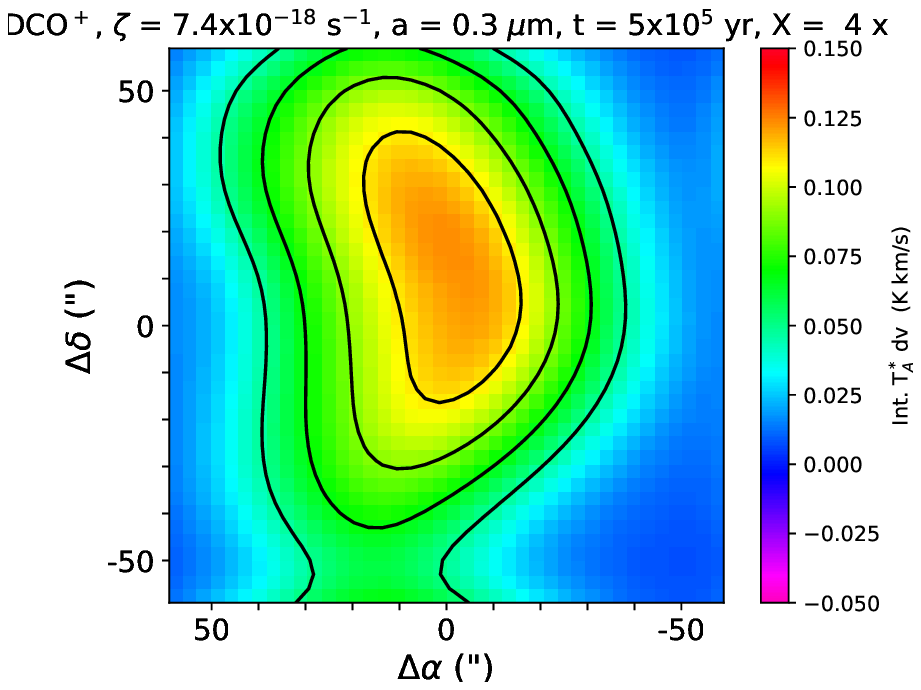}
\end{picture}}

\put(10,35){\bf \Large a}
\put(75,35){\bf \Large b}
\put(137,35){\bf \Large c}
\end{picture}

\caption{ Integrated o$\htwodplusline$, $\diaz(4-3)$ and $\dcoplus(5-4)$ maps from a model where the grain radius is assumed to be $a=0.3\,\mu$m. This model approximately reproduces the observed o$\htwodplus$ and $\diaz$ line intensities, but under-predicts the $\dcoplus$ intensity. In the core model shown in the figure, the fractional $\dcoplus$ abundance is multiplied by four. The simulation times are shown in the titles of the images.}
\label{model_maps_a3}
\end{figure*}

{ The integrated intensity maps calculated from the model with the grain radius $a=0.3\,\mu$m are shown in Fig.~\ref{model_maps_a3}. The o$\htwodplusline$ and $\diaz(4-3)$ maps are as predicted by the model, but the fractional $\dcoplus$ abundance is multiplied by four to achieve the observed peak intensity of the $J=5-4$ line. Moreover, the simulated $\htwodplusline$ and $\diaz(4-3)$ maps did not agree with the observations at the same time,  and the $\diaz$ map represents a later stage of evolution.} 

\section{Comparison with the methanol map}

\begin{figure*}
\unitlength=1mm
\begin{picture}(160,70)(0,0)
\put(110,0){
\begin{picture}(0,0) 
\includegraphics[width=8cm,angle=0]{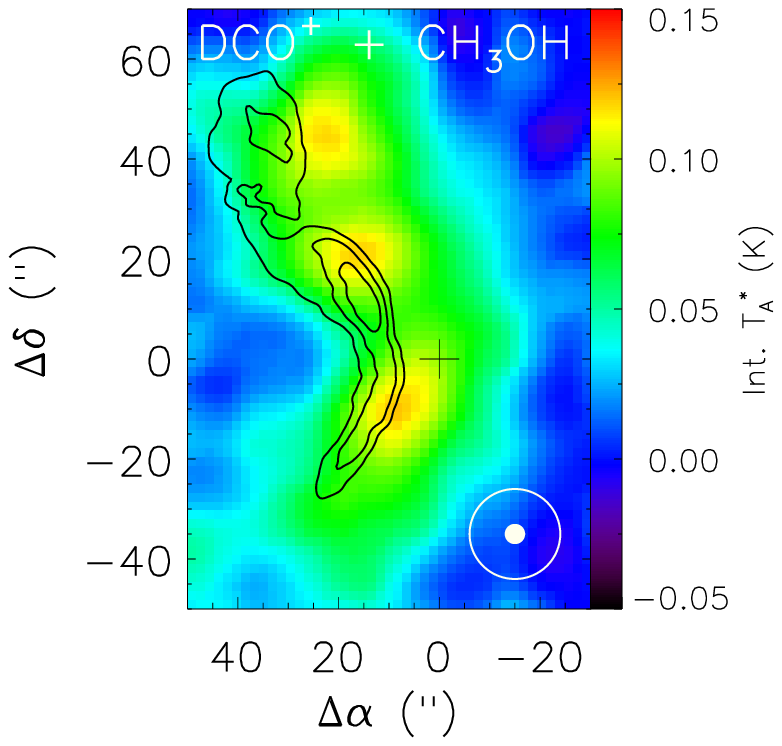}
\end{picture}}
\put(50,0){
\begin{picture}(0,0) 
\includegraphics[width=8cm,angle=0]{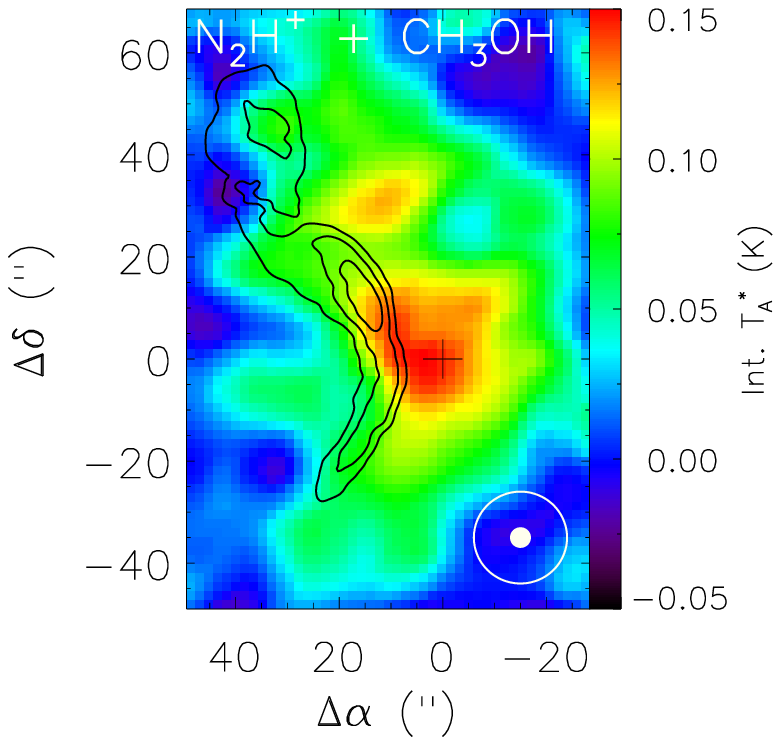}
\end{picture}}
\put(-10,0){
\begin{picture}(0,0) 
\includegraphics[width=8cm,angle=0]{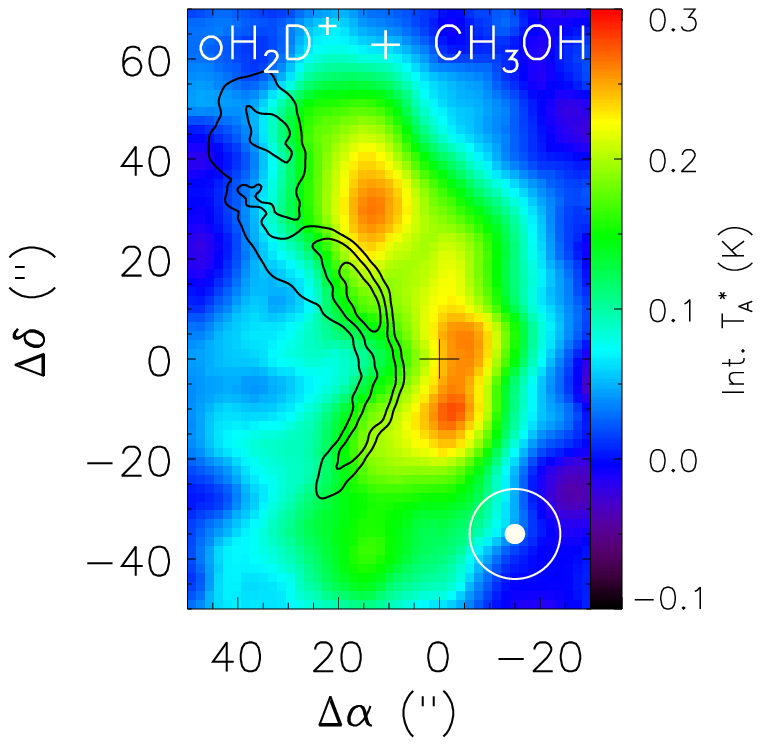}
\end{picture}}

\end{picture}  
\caption{Integrated $\meth(2_k-1_k)$ line intensity contours superposed on the o$\htwodplusline$, $\diaz(4-3)$, and $\dcoplus(5-4)$ maps of H-MM1. The contour levels are 2, 3, and 4 K\,$\kms$ (on the $T_{\rm B}$ scale). The methanol observations at 96.7\,GHz are from ALMA and have an angular resolution of $4\arcsec$. The colour scales of the pixel images are the same as in Fig.~\ref{line_maps}. The APEX and ALMA beams are shown in the bottom right.}
\label{line_maps+meth}
\end{figure*}

The distribution of strong methanol emission observed with ALMA is shown as contour plots on the o$\htwodplus$, $\diaz$, and $\dcoplus$ maps in Fig.~\ref{line_maps+meth}. The methanol data is from \cite{2020ApJ...895..101H}.  

\section{Predicted fractionation ratios}

\begin{figure*}
\unitlength=1mm

\begin{picture}(160,120)(0,0)
\put(10,60){
\begin{picture}(0,0) 
\includegraphics[width=8cm,angle=0]{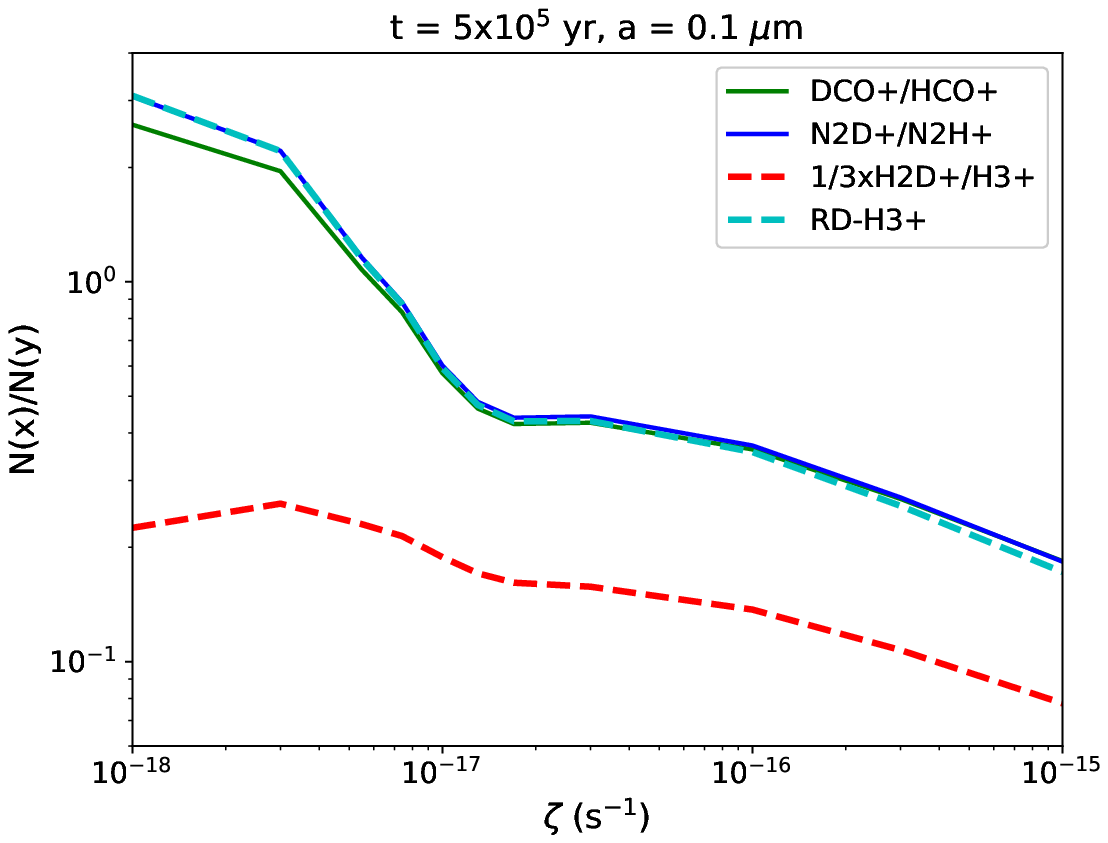}
\end{picture}}

\put(90,60){
\begin{picture}(0,0) 
\includegraphics[width=8cm,angle=0]{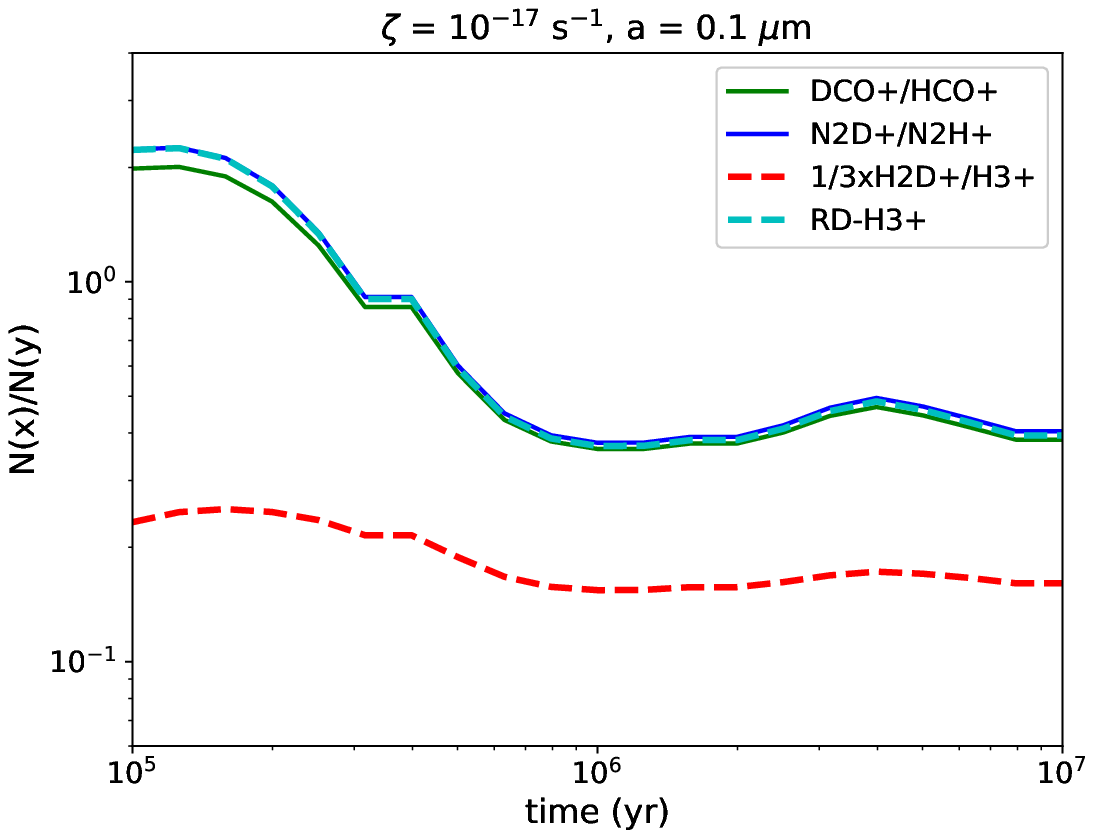}
\end{picture}}

\put(10,0){
\begin{picture}(0,0) 
\includegraphics[width=8cm,angle=0]{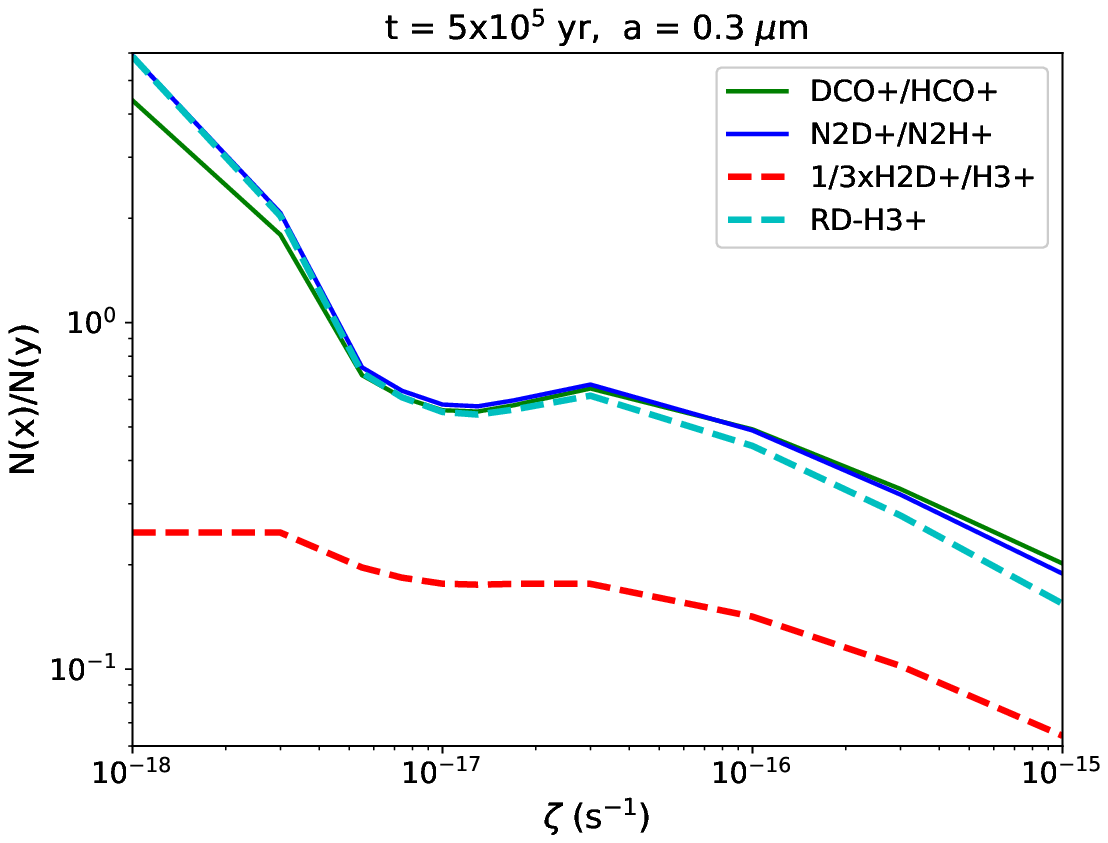}
\end{picture}}

\put(90,0){
\begin{picture}(0,0) 
\includegraphics[width=8cm,angle=0]{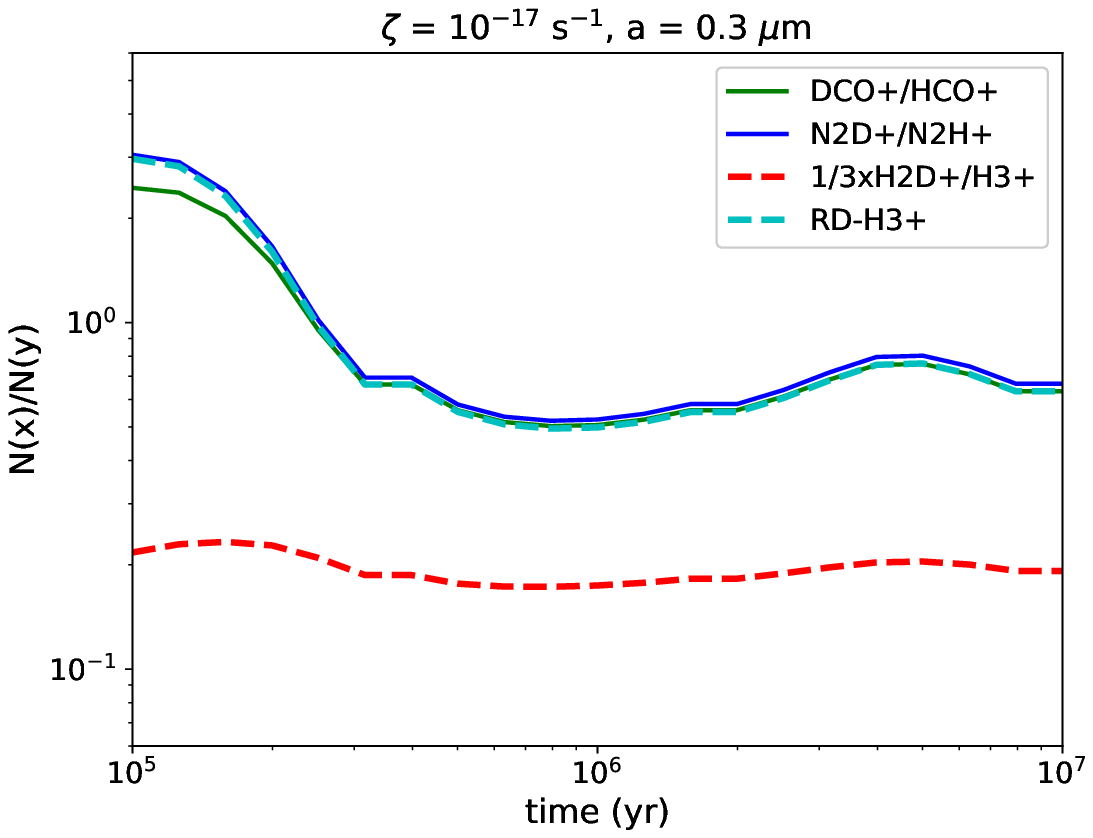}
\end{picture}}
\put(25,107){\bf \Large a}
\put(105,107){\bf \Large b}
\put(25,47){\bf \Large c}
\put(105,47){\bf \Large d}
\end{picture}
\caption{Fractionation ratios $\dcoplus/\hcoplus$ and $\ddiaz/\diaz$ as functions of $\zeta_{\htwo}$ (left) and time (right) according to our chemistry model. { The upper row ({\bf a} and {\bf b}) corresponds to a model with the grain radius $a=0.1\,\mu$m, whereas the simulation used for the bottom row ({\bf c} and {\bf d}) assumes $a=0.3\,\mu$m.} The cyan curve shows the prediction from a steady-state formula derived for chemical network including multiply deuterated forms (see Sect.~\ref{discussion}). The red dashed curve shows the ratio $1/3\times \htwodplus/\hthreeplus.$ }
\label{RD}
\end{figure*}

The modelled fractionation ratios $\dcoplus/\hcoplus$ and $\ddiaz/\diaz$ as functions of the cosmic-ray ionisation rate $\zeta_{\htwo}$ and the simulation time are shown in Fig~\ref{RD}, together with the prediction from { the steady-state formula accounting for multiply deuterated forms of $\hthreeplus$ (\citealt{2008A&A...492..703C}; \citealt{2009A&A...494..623P}; see Sect.~\ref{discussion})}. Also shown is the ratio $1/3\times\htwodplus/\hthreeplus$, which is often assumed to be equal to $\dcoplus/\hcoplus$ or $\ddiaz/\diaz$ in analytical methods of deriving $\zeta_{\htwo}$. The quantities presented in the figure are column density ratios along the line of sight going through the core centre. { The diagrams are shown for two models: one assuming a unique grain radius of $a=0.1\,\mu$m and another assuming $a=0.3\,\mu$m.}

\end{appendix}

\end{document}